\documentclass[12pt,letterpaper]{article}

\usepackage{graphicx, color}
\usepackage{dcolumn}
\usepackage{bm}
\usepackage[mathlines]{lineno}
\usepackage{amsmath}
\usepackage{amssymb}
\usepackage[numbers,sort&compress]{natbib}
\usepackage{bm}
\usepackage{float}
\usepackage{subcaption}
\usepackage{mathrsfs}
\usepackage{lipsum}
\usepackage{hyperref}
\usepackage[normalem]{ulem}
\hypersetup{
    colorlinks=true,       
    linkcolor=red,          
    citecolor=blue,        
    filecolor=magenta,      
    urlcolor=blue           
}
\usepackage[all]{hypcap} 
\usepackage{cleveref}

\newcommand{\beq}{\begin{equation}}
\newcommand{\eeq}{\end{equation}}

\def\lap{\lower.5ex\hbox{$\; \buildrel < \over \sim \;$}}
\def\gap{\lower.5ex\hbox{$\; \buildrel > \over \sim \;$}}

\def\be{\begin{equation}}
\def\ee{\end{equation}}
\def\ba{\begin{eqnarray}}
\def\ea{\end{eqnarray}}
\newcommand{\Acal}{\mathcal{A}}
\newcommand{\cmb}{{\mathrm{cmb}}}

\newcommand{\eref}[1]{Eq.~(\ref{#1})}

\newcommand{\fref}[1]{Fig.~\ref{#1}}
\newcommand{\sref}[1]{Sec.~\ref{#1}}
\newcommand{\aref}[1]{Appendix~\ref{#1}}

\newcommand{\rref}[1]{Ref.~\cite{#1}}

\newcommand{\pref}[1]{(\ref{#1})}

\title{\bf CMB birefringence from \\ ultralight-axion string networks}
\author{Mudit Jain, \ Andrew J. Long, \ and \ Mustafa A. Amin}
\date
{\small \it Department of Physics and Astronomy, Rice University, Houston, Texas 77005, U.S.A.}

\begin{document}

\maketitle

\begin{abstract}
The polarization of Cosmic Microwave Background (CMB) photons is rotated as they pass through (ultralight-) axion string loops.  Studying this birefringence can reveal valuable information about the axion-photon coupling and the structure of the string network. We develop an approximate analytic formalism and identify a kernel function that can be used to calculate the two-point correlation function for CMB birefringence induced by an arbitrary axion string network. Using this formalism, we evaluate the birefringence signal for some simple loop distributions (including scaling and network collapse). We find that the angular correlation function has a characteristic angular scale set by $\theta_\mathrm{min}$, which corresponds to the angular extent of the loops at the time of recombination. This results in a peak in the birefringence power spectrum around $\ell_p \sim 1/\theta_\mathrm{min}$.  An additional scale, controlled by the axion's mass, is introduced if the network collapses before today.  
\end{abstract}

\newpage
\begingroup
\hypersetup{linkcolor=black}
\tableofcontents
\endgroup

\setlength{\parindent}{25pt}
\setlength{\parskip}{1.2ex}

\section{Introduction}\label{sec:intro}

In this article we are interested in exploring cosmological signatures of ultralight axion-like particles.  
Originally, the axion was proposed to solve the strong CP problem~\cite{Wilczek:1977pj}, and it still provides one of the most compelling explanations for the tiny size of neutron's electric dipole moment~\cite{Crewther:1979pi,Afach:2015sja}.  
It also serves as a viable dark matter candidate~\cite{Preskill:1982cy,Abbott:1982af,Dine:1982ah}. 
More generally, theories with compactified extra dimensions such as string theory generally predict a wide spectrum of axion-like particles, which are associated with the size and shape of the compactified dimensions from the four-dimensional perspective~\cite{Witten:1984dg,Svrcek:2006yi}.  
This idea has been called the string axiverse~\cite{Arvanitaki:2009fg} (also see~\cite{Acharya:2010zx,Cicoli:2012sz}).  
Since the axions' masses are induced by non-perturbative effects, $m_a \propto e^{-S_E}$, one expects to find different species of axion particles whose masses vary over many orders of magnitude~\cite{Hui:2016ltb}.  
In this paper, we are primarily interested in \textit{ultralight} axions (ULAs) with masses below the Hubble parameter at recombination $m_a \lesssim H_\cmb$ where $H_\cmb\simeq 3 \times 10^{-29} \ \mathrm{eV}$.  

A variety of experiments are currently underway that seek to discover ultralight axions and to study their interactions with ordinary matter.  
Laboratory-based efforts~\cite{Spector:2016vwo,Beyer:2020dag}  are attempting to produce axions from intense electromagnetic fields, while axion helioscopes~\cite{Inoue:2008zp,Arik:2008mq,Anastassopoulos:2017ftl,Armengaud:2019uso} are searching for axions produced in our own Sun.  
Both of these techniques are sensitive to arbitrarily light axions, including ultralight axions.  
In fact the XENON1T experiment has recently reported an excess of electronic recoil events, which can be interpreted as evidence for an axion coupling to photons and electrons~\cite{Aprile:2020tmw}.  
Other experimental strategies, such as the production of axions at high-energy collider facilities~\cite{Bauer:2018uxu}, lose sensitivity as $m_a \to 0$ since the axion does not decay within the detector.  
Axions can also be tested in astrophysical `laboratories', and searches for axion emission from stellar environments has led to some of the strongest constraints on ultralight axions~\cite{Raffelt:2006cw,Fortin:2018aom,Fortin:2018ehg,Dessert:2019sgw,Dessert:2019dos,Buschmann:2019pfp,Dessert:2020lil}.  
Additionally, the phenomenon of gravitational superradiance causes axions to be created around rapidly-spinning black holes, leading to strong constraints in particular mass windows~~\cite{Arvanitaki:2009fg,Arvanitaki:2010sy,Kitajima:2018zco,Stott:2018opm}.  
Other experimental strategies are available for axion detection if a population of cold axion particles makes up the dark matter~\cite{Du:2018uak,Braine:2019fqb}, but ultralight axions can only provide a tiny fraction of the total dark matter abundance~\cite{Hu:2000ke}, since they would not gravitationally cluster on small length scales, leading to conflict with Lyman-$\alpha$ forest observations~\cite{Hui:1998hq,Irsic:2017yje}.
See Refs.~\cite{Marsh:2015xka,Grin:2019mub,Graham:2015ouw} for comprehensive reviews of experimental searches.  

Theories with ultralight axions can also predict a cosmological relic surviving in the Universe today -- namely, a network of cosmic strings (and possibly also domain walls) -- that provides an additional handle for axion detection~\cite{Vilenkin:1982ks}.  
Generally speaking, the axion is a Goldstone boson arising from the breaking of a global $\mathrm{U}(1)$ symmetry called the Peccei-Quinn (PQ) symmetry~\cite{Peccei:1977hh,Peccei:1977ur}. If the PQ symmetry was broken after inflation, then the corresponding cosmological phase transition would have filled the universe with cosmic strings~\cite{Kibble:1976sj,Kibble:1980mv,VilenkinShellard:1994}.  
The strings' gravitational influence on light and matter can be probed through distortions of the cosmic microwave background (CMB) radiation~\cite{Kaiser:1984iv,Lopez-Eiguren:2017dmc} and gravitational wave measurements (e.g., pulsar timing array or interferometry)~\cite{Chang:2019mza,Ramberg:2020oct,Blanco-Pillado:2013qja,Blanco-Pillado:2017rnf,Blanco-Pillado:2017oxo,Auclair:2019wcv,Gorghetto:2021fsn,Figueroa:2020lvo,Gelmini:2021yzu}.  
Additional signatures can arise in specific models~\cite{Fukuda:2020kym,Abe:2020ure,Agrawal:2020euj}.  

Beyond their gravitational signatures, the presence of cosmic axion strings in our Universe today opens a new channel for the discovery of ultralight axions through the phenomenon of cosmological birefringence.  
The ultralight axion can couple to electromagnetism through a topological Chern-Simons interaction term in the Lagrangian, $\mathscr{L} \sim g_{a\gamma\gamma} a F \tilde{F}$.  
Consequently, a classical axion field $a(\vec{x},t)$ causes the polarization axis of a propagating electromagnetic wave to rotate by an angle $\Delta \Phi \sim g_{a\gamma\gamma} \Delta a$, which is the phenomenon of axion-induced birefringence~\cite{Carroll:1989vb,Carroll:1991zs,Harari:1992ea,Carroll:1998bd}.  
The cosmic microwave background (CMB) radiation, since it is a diffuse source that can be measured with high precision, provides an excellent probe of axion-induced birefringence.  

The effect of axion-induced birefringence on the CMB can be inferred from measurements of the CMB polarization power spectra.  
It is not possible to directly measure $\Delta\Phi$ at any given point on the sky, since the polarization axis of a CMB photon at its time of emission is not known (although it is correlated with the local temperature quadrupole anisotropy).  
Nevertheless, the phase shifts resulting from axion-induced birefringence are correlated across the sky, and they leave their imprint on the CMB polarization anisotropies.  
For instance, birefringence converts E-mode polarization into B-mode polarization~\cite{Lue:1998mq,Liu:2006uh,Gluscevic:2012me,Contreras:2017sgi}, leading to an enhanced B-mode power spectrum as well as EB and TB cross correlations.  In particular, these parity-odd cross correlations are the result of the parity-violating axion field background, whereas other sources of B-mode power, such as primordial gravitational waves, need not be accompanied by parity-odd cross correlations.  Then ratios, such as EB/EE or TB/TE, have a linear dependence on the birefringence angle.  
More generally, probing cosmic birefringence is an important science driver for the next generation of CMB telescopes~\cite{Abazajian:2016yjj}. 
In fact recently, the authors of~\cite{Minami:2020odp} pointed out that Planck 2018 data~\cite{Aghanim:2018eyx} contains weak evidence ($2.4\sigma$) for isotropic birefringence; see also Refs.~\cite{Takahashi:2020tqv,Fujita:2020ecn,Nakagawa:2021nme} for some plausible explanations using ULAs.

The effect of axion strings on CMB birefringence was studied recently by Agrawal, Hook, and Huang~\cite{Agrawal:2019lkr}.
A CMB photon passing through an axion string loop has its polarization axis shifted by an angle $\Delta \Phi \sim g_{a\gamma\gamma} f_a$ where $g_{a\gamma\gamma}$ is the axion-photon coupling parameter and $f_a$ is the axion decay constant. 
In most models $g_{a\gamma\gamma} \propto \Acal / f_a$, and so it is remarkable to note that the birefringence signal is insensitive to the PQ scale $f_a$, and thus the string tension $\mu$, but rather directly probes the anomaly coefficient $\Acal$!
The authors of Ref.~\cite{Agrawal:2019lkr} calculated the birefringence power spectrum for a simple model of the axion string network, and they identified that axion string-induced birefringence is within reach of the next generation of CMB telescopes including CMB-S4.

The string-induced birefringence signal also depends upon the structure of the axion string network, which is not understood very well. In general, a string network contains long strings that cross the Hubble volume and sub-Hubble-scale string loops.  The number of long strings and the length distribution of string loops both evolve with time, and this evolution has been extensively studied using analytical methods and numerical simulations.  However a controversy has developed recently in regard to whether the string network's evolution exhibits a property called scaling~\cite{VilenkinShellard:1994} in which the length scale of the string network tracks the slowly-evolving cosmological Hubble scale.  
Some studies~\cite{Yamaguchi:1998gx,Yamaguchi:2002sh,Hiramatsu:2010yu,Hiramatsu:2012gg,Kawasaki:2014sqa,Lopez-Eiguren:2017dmc,Hindmarsh:2019csc,Hindmarsh:2021vih} conclude that these string networks exhibit scaling, while other studies~\cite{Gorghetto:2018myk,Vaquero:2018tib,Kawasaki:2018bzv,Martins:2018dqg,Buschmann:2019icd,Klaer:2019fxc,Gorghetto:2020qws} find a logarithmic (or a milder) deviation away from scaling. 
Even a small deviation away from scaling can have a large impact on the number of strings in the Universe today.  
While the community works to settle this controversy, it is important to assess the impact of different string network models on the birefringence signal.

Building upon the work of~\cite{Agrawal:2019lkr}, our goal is to develop an analytical formalism to calculate cosmic birefringence due to string loop networks, and also accommodate a broader class of them.
Here we provide a brief summary of some of our key results: 
(1) There exists a `universal' kernel function that can be used to compute the signal due to different string networks (in particular different string loop length distributions); 
(2) The typical size of loops in the string network present at recombination, results in an approximately constant correlation function $\langle \Delta\Phi\,\Delta\Phi\rangle$ as the opening angle becomes smaller than some characteristic angular scale $\theta_\textrm{min}$ set by such loop radii divided by $H^{-1}_\cmb$. This leads to a peak in the birefringence power spectrum at the corresponding multipoles $\ell_p \sim \pi/
\theta_\mathrm{min}$; 
(3) We find that the correlation function and angular power spectrum have an approximately universal shape (for approximately single-scale models), with the amplitude and width determined by the characteristic loop length;
(4) For reasonable string network models, the largest contribution to the total birefringence comes from nearly Hubble-scale loops (and also `long' strings to the extent that they can be modelled as Hubble-scale loops);
(5) For string networks that collapse sometime between recombination and today, there exists another scale $\ell_{m_a} \propto 1/\theta_c$ in the birefringence power spectrum. The angle $\theta_c$ corresponds to the angular extent of typical loops present at the time of collapse. For $\ell < \ell_{m_a}$ the power spectrum increases like $\sim \ell^2$, while it has a similar behavior as in (2) for higher $\ell$.

The rest of the paper is organized as follows.  
We begin in \sref{sec:framework} by discussing the model and system of interest: an ultralight axion that couples to photons, forms a cosmological axion-string network, and induces birefringence in the CMB radiation.  
Next in \sref{sec:analytic} we review the loop-crossing model formalism for calculating the two-point correlation function of the birefringence signal.  
In particular we identify a kernel function that appears in this calculation, and we derive an analytic estimate to the kernel function, which is also validated against direct numerical integration.  
Our analytic results for the kernel function are used in \sref{sec:models} to evaluate the birefringence signal for several different models of the axion string network, compared against the direct respective numerical results. We also compare the predicted signal against current and projected constraints on CMB anisotropic birefringence.  
Finally in \sref{sec:conclusion} we discuss and summarize our key results.  
The article includes three appendices: \aref{app:derive_Delta_Phi} provides a derivation of the axion-induced CMB birefringence effect; \aref{app:biref_washout} provides an estimate of CMB birefringence due to axions particles produced by domain wall collapse; and \aref{app:domain_walls} provides a discussion of CMB birefringence in a model with stable domain walls.  

\section{Theoretical framework}\label{sec:framework}

In this section we discuss the axion model under consideration, the phenomenon of axion-induced birefringence, and the cosmological network of axion strings.  

\subsection{An ultralight axion coupled to light}\label{sub:axion}

The axion's interaction with electromagnetism is captured by a term in the Lagrangian 
\begin{align}
    \mathscr{L}_\mathrm{int} = - \frac{1}{4} \, g_{a\gamma\gamma} \, a \, F_{\mu\nu} \tilde{F}^{\mu\nu}
\end{align}
where $a(x)$ is the pseudoscalar axion field, $F_{\mu\nu}(x)$ is the electromagnetic field strength tensor, and $\tilde{F}^{\mu\nu}(x) = 1/2 \, \epsilon^{\mu\nu\rho\sigma} F_{\rho\sigma}$ is the dual tensor. 
In the simplest models, the axion's interaction with  electromagnetism is induced by an anomaly, and for these models we can write the coupling as $g_{a\gamma\gamma} = - \Acal \alpha_\mathrm{em} / \pi f_a$ where $\Acal = C_\gamma/2$ is the anomaly coefficient, $\alpha_\mathrm{em} \simeq 1/137$ is the electromagnetic fine structure constant, and $f_a$ is the Peccei-Quinn scale.  
Typically $\Acal$ is a $\mathcal{O}(1)$ rational number.  
A variety of probes constrain the axion-photon coupling $g_{a\gamma\gamma}$.  
Most notable are limits from the CAST helioscope which imposes $|g_{a\gamma\gamma}| \lesssim 0.66 \times 10^{-10} \ \mathrm{GeV}^{-1}$ for axion masses below roughly $10^{-2} \ \mathrm{eV}$~\cite{Anastassopoulos:2017ftl}.  

\begin{figure}[t]
\centering
\includegraphics[width=1\columnwidth]{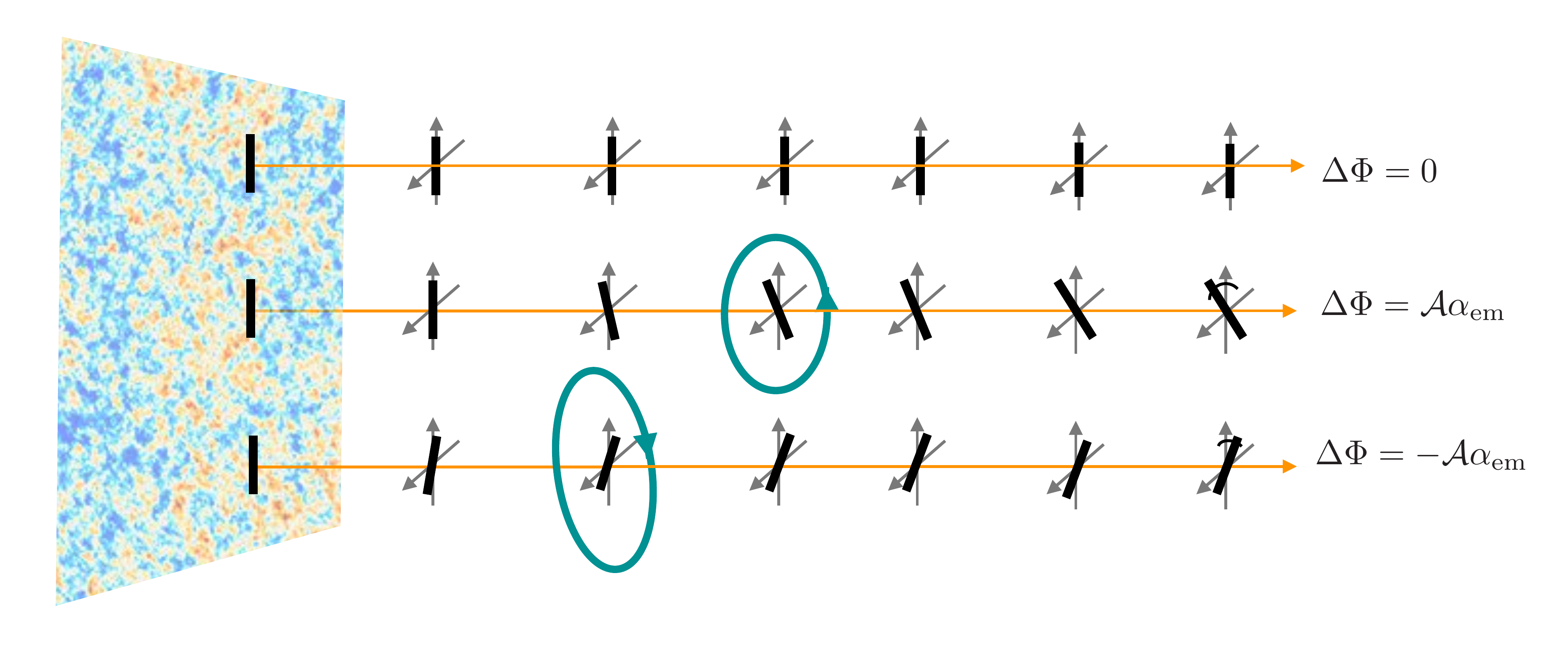}\vspace{0.1cm}\\
\caption{The polarization of CMB photons (indicated by the black bar) passing through axionic cosmic string loops (shown in green) is rotated by an angle $\Delta \Phi$. The rotation accumulates along the photon's path (orange arrow) as $\Delta \Phi=\mathcal{A}\alpha_{\rm em}/(2\pi f_a)\int_{x_i}^{x_f}\mathrm{d}x \cdot \partial_x a$, where $a$ is the axion field, $\mathcal{A}=\mathcal{O}(1)$ and $\alpha_{\rm em}\approx 1/137$. Along the photon path passing through the loop,  $|a(x_f)-a(x_i)|=2\pi f_a$ when the points $x_i$ and $x_f$ are sufficiently far from the loop. }
\label{fig:birefringence}
\end{figure}

\subsection{Axion-induced birefringence}\label{sub:birefringence}

Since the axion is an ultralight particle, it is easy to arrange systems where the occupation number of the field is high, and the field admits a classical description. 
As a photon passes through a classical axion field, it will experience a rotation of its polarization axis, a phenomenon known as birefringence. The polarization rotation angle that results from this axion-induced birefringence effect is~\cite{Carroll:1989vb,Carroll:1991zs,Harari:1992ea,Carroll:1998bd,Fedderke:2019ajk}
\begin{align}\label{eq:Delta_Phi}
    \Delta \Phi 
    & = \frac{g_{a\gamma\gamma}}{2} \, \int_C \mathrm{d}X^\mu \, \partial_\mu a(x)
    \;.
\end{align}
To evaluate $\Delta \Phi$ one integrates the axion's spacetime gradient $\partial_\mu a$ along the photon's worldline $X^\mu$ that connects the point of photon emission with the point of photon detection.  
We include a derivation of \eref{eq:Delta_Phi} in \aref{app:derive_Delta_Phi}.  

If the axion field has a trivial topology, with $-\pi f_a \ll a(x) \ll \pi f_a$ throughout spacetime, then the integral in \eref{eq:Delta_Phi} gives simply $\Delta \Phi = g_{a\gamma\gamma} (a_d - a_e)/2$ where $a_e$ and $a_d$ are the values of the axion field at the photon's point of emission and detection, respectively.  
For example, in models with a larger value of $m_a$ than what we are interested in this paper, it is possible for the axion to make up some or all of the dark matter.  
Then the axion field value varies in space and time with the local dark matter density as $|a| \sim \sqrt{\rho_\mathrm{dm}} / m_a$.  
The associated CMB birefringence effect is on the order of $|\Delta \Phi| \sim |g_{a\gamma\gamma}| \sqrt{\rho_\mathrm{dm}} / m_a$~\cite{Fedderke:2019ajk}.
Generally $|a| \ll f_a$ for axion dark matter implying a relatively small birefringence signal $|\Delta \Phi| \ll |g_{a\gamma\gamma}| f_a \sim |\Acal| \alpha_\mathrm{em}$. 

The same theories that describe axion-like particles generally also have topological defects in the spectrum of the theory.  
The existence of these defect solutions is a consequence of the vacuum's nontrivial topology, and our work focuses on the simplest theories that admit two types of defects: one-dimensional strings and two-dimensional domain walls. 
In such cases, the birefringence signal can be much larger.
An axion string is a configuration of the axion field that has a local cylindrical symmetry~\cite{VilenkinShellard:1994}. For a winding number $w = \pm 1$ string the axion field changes by $|\Delta a| = 2\pi f_a$ along a closed path that encircles the string.
In particular, consider a photon that passes through an axion string \textit{loop}, as shown \fref{fig:birefringence}. 
If the points of photon emission and detection are far away from the loop, then the photon experiences the full $|\Delta a| = 2 \pi f_a$ change in the axion field (for a winding number $w=\pm1$ loop, with a plus or a minus sign depending on the orientation of the loop), and the birefringence angle is\footnote{For a photon that traverses an axion domain wall, the change in the field amplitude is $|\Delta a| = 2\pi f_a/N_{\mathrm{dw}}$ where the positive integer $N_\mathrm{dw}$ is called the domain wall number.  The corresponding birefringence is $|\Delta \Phi| = \mathcal{A} \alpha_\mathrm{em} / N_\mathrm{dw}$.}~\cite{Agrawal:2019lkr}
\begin{equation}\label{eq:Delta_Phi_string}
    \Delta \Phi = \pm g_{a\gamma\gamma} \pi f_a = \pm \Acal \alpha_\mathrm{em}
    \;.
\end{equation}
Here we must make two important observations~\cite{Agrawal:2019lkr}.  
First the birefringence effect induced by axion strings is generally much larger than the effect induced by axion dark matter, since the field excursion is necessarily larger for the topological defect.  
Second the string-induced birefringence is insensitive to the Peccei-Quinn scale for $g_{a\gamma\gamma} \propto 1/f_a$, which is the expected scaling in the simplest and most compelling axion theories (but see also \rref{Farina:2016tgd}).  
In this sense $\Delta \Phi$ is a direct probe of the anomaly coefficient, $\Acal$.  
Moreover, models with $\Acal = \mathcal{O}(1)$ provide natural targets!

\subsection{A network of axion strings}\label{sub:strings}

The axion is the Goldstone boson of a U(1) complex Peccei-Quinn field with a symmetry breaking potential. At the PQ phase transition, the Universe is populated with a network of axion strings~\cite{Kibble:1980mv,Kibble:1976sj,Kibble:1976sj,VilenkinShellard:1994}.\footnote{This must happen after inflation since otherwise the network would be `inflated out'. This imposes an upper limit on the Peccei-Quinn scale, $f_a \lesssim f_{a,\mathrm{max}} \equiv \sqrt{H_\mathrm{inf} M_\mathrm{pl}}$.} The network's evolution consists of long strings intersecting and reconnecting to form loops, and loops oscillating and radiating axion particles.  Axion radiation from the network is efficient, and a loop typically collapses in a time scale set by its light-crossing time; i.e., less than $\mathcal{O}(1)$ Hubble time.\footnote{Rapid loop collapse is a general property of global string networks, such as axion strings.  By contrast, gauge string loops collapse slowly by gradual gravitational wave emission, and the network contains many small loops with an abundance controlled by the string tension $G\mu$~\cite{Blanco-Pillado:2013qja}.}  The network soon converges to an attractor solution (i.e. independent of the exact initial conditions). A property of this attractor solution is that the energy density in the string network, to leading order, scales like the total dominant energy density in the Universe and is said to be in scaling~\cite{Yamaguchi:1998gx,Yamaguchi:2002sh,Hiramatsu:2010yu,Hiramatsu:2012gg,Kawasaki:2014sqa,Lopez-Eiguren:2017dmc,Hindmarsh:2019csc,Hindmarsh:2021vih}. With the string tension $\mu \simeq \pi\,f_a^2\log(f_a/H)$, under scaling the energy density of the string network is written as $\rho = \xi\,\mu\,H^2$ where $\xi$ counts the total string length (in units of inverse Hubble) in a Hubble volume and is a constant in time. However, some recent simulations have suggested that there is deviation from scaling, in that $\xi$ grows logarithmically (or milder) in time~\cite{Gorghetto:2018myk,Vaquero:2018tib,Kawasaki:2018bzv,Martins:2018dqg,Buschmann:2019icd,Klaer:2019fxc,Gorghetto:2020qws}.

Generally in many models of axions, the PQ $U(1)$ symmetry is only approximate and the axion has an effective potential. Depending upon the exact details of the UV theory, this potential has a shift symmetry $\mathbb{Z}_{N_{\mathrm{dw}}}$ where $N_{\mathrm{dw}}$ is called the domain wall number~\cite{VilenkinShellard:1994}. Since axion is a compact field with period $2\pi f_a$ ($a \rightarrow a + 2\pi\,f_a$ corresponds to the same point in the U(1) complex field space), $N_\mathrm{dw}$ counts the number of \textit{distinct} vacua in its potential. Once $3H \lesssim m_a$, the axion field in different regions of space relaxes towards the local minimum of the axion potential, resulting in the formation of $N_{\mathrm{dw}}$ 'distinct' domain walls that generally share a common string at their intersections. Two cases are in order here: If $N_{\mathrm{dw}} = 1$ meaning a unique vacuum, there is only one `type' of wall and therefore the network soon collapses under the walls' tension, resulting in a bath of barely-relativistic axion excitations~\cite{Chang:1998tb,Hiramatsu:2012gg,Hiramatsu:2012sc,Hiramatsu:2010yn}. While on the other hand if $N_\mathrm{dw} > 1$, the network is stable since we have `distinct' walls that balance each other's tension. Since the energy density of the domain wall network dilutes like $\rho \propto H$ to leading order, it can come to dominate the Universe quite quickly.
In this latter scenario therefore, the problem of overclosure of the Universe must be mitigated~\cite{Sikivie:1982qv}. However in the case of ULAs with sufficiently low masses, there is no problem of overclosure as long as the Peccei-Quinn scale is taken to be small enough, and therefore there can be an induced birefringence due to such domain wall networks. For instance in a recent work~\cite{Takahashi:2020tqv}, the authors consider cosmic birefringence due to a domain wall network without strings, however for the case when $m_a \gg H_\cmb$.

In this work we primarily focus on string networks for ULAs with $m_a \lesssim 3H_\cmb$, so that the network can be present till later than recombination to have induced a birefringence on the CMB. In~\sref{sec:analytic} we develop our formalism for calculating this birefringence. Depending upon how small the mass is and the domain wall number $N_{\mathrm{dw}}$, there are two broad cases that we study in~\sref{sec:models}: (1) $m_a \lesssim 3H_0$. In this case the string network would have survived till today, resulting in the maximum amount of birefringence possible. We study two different string networks for this scenario in~\sref{sub:model_1} and~\sref{sub:model_2}. (2) $3H_0 \lesssim m_a \lesssim 3H_\cmb$. In this case the string network can survive only until $3H \sim m_a$, resulting in a smaller induced birefringence. We calculate this in~\sref{sub:model_3}. 

Depending upon the domain wall number $N_\mathrm{dw}$, there are further two subcategories for case (2): First when $N_\mathrm{dw}=1$ and the string network collapses into a bath of barely-relativistic axions. In~\aref{app:biref_washout} we provide an estimate of birefringence due to this bath of axions and show that it is subdominant for a power spectrum suggested by simulations, as compared to that induced from the string network. Second is when $N_\mathrm{dw} > 1$ and a network of stable domain walls form. In this case there can be again be an induced birefringence comparable to that of the string network, since the birefringence due to each domain wall crossing is the same as that due to a string loop crossing, modulo $N_\mathrm{dw}$. In~\aref{app:domain_walls} we discuss how our formalism can be extended for the case of domain wall networks. 

\subsection{The loop length distribution function}\label{sub:length}

An axion string network consists of long strings that cross a Hubble volume and sub-Hubble scale loops.  
The loops can have a range of sizes, shapes, and velocities.  
In this analysis, we assume that all loops are circular and moving with a negligibly small speed in the cosmic rest frame.
We define the loop length distribution function $\nu(r,\tau)$ by writing 
\begin{align}\label{eq:dn_from_nu}
    \mathrm{d}n = \nu(r,\tau) \, \mathrm{d}r 
    \;,
\end{align}
such that $\mathrm{d}n$ is the comoving number density of circular loops with comoving radius between $r$ and $r + \mathrm{d}r$ at conformal time $\tau$, and $L = 2 \pi a(\tau) r$ is the corresponding physical, invariant loop length.  
The comoving number density of all loops at conformal time $\tau$ is then calculated as 
\begin{align}\label{eq:n_def}
    n(\tau) = \int_0^\infty \! \! \mathrm{d}r \ \nu(r,\tau) 
    \;.
\end{align}
The integrand $\nu(r,\tau)$ only has support for sub-Hubble scale loops, corresponding to roughly $r \lesssim d_H(\tau) / a(\tau) \sim 1 / a(\tau)H(\tau)$.  
A loop of comoving radius $r$ and tension $\mu(\tau)$ has an energy of $E(r,\tau) = 2\pi a(\tau) r \mu(\tau)$, and the energy density per comoving volume at conformal time $\tau$ is 
\begin{align}\label{eq:rho_def}
    \rho(\tau) 
    & = \int \! \mathrm{d} n \,  E(r,\tau) 
    = \int_0^{\infty} \! \! \mathrm{d} r \  2\pi a(\tau) r \mu(\tau) \, \nu(r,\tau) 
    \;.
\end{align}
In order to characterize how the energy is distributed across loops of different lengths, we also define the dimensionless variables
\begin{align}\label{eq:xi_def}
    \xi(r,\tau) 
    = a(\tau)^{-2} H(\tau)^{-2} \int_0^r \! \! \mathrm{d} r^\prime \, (2\pi r^\prime) \, \nu(r^\prime,\tau) 
    \qquad \text{and} \qquad 
    \xi_\infty(\tau) 
    = \lim_{r \to \infty} \xi(r,\tau) 
    \;.
\end{align}
It follows that $\rho(r,\tau) = \xi(r,\tau) \, a(\tau)^3 \, \mu(\tau)  \,  H(\tau)^2$
is the energy density of all loops with comoving radius less than $r$ at conformal time $\tau$.
Sending $r \to \infty$ gives the total energy density
\begin{align}\label{eq:rho_from_xi}
    \rho(\tau) 
    = \xi_\infty(\tau) \, a(\tau)^3 \, \mu(\tau)  \,  H(\tau)^2 
    \;.  
\end{align}
Since a Hubble-scale loop has an energy $E_H \sim \mu d_H \sim \mu / H$, then the effective number of Hubble-scale loops per Hubble volume is $\rho / E_H a^3 H^3 = \xi_\infty$.  
That is to say, if we could rearrange all sub-Hubble loops into a minimal number of Hubble-scale loops while conserving energy, then we would find $\xi_\infty(\tau)$ such Hubble-scale loops at conformal time $\tau$.  
If the string network is in scaling, then $\xi(r,\tau)$ and $\xi_\infty(\tau)$ are time-independent.  
In \sref{sub:kernel} we will introduce a convenient parametrization of the loop length distribution function \pref{eq:nu_from_chi} that simplifies our birefringence calculation considerably in the scaling regime.  
In this work, we study several different models for the axion string network, which corresponds to different choices for the function $\nu(r,\tau)$.  

\section{Analytic formalism \& numerical validation}\label{sec:analytic}

\subsection{The loop-crossing model}\label{sub:loop_crossing}

We seek to calculate the two-point angular correlation function for the axion-induced birefringence $\langle \Delta \Phi \, \Delta \Phi \rangle$.  
To do so, we work with a simple model for the propagation of CMB photons through the string network that was proposed in Ref.~\cite{Agrawal:2019lkr}.  
We call this the \textit{loop-crossing model}.  

We imagine that the CMB photons propagate through a network of string loops on their way from the surface of last scattering to a telescope on Earth.  
Simulations suggest that a good fraction of the total string length is in long (``infinite") strings in the network. Such long strings have typical curvature radius that goes like $H^{-1}$. This motivates us to consider such large strings as string loops of radii $\sim H^{-1}$.
Therefore in general, we allow the possibility of string loops that can have maximum radii of the order of the inverse Hubble.
To calculate $\Delta \Phi$ the loop-crossing model instructs us to count the number of loops that a photon intersects as it travels to Earth.  
An intersection occurs when the photon's trajectory pierces the surface that's bounded by the loop.  
Each such intersection leads to a phase shift of $\Delta \Phi = \pm \Acal \, \alpha_\mathrm{em}$ where the sign is assumed to be random and equally distributed between positive and negative values.  
As a photon passes through multiple loops, the accumulated phase shift evolves like a random walk with $\langle (\Delta \Phi)^2 \rangle = (\Acal \alpha_\mathrm{em})^2 N$ growing proportionally to the number of loops encountered $N$.  

To calculate the two-point correlation function the loop-crossing model instructs us to consider a pair of photons with orientations $\hat{\gamma}$ and $\hat{\gamma}^\prime$ and to count the number of loops $N_\text{both}(\hat{\gamma}_1, \hat{\gamma}_2)$ that both photons intersect [see eq. \pref{eq:N_both}].  
Then the two-point correlation function is given by 
\begin{align}\label{eq:DPhi_from_Nboth}
    \langle \Delta \Phi(\hat{\gamma}_1) \, \Delta \Phi(\hat{\gamma}_2) \rangle = \bigl( \Acal \, \alpha_\mathrm{em} \bigr)^2 \, N_\text{both}(\hat{\gamma}_1, \hat{\gamma}_2) 
    \;.
\end{align}
Due to the statistical isotropy of the string network, $N_\mathrm{both}$ only depends on the angle between the two photons.  

\begin{figure}[t]
\centering
\includegraphics[width=1\columnwidth]{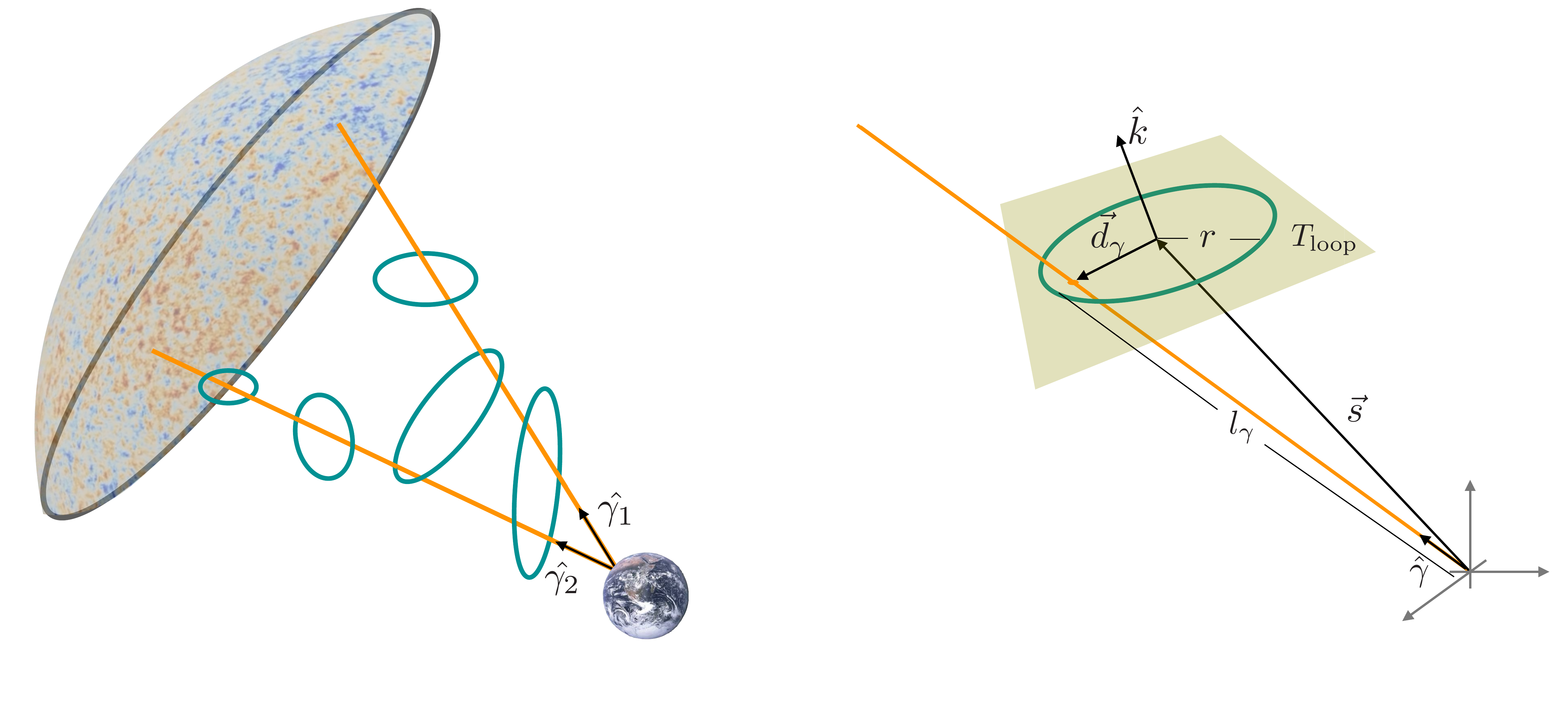}\vspace{0.1cm}\\
\caption{Left: An illustration of two photons passing through several axion string loops as they travel from the surface of last scattering to Earth. Heuristically, non-zero birefringence correlation arises from all the loops intersected by both photons.  Right: The geometry of a photon ray as it passes close to an axion string loop.}
\label{fig:loop_geometry}
\end{figure}

The geometry of a photon and a string loop is shown in \fref{fig:loop_geometry}.  
Consider a photon that reaches Earth today and that has a direction of origin indicated by the unit vector $\hat{\gamma}$.  
Consider also a cosmic string loop between Earth and the CMB's surface of last scattering.  
Recall that loops are assumed to be circular with comoving radius $r$, and additionally the plane of the loop is denoted by $T_\mathrm{loop}$.  
At some time, the photon passed through the plane of this loop at a point $p_\gamma$ and we are interested in the loop's configuration at this time.  
Let $\vec{s}$ be the comoving distance from the Earth to the loop, $\hat{k}$ be the unit vector normal to the plane of the loop, and $\vec{d}_\gamma$ be the comoving distance from the center of the loop to the point $p_\gamma$ at the time of the photon's crossing.  
We can write $\vec{d}_{\gamma} = l_{\gamma} \hat{\gamma} - \vec{s}$ where $l_\gamma$ is the comoving distance that the photon travels from $p_\gamma$ to Earth.  
Since $\vec{d}_\gamma$ lies in the plane $T_\mathrm{loop}$ while $\hat{k}$ is normal to the plane, it follows that $\vec{d}\cdot\hat{k}=0$ and 
\begin{equation}\label{eq:l_gamma}
    l_{\gamma} = \frac{\vec{s}\cdot\hat{k}}{\hat{\gamma}\cdot\hat{k}}
    \;.
\end{equation}
Finally we have the comoving distance from the center of the loop to the point $p_\gamma$ where the photon crosses the plane of the loop:
\begin{equation}\label{eq:d_gamma}
    \vec{d}_{\gamma}
    = \dfrac{\hat{k}\times(\hat{\gamma}\times\vec{s})}{\hat{\gamma}\cdot\hat{k}}
    \;,
\end{equation}
and $d_\gamma = |\vec{d}_\gamma|$.  
If $d_\gamma < r$ then the photon passes through the loop, but it misses the loop if $r < d_\gamma$.  

Using the geometrical variables from \fref{fig:loop_geometry} the number of loops intersected by both photons arriving from directions $\hat{\gamma}_1$ and $\hat{\gamma}_2$ is written as 
\begin{align}\label{eq:N_both}
    N_\mathrm{both}(\hat{\gamma}_1, \hat{\gamma}_2) 
    = \int \! \mathrm{d} N \, \Theta\left(r - d_1\right) 
    \Theta\left(r - d_2\right) \Theta\left(l_1\right) \Theta\left(l_2\right) 
    \;.
\end{align}
Here $d_1$ and $d_2$ correspond to $d_\gamma$ for $\hat{\gamma}_1$ and $\hat{\gamma}_2$, respectively.  
Similarly $l_1$ and $l_2$ correspond to $l_\gamma$.  
The distribution of loops is encoded by $\mathrm{d}N$ (see below).  
The step functions in \eref{eq:N_both} ensure that both photons pass through the loop.  
In addition to imposing $d_\gamma < r$ for both photons, it is also necessary that $0 < l_\gamma$ for each photon.  
This is to exclude geometries, for example, in which the loop is located in the northern hemisphere while the photon arrives from the southern hemisphere.  
If $l_\gamma < 0$ then it no longer has the interpretation as the comoving distance from $p_\gamma$ to Earth.   
The condition $0 < l_\gamma$ was not written explicitly in \rref{Agrawal:2019lkr}.  

Finally, the distribution of loops is given by 
\begin{align}
    \mathrm{d}N = \mathrm{d}^3 \vec{s} \ \frac{\mathrm{d}^2 \hat{k}}{4\pi} \ \mathrm{d}r \ \nu(r,\tau)
    \;,
\end{align}
which equals the number of loops with comoving distance between $\vec{s}$ and $\vec{s} + \mathrm{d}\vec{s}$, orientation between $\hat{k}$ and $\hat{k} + \mathrm{d}\hat{k}$, and comoving radius between $r$ and $r + \mathrm{d}r$ at conformal time $\tau$.  
The loops are assumed to be statistically homogeneous and isotropically oriented so that $\nu$ does not depend on $\vec{s}$ or $\hat{k}$ directly.  
However, since we are only interested in the loops that lie along our past light cone, we must restrict our attention to $\tau = \tau_0 - s$ where $\tau_0$ is the conformal time today and $s = |\vec{s}|$ so that $\tau$ is the conformal time when the photon encountered the loop.  
It is also convenient to change the time variable from conformal time $\tau$ to redshift $z$, using the relations 
\begin{align}\label{eq:tau_from_z}
    a(z) = \frac{a_0}{1+z}
    \ , \qquad 
    \tau(z) = \int_z^\infty \! \frac{\mathrm{d}z}{a_0 H(z)}
    \ , \qquad \text{and} \qquad
    s(z) = \tau(0) - \tau(z) 
\end{align}
where $a_0$ is the FRW scale factor today.   
For our numerical work below, the Hubble parameter $H = H(z)$ is calculated for the $\Lambda$CDM cosmology.\footnote{Recall that $H(z) = H_0 \sqrt{\Omega_\Lambda + \Omega_m (1+z)^3 + \Omega_r (1+z)^4}$.  We take $H_0 = 70 \ \mathrm{km}/\mathrm{Mpc}/\mathrm{sec}$, $\Omega_\Lambda = 0.7$, $\Omega_m = 0.3$ and $\Omega_r = 9\times10^{-5}$ (including neutrinos)~\cite{Aghanim:2018eyx}.}  
For instance, during the matter-dominated epoch we have the scaling $a \sim \tau^2$ and $H \sim \tau^{-3}$.  
With this identification we can write 
\begin{align}\label{eq:dN}
    \mathrm{d}N = 
    \mathrm{d}z \, 
    \mathrm{d}^2 \hat{s} \, 
    \frac{\mathrm{d}^2 \hat{k}}{4\pi} \, 
    \mathrm{d}r \, 
    \frac{s(z)^2}{a_0 H(z)} \, 
    \nu(r,\tau(z))
    \;,
\end{align}
where $s(z)$ and $\tau(z)$ are calculated using \eref{eq:tau_from_z}.  

Combining Eqs.~(\ref{eq:DPhi_from_Nboth}),~(\ref{eq:N_both}),~and~(\ref{eq:dN}) yields the loop-crossing model's two-point correlation function~\cite{Agrawal:2019lkr}
\begin{equation}\label{eq:DPhi_DPhi}
\begin{split}
    \langle \Delta \Phi(\hat{\gamma}_1) \, \Delta \Phi(\hat{\gamma}_2) \rangle 
    & = \bigl( \Acal \, \alpha_\mathrm{em} \bigr)^2 
    \int_0^{z_\cmb} \! \! \mathrm{d} z 
    \int \! \mathrm{d}^2 \hat{s}
    \int \! \frac{\mathrm{d}^2 \hat{k}}{4\pi} 
    \int_0^\infty \! \mathrm{d}r \ 
    \frac{s(z)^2}{a_0 H(z)} \, 
    \nu(r,\tau(z)) 
    \\ & \hspace{3cm} \times 
    \Theta\left(r - d_1\right) 
    \Theta\left(r - d_2\right) 
    \Theta\left(l_1\right) 
    \Theta\left(l_2\right) 
    \;.
\end{split}
\end{equation}
The integral over redshift extends from recombination at $z_\cmb \simeq 1100$ until today at $z=0$.
The integral over comoving loop radius $r$ extends over all possible values, but the loop length distribution function $\nu(r,\tau)$ restricts $r \lesssim d_H/a \sim 1 / aH$.  
Once a string network model is assumed and a loop length distribution function $\nu(r,\tau)$ is specified, the integrals can be evaluated, which we discuss further in the next section.  
Since the distribution is statistically isotropic, the two-point correlation function only depends on the opening angle $\hat{\gamma}_1 \cdot \hat{\gamma}_2 = \cos \theta_o$.  
Then, using the Legendre function decomposition
\begin{align}
    \langle \Delta \Phi(\hat{\gamma}_1) \, \Delta \Phi(\hat{\gamma}_2)\rangle 
    = \sum^{\infty}_{\ell=0} 
    \frac{(2\ell+1)}{4\pi} 
    C_\ell^{\Phi\Phi} \, 
    P_\ell(\cos\theta_o)
\end{align}
the angular power spectrum is calculated as
\begin{equation}\label{eq:Cl_def}
    C_\ell^{\Phi\Phi} = 2\pi \int^{1}_{-1} \mathrm{d}(\cos\theta_o) \, P_\ell(\cos\theta_o) \, \langle\Delta\Phi_1(\hat{\gamma}_1)\Delta\Phi_2(\hat{\gamma}_2)\rangle
\end{equation}
for positive integer $\ell$.

We close this section by remarking upon a limitation of the loop-crossing model.  
This framework assumes that each loop induces a $\Delta \Phi = \pm \Acal \alpha_\mathrm{em}$ phase shift, which corresponds to a complete cycle in the axion's field space $\Delta a = \pm 2 \pi f_a$.  
This is a reasonable assumption when the point of photon emission is very far ``behind'' the loop, and the photon absorption occurs far ``in front'' of the loop.  
Specifically, these distances are assumed to be large compared to the radius of the loop.  
This assumption breaks down for ``nearby'' loops, such as Hubble-scale loops within our Hubble volume today. 
The birefringence effect from these loops will be smaller than \eref{eq:DPhi_DPhi} suggests, since $|\Delta a| < 2\pi f_a$.  
To assess the impact of this effect, we have calculated the birefringence signal by evaluating the integral from \eref{eq:DPhi_DPhi} with $z$ cut off at $s(z) = r$ and compared with the calculation when $z$ is allowed to go down to zero.  
We find that the two calculations agree to better than $\mathcal{O}(1\%)$ accuracy at $\ell \gtrsim 5$ in general.  
If the network contains Hubble-scale loops, then the discrepancy can be as large as $\mathcal{O}(20\%)$ at $\ell \simeq 2$, and the agreement is improved for sub-Hubble-scale loops.  

\subsection{Kernel function}\label{sub:kernel}

None of the $6$ integrals in \eref{eq:DPhi_DPhi} are especially easy to evaluate, particularly because the $r$ and $s$ integrals become entangled by the loop length distribution function $\nu(r,\tau)$.  
However, here we point out that the calculation factorizes into the evaluation of a kernel function, which is independent of the string network model, and the integration of this kernel against a rescaled loop length distribution.  
To see how this simplification arises, we first introduce an alternative parameterization of the loop length distribution function $\nu(r,\tau)$ by writing 
\begin{align}\label{eq:nu_from_chi}
    \nu(r,\tau(z)) & = \int_0^\infty \! \! \mathrm{d} \zeta \ \chi(\zeta,z) \, \frac{a(z)^2 H(z)^2}{2\pi r} \ \delta\Bigl( r - \frac{\zeta}{a(z) H(z)} \Bigr) 
    \;,
\end{align}
where $\zeta$ is a dimensionless integration variable.  
Owing to the Dirac delta function in the integrand, any $\nu(r,\tau)$ can be constructed from an appropriately-chosen $\chi(\zeta,z)$, and so there is no loss of generality in writing \eref{eq:nu_from_chi}.  
With this parametrization, note that the dimensionless variables $\xi(r,\tau)$ and $\xi_\infty(\tau)$ from \eref{eq:xi_def} are simply 
\begin{align}\label{eq:xi_from_chi}
    \xi(r,\tau(z)) 
    = \int_0^{r a(\tau) H(\tau)} \! \! \mathrm{d} \zeta \ \chi(\zeta,z) 
    \qquad \text{and} \qquad 
    \xi_\infty(\tau(z)) 
    = \int_0^\infty \! \! \mathrm{d} \zeta \ \chi(\zeta,z) 
    \;,
\end{align}
implying $\chi(\zeta,z) = d\xi/d\zeta$.  

This parametrization \pref{eq:nu_from_chi} is particularly useful to study a string network in the scaling regime, which now corresponds to 
\begin{align}\label{eq:scaling_regime}
    \chi(\zeta,z) \to \chi(\zeta) 
    \qquad \text{and} \qquad 
    \xi(\zeta,\tau) \to \xi(\zeta) 
    \qquad \text{(scaling regime)} 
    \;.
\end{align}
When we discuss specific string network models in \sref{sec:models}, we will see that the calculation of the two-point correlation function simplifies if the string network is in scaling.\footnote{It is also important to note that we only need to assume scaling from recombination until today.  If the string network has a logarithmic deviation from scaling, as some recent string network simulations have suggested~\cite{Gorghetto:2018myk}, then the associated error is expected to be no more than $1 - (\log f_a/H_\cmb)/(\log f_a/H_0) \sim 10\%$.}

Upon using the parametrization in \eref{eq:nu_from_chi}, the two-point function from \eref{eq:DPhi_DPhi} becomes 
\begin{equation}\label{eq:DPhi_numerical}
\begin{split}
    \langle \Delta \Phi(\hat{\gamma}_1) \, \Delta \Phi(\hat{\gamma}_2) \rangle 
    & = \bigl( \Acal \, \alpha_\mathrm{em} \bigr)^2 
    \int_0^{z_\cmb} \! \mathrm{d} z 
    \int \! \mathrm{d}^2 \hat{s} 
    \int \! \frac{\mathrm{d}^2 \hat{k}}{4\pi} 
    \int_0^\infty \! \! \mathrm{d} \zeta \ 
    \frac{a^3 s^2 H^2}{2\pi \zeta a_0} \ 
    \chi(\zeta,z) \, 
    \\ & \hspace{2.5cm} \times 
    \Theta\left(\zeta - aH d_1\right) 
    \Theta\left(\zeta - aH d_2\right) 
    \Theta\left(l_1\right) 
    \Theta\left(l_2\right) 
    \;. 
\end{split}
\end{equation}
We have traded the $r$ integral for a $\zeta$ integral, and the string network model is now encoded in $\chi(\zeta,z)$.
We now factorize the integrals to write 
\begin{align}\label{eq:DPhi_from_Q}
    \langle \Delta \Phi(\hat{\gamma}_1) \, \Delta \Phi(\hat{\gamma}_2) \rangle 
    & = 
    \bigl( \Acal \, \alpha_\mathrm{em} \bigr)^2 
    \int_0^{z_\cmb} \! \! \mathrm{d} z 
    \int_0^\infty \! \! \mathrm{d} \zeta \
    Q(\zeta,z,\theta_o) \ 
    \chi(\zeta,z)
    \;,
\end{align}
where we've defined the dimensionless kernel function, 
\begin{align}\label{eq:Q_def}
    Q(\zeta,z,\theta_o)
    & = 
    \int \! \mathrm{d}^2 \hat{s} 
    \int \! \frac{\mathrm{d}^2 \hat{k}}{4\pi} \ 
    \frac{a^3 s^2 H^2}{2\pi \zeta a_0} \ 
    \Theta\left(\zeta - aH d_1\right) 
    \Theta\left(\zeta - aH d_2\right) 
    \Theta\left(l_1\right) 
    \Theta\left(l_2\right) 
    \;,
\end{align}
which only depends on $\zeta$, $z$, and the opening angle $\theta_o$.  
The advantage here is that the kernel function $Q(\zeta,z,\theta_o)$ is independent of the string network model, which is captured by $\chi(\zeta,z)$.  
So the kernel function only needs to be evaluated once, and then the two-point correlation function is obtained by evaluating the remaining integrals in \eref{eq:DPhi_from_Q}.  
In the following subsections we derive approximate analytic expressions for the kernel function, and in \sref{sec:models} we use these results to calculate the birefringence signal for specific string network models.  

It is useful to introduce Cartesian and polar coordinates.  
The trajectories of the two photons, $\hat{\gamma}_1$ and $\hat{\gamma}_2$, define the basis vectors of a Cartesian coordinate system 
\begin{subequations}\label{eq:coordinates}
\begin{align}
    \hat{x} = \frac{\hat{\gamma}_1 - \hat{\gamma}_2}{|\hat{\gamma}_1 - \hat{\gamma}_2|} 
    \ , \quad 
    \hat{y} = \frac{\hat{\gamma}_2 \times \hat{\gamma}_1}{|\hat{\gamma}_2 \times \hat{\gamma}_1|} 
    \ , \quad 
    \hat{z} = \frac{\hat{\gamma}_1 + \hat{\gamma}_2}{|\hat{\gamma}_1 + \hat{\gamma}_2|} 
    \;,
\end{align}
where $\hat{\gamma}_2 \neq \pm \hat{\gamma}_1$.  
Here $\hat{x}$ and $\hat{z}$ lie in the plane spanned by $\hat{\gamma}_1$ and $\hat{\gamma}_2$, and in particular $\hat{z}$ is the midpoint between the photons.  
The various unit vectors can be expressed using polar coordinates 
\begin{align}
    \hat{k} & = \sin \theta_k \cos \phi_k \, \hat{x} + \sin \theta_k \sin \phi_k \, \hat{y} + \cos \theta_k \, \hat{z} \\ 
    \hat{s} & = \sin \theta_s \cos \phi_s \, \hat{x} + \sin \theta_s \sin \phi_s \, \hat{y} + \cos \theta_s \, \hat{z} \\ 
    \hat{\gamma}_1 & = \sin \frac{\theta_o}{2} \, \hat{x} + \cos \frac{\theta_o}{2} \, \hat{z} \\ 
    \hat{\gamma}_2 & = - \sin \frac{\theta_o}{2} \, \hat{x} + \cos \frac{\theta_o}{2} \, \hat{z} 
\end{align}
\end{subequations}
where the polar angles, $\theta_k$ and $\theta_s$, are measured from the $\hat{z}$ axis, and the azimuthal angles, $\phi_k$ and $\phi_s$, from the $\hat{x}$ axis.  
The opening angle between the two photons obeys $\hat{\gamma}_1 \cdot \hat{\gamma}_2 = \cos \theta_o$. 

\subsection{Kernel function at zero opening angle}\label{sub:zero_opening}

In this subsection and the next one, we obtain analytic approximations for the kernel function $Q(\zeta,z,\theta_o)$ by estimating the integrals in \eref{eq:Q_def}.  
Here we first consider the autocorrelation function with zero opening angle between the two photons, $\theta_o=0$.  
Using the coordinate system in \eref{eq:coordinates} implies
\begin{align}
    l_1 = l_2 = l_\gamma & = s \, \bigl( \tan\theta_k \sin\theta_s \cos(\phi_s-\phi_k) + \cos\theta_s \bigr) \\ 
    d_1 = d_2 = d_\gamma & = s\,\sin\theta_s\sqrt{1 + \cos^2\left(\phi_s-\phi_k\right)\tan^2\theta_k}
\end{align} 
for $l_\gamma$ from \eref{eq:l_gamma} and $d_\gamma = |\vec{d}_\gamma|$ from \eref{eq:d_gamma}.  
The kernel function \pref{eq:Q_def} at $\theta_o = 0$ becomes 
\begin{equation}\label{eq:Q_zeta0_0}
\begin{split}
    Q(\zeta,z,0)
    & = 
    \frac{a^3 s^2 H^2}{8\pi^2\zeta a_0} 
    \int_0^\pi \! \mathrm{d}\theta_s \, \sin \theta_s 
    \int_0^{2\pi} \! \mathrm{d}\phi_s 
    \int_0^{2\pi} \!
    \mathrm{d}\phi_k 
    \int_0^\pi \! \mathrm{d}\theta_k \, \sin \theta_k 
    \\ & \hspace{3cm} \times 
    \Theta\Bigl(\zeta - aHs \,\sin\theta_s\sqrt{1 + \cos^2\left(\phi_s-\phi_k\right)\tan^2\theta_k} \Bigr) 
    \\ & \hspace{3cm} \times 
    \Theta\Bigl(\tan\theta_k \sin\theta_s \cos(\phi_s-\phi_k) + \cos\theta_s\Bigr) 
    \;.
\end{split}
\end{equation}
In these polar coordinates, both photons arrive from the ``north pole'' and it is very unlikely for them to pass through a loop centered in the ``southern hemisphere'' where $\pi/2 < \theta_s < \pi$, which lets us approximate\footnote{It's possible that photons from the north pole can pass through loops that are centered in the southern hemisphere, particularly if the loops are close (small $s$), but from a direct numerical evaluation, we find that these configurations have a negligible contribution to the integral here.  As such we restrict $0 < \theta_s < \pi/2$.  In the restricted phase space we have also $l_\gamma > 0$, and the second step function evaluates to $1$.}
\begin{equation}\label{eq:Q_zeta0_1}
\begin{split}
    Q(\zeta,z,0)
    & \approx 
    \frac{a^3 s^2 H^2}{8\pi^2\zeta a_0} 
    \int_0^{\pi/2} \! \mathrm{d}\theta_s \, \sin \theta_s 
    \int_0^{2\pi} \! \mathrm{d}\phi_s
    \int_0^{2\pi} \! \mathrm{d}\phi_k
    \int_0^\pi \! \mathrm{d}\theta_k \, \sin \theta_k \ 
    \\ & \hspace{3cm} \times 
    \Theta\Bigl(\zeta - aH s \,\sin\theta_s\sqrt{1 + \cos^2\left(\phi_s-\phi_k\right)\tan^2\theta_k} \Bigr)
\end{split}
\end{equation}
First, evaluating the $\theta_k$ integral gives 
\begin{equation}\label{eq:Q_zeta0_2}
\begin{split}
    Q(\zeta,z,0)
    & \approx 
    \frac{a^3 s^2 H^2}{4\pi^2\zeta a_0} 
    \int_0^{\pi/2} \! \mathrm{d}\theta_s \, \sin \theta_s 
    \int_0^{2\pi} \! \mathrm{d}\phi_s 
    \int_0^{2\pi} \! \mathrm{d}\phi_k \
    \Theta\Bigl(\zeta - aH s \,\sin\theta_s\Bigr)
    \\ & \hspace{4cm} \times 
    \left(1 - \dfrac{aH s \, |\cos(\phi_k-\phi_s)| \sin\theta_s}{\sqrt{\zeta^2 - a^2 H^2 s^2 \, \sin^2(\phi_k-\phi_s) \sin^2\theta_s}}\right)
    \;.
\end{split}
\end{equation}
Second, evaluating the $\phi_s$ and $\phi_k$ integrals gives 
\begin{equation}\label{eq:Q_zeta0_3}
\begin{split}
    Q(\zeta,z,0)
    & \approx 
    \frac{a^3 s^2 H^2}{\zeta a_0} 
    \int_0^{\pi/2} \! \mathrm{d}\theta_s \, \sin \theta_s \ 
    \Theta\Bigl(\zeta - aH s \, \sin\theta_s\Bigr) 
    \\ & \hspace{3cm} \times 
    \left(1 - \dfrac{2}{\pi} \, \mathrm{arctan}\left(\dfrac{aH s \, \sin\theta_s}{\sqrt{\zeta^2 - a^2 H^2 s^2 \, \sin^2\theta_s}}\right)\right)
    \;.
\end{split}
\end{equation}
Third, the $\theta_s$ integral can be evaluated in terms of a hyper-geometric function, but it is illuminating to study the limiting regimes $\zeta \ll aH s$ and $aH s \ll \zeta$.  
Evaluating the $\theta_s$ integral in these regimes gives 
\begin{align}\label{eq:Q_zeta0_4}
    Q(\zeta,z,0)
    & \approx 
    \begin{cases} 
    \frac{a^3 s^2 H^2}{\zeta a_0} & , \quad aH s \ll \zeta \\ 
    \frac{\zeta a}{4 a_0} & , \quad \zeta \ll aH s 
    \end{cases} 
    \;.
\end{align}
This expression is one of the main results of our work; it gives the kernel function at zero opening angle $\theta_o=0$ as a function of the dimensionless loop radius $\zeta=r/aH$ and redshift $z$.  
Recall that $s = \tau_0 - \tau(z)$ and $\tau(z)$ is given by \eref{eq:tau_from_z}.  

To further evaluate \eref{eq:Q_zeta0_4} we must specify a cosmological model.  
For instance, if we neglect dark energy and treat the universe as matter-dominated between recombination and today\footnote{In a matter-dominated universe recall that $H(z) = H_0 (1+z)^{3/2}$, $a(z) = a_0 (1+z)^{-1}$, $\tau(z) = \tau_0 (1+z)^{-1/2}$, $\tau_0 = 2/a_0H_0$, and $s(z) = \tau_0[ 1 - (1+z)^{-1/2}]$.} then we find $Q \approx z^2 / \zeta$ for $aHs \ll \zeta < 1$ and $Q \approx \zeta/4(1+z)$ for $\zeta \ll aHs$.  
For the $\Lambda$CDM cosmology ($\Omega_m = 1 - \Omega_\Lambda = 0.3$), the product $aHs$ is calculated from the integral in \eref{eq:tau_from_z}, which evaluates to a hypergeometric function.  
However, it is useful to note that this quantity admits an empirical fitting function $aHs \approx \log(1+z)$, which agrees to better than $5\%$ precision over $0 < z < 1$.  
Consequently, in the $\Lambda$CDM cosmology the kernel function at zero opening angle \pref{eq:Q_zeta0_3} is well-approximated by 
\begin{align}\label{eq:Q_zeta0_lcdm2}
    Q(\zeta,z,0)
    & \approx 
    \begin{cases} 
    \frac{\zeta}{4} \, (1+z)^{-1} & , \quad z_t(\zeta) < z \\ 
    \frac{1}{\zeta} \, (1+z)^{-1}\log(1+z)^2 & , \quad z < z_t(\zeta) 
    \end{cases}
    \;.
\end{align}
To ensure that $Q(\zeta,z,0)$ is continuous as $\zeta$ and $z$ are varied, we match the two limiting regimes at the `transition' redshift $z_t(\zeta) \equiv e^{\zeta/2}-1$ such that $aHs|_{z=z_t} \approx \zeta/2$.  
We will focus on string networks that only contain sub-Hubble scale loops; this corresponds to $\zeta < 1$, which implies $z_t < 0.6$, and in practice the $z_t < z$ case is more relevant.  

\begin{figure}[t!]
\centering
\includegraphics[width=0.74\columnwidth]{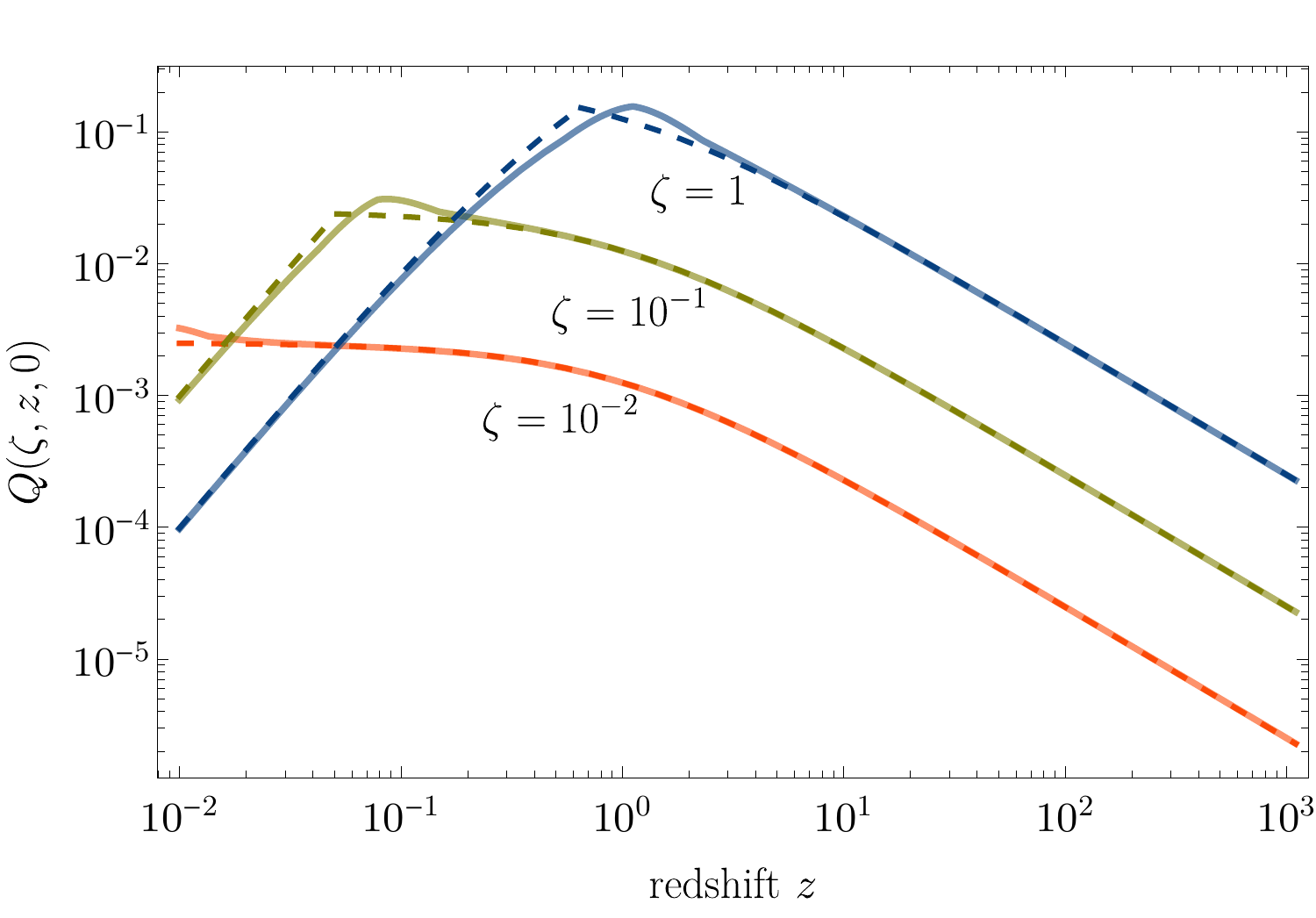}\vspace{0.1cm}
\caption{The kernel function $Q(\zeta,z,\theta_o)$ at zero opening angle $\theta_o = 0$ (where $\theta_0$ is the angle between two directions of CMB photons).  We show a few values of the dimensionless loop radius $\zeta = raH$. The ``exact'' kernel function (solid) is obtained by evaluating the integrals in \eref{eq:Q_zeta0_0} numerically.  The ``approximate'' kernel function (dashed) corresponds to \eref{eq:Q_zeta0_lcdm2}. The kernel function (including its finite opening angle extension) is model independent, and can be used with input from different string-network models, $\chi(\zeta,z)$, to calculate the two-point correlation function for birefringence phase shifts via integrals over $\zeta$ and $z$.}
\label{fig:ztimesintegrandvsz}
\end{figure}

To validate the approximate kernel function in \eref{eq:Q_zeta0_lcdm2}, we have also calculated the kernel function ``directly'' by evaluating the integrals in \eref{eq:Q_zeta0_0} numerically without any approximation. We compare these two approaches in Figure~\ref{fig:ztimesintegrandvsz}, which shows the kernel functions $Q(\zeta,z,0)$ for several values of $\zeta$ as a function of redshift $z$. At asymptotically small and large redshift our approximation agrees arbitrarily well, while around $z = z_t(\zeta) = e^{\zeta/2}-1$ there is a slight disagreement, which is of-course an artifact of our piece-wise approximation to the actual hypergeometric function. Overall, the figure shows a good agreement between the numerically-estimated integral \pref{eq:Q_zeta0_0} and our analytic approximation \pref{eq:Q_zeta0_lcdm2}.  

\subsection{Mapping to finite opening angles}\label{sub:finite_opening}

Having calculated the kernel function at zero opening angle, we can now use this quantity to construct an estimate of the kernel function at $\theta_o \neq 0$.  
This construction relies on a correspondence between photon-pair opening angles and the angular extent of loops on the sky.  

\begin{figure}[t]
\centering
\includegraphics[width=1.02\columnwidth]{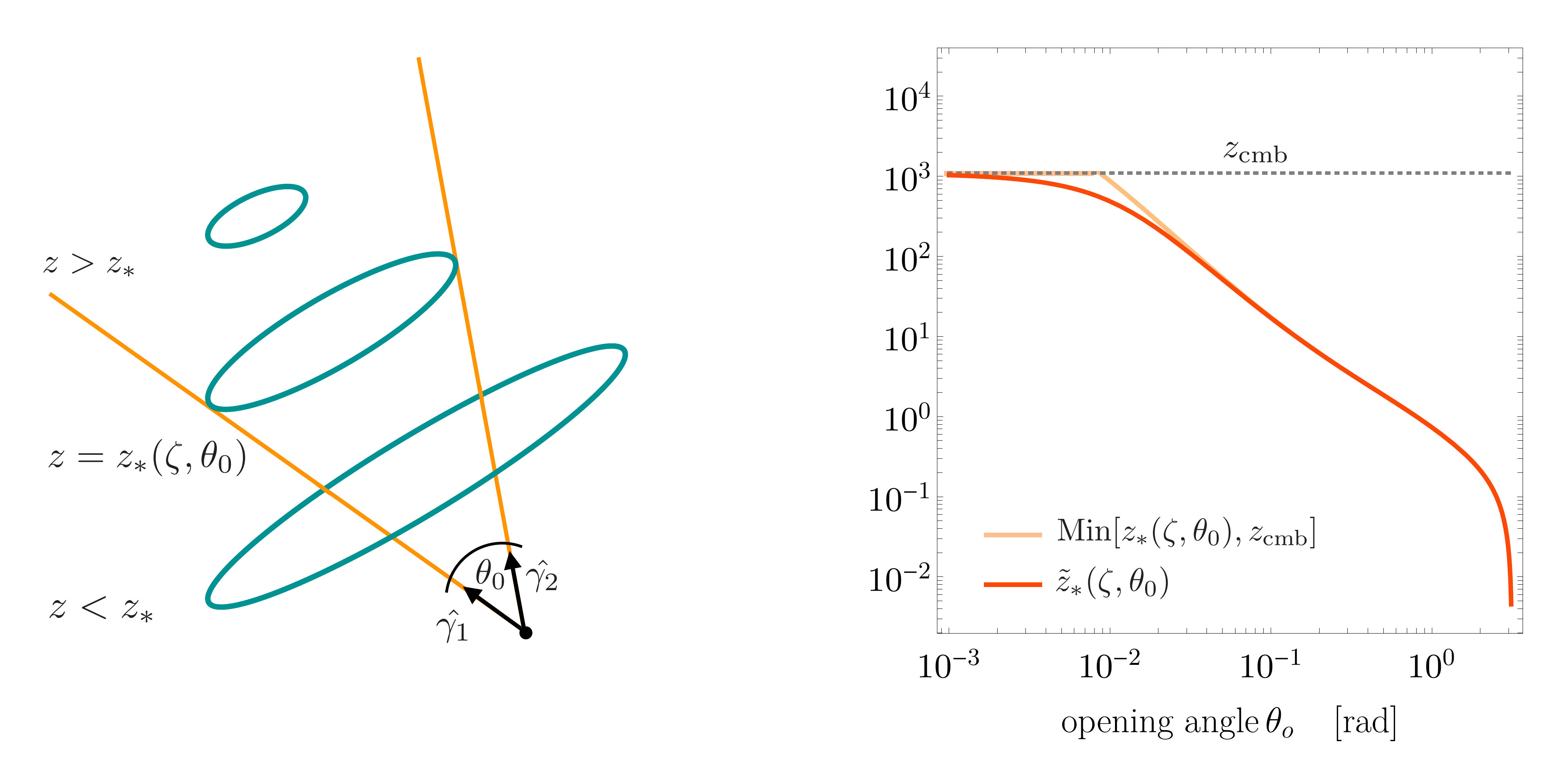}\vspace{0.1cm}\\
\caption{Left:  An illustration of the special redshift $z_\ast(\zeta,\theta_o)$.  Loops with a given dimensionless radius $\zeta = r/aH$ are smaller at higher redshift $z$.  For a given pair of photons there is a special redshift $z = z_\ast(\zeta,\theta_o)$ at which the opening angle $\theta_o$ matches the angular scale of the effective circular loop.  We obtain our kernel function $Q$ for a non-zero opening angle $\theta_o$, via $Q(\zeta,z,\theta_o)\approx \Theta(z_*-z)Q(\zeta,z,0)$.  Right:  A plot of $\textrm{Min}[z_\ast(\zeta,\theta_o),z_\textrm{cmb}]$ and $\tilde{z}_\ast(\zeta,\theta_o)$ where the abrupt transition has been smoothed in the latter. This is for LCDM cosmology and $\zeta = 1$. }
\label{fig:opening_angle}
\end{figure}

Loops with dimensionless radius $\zeta$ have comoving radius $r(\zeta,z) = \zeta/(a(z) H(z))$ at redshift $z$, and the angular diameter of face-on loops (i.e. $\hat{\bm k}\cdot\hat{\bm s}=1$) is
\begin{align}\label{eq:angular_diameter}
    \theta_d(\zeta,z) 
    = 2 \, \mathrm{arctan} \left( \frac{r(\zeta,z)}{s(z)}  \right) = 2 \, \mathrm{arctan} \left( \frac{\, \zeta}{a(z) H(z) s(z)}  \right) 
    \;.
\end{align}
For example, in a matter-dominated universe $\theta_d(\zeta,z) = 2 \, \mathrm{arctan}[(\zeta/2) / (\sqrt{1+z}-1)]$.   
The angle $\theta_d(\zeta,z)$ is a monotonically decreasing function of $z$, as shown in \fref{fig:opening_angle}, meaning that the angular extent of a loop with dimensionless radius $\zeta$, is small at large $z$ but grows larger at small $z$. 

To calculate the two-point correlation function with the loop-crossing model, we are only interested in pairs of photons that pass through the same loops.  
If a given pair of photons has an opening angle of $\theta_o$, the two photons will only become correlated starting from the time when $\theta_o \sim \theta_d(\zeta,z)$, because it is impossible for both photons to pass through a smaller loop at earlier times. 
This construction will allow us to map between the zero opening angle kernel function $Q(\zeta,z,0)$ and the finite opening angle kernel function $Q(\zeta,z,\theta_o)$.  
Before doing so, we first need to discuss corrections that cause the effective angular scale of a loop $\theta_\mathrm{eff}(\zeta,z)$ to be slightly smaller than its angular diameter $\theta_d(\zeta,z)$.  
 
For a face-on circular loop the maximal angular extent on the sky is $\theta_d(\zeta,z)$ along a diameter, while the angular extent is smaller along a chord.  
In addition if the loop is not face on (i.e. $\hat{\bm k} \cdot \hat{\bm s} \neq 1$), then it will not appear circular on the sky, and its angular extent is also reduced.  
In order to accommodate these corrections in our mapping, we identify the effective angular extent of a loop with dimensionless radius $\zeta$ at redshift $z$ as 
\begin{equation}\label{eq:angle_scale_map}
    \theta_\mathrm{eff}(\zeta,z) 
    = 2 \, \mathrm{arctan} \left( \frac{\lambda \, \zeta}{a(z) H(z) s(z)}  \right) 
    \;.
\end{equation}
Notice that $\theta_\mathrm{eff}(\zeta,z)$ differs from $\theta_d(\zeta,z)$ through the factor $0 < \lambda < 1$, which accounts for the loop's reduced angular extent.  
We choose $\lambda = 0.3$ since we find that it improves the agreement between our analytical and numerical results.  

Notice that $\theta_\mathrm{eff}(\zeta,z)$, like $\theta_d(\zeta,z)$, is a monotonically decreasing function of $z$.  
Thus for a given photon pair opening angle $\theta_o$ we have $\theta_\mathrm{eff}(\zeta,z) = \theta_o$ when $z = z_\ast(\zeta,\theta_o)$.  
Here $z_\ast(\zeta,\theta_o)$ is the solution of
\begin{equation}\label{eq:z_ast_def}
    a(z_\ast) H(z_\ast) s(z_\ast)
    = 
    \lambda \zeta \, \cot\left(\dfrac{\theta_o}{2}\right) 
    \;.
\end{equation}
For example, in a matter-dominated universe we have
\begin{align}\label{eqn:z_star_effective}
    z_\ast(\zeta,\theta_o) = \lambda \zeta \cot \left( \frac{\theta_o}{2} \right) + \frac{\lambda^2 \zeta^2}{4}  \cot^2 \left( \frac{\theta_o}{2} \right) 
    \;,
\end{align}
and in fact this expression also provides a good approximation for the $\Lambda$CDM cosmology for $\theta \gtrsim \theta_\cmb \approx 0.015$.  
The behavior of $z_\ast(\zeta,\theta_o)$ is illustrated in \fref{fig:opening_angle}, which shows how $z_\ast$ decreases with increasing $\theta_o$ for a given dimensionless loop radius $\zeta$.  
A larger photon pair opening angle $\theta_o$ requires a larger loop to encircle it, and for a given $\zeta$ such larger loops are only present at later times when $z$ is smaller.  

It is also important to consider string loops at recombination, since this represents the earliest time at which CMB photons could possibly pass through loops and acquire correlated phase shifts.
By evaluating \eref{eq:angle_scale_map} at $z = z_\cmb$ we obtain 
\begin{equation}\label{eq:theta_cmb}
    \theta_\mathrm{min}(\zeta) 
    = \theta_\mathrm{eff}(\zeta,z_\cmb) 
    = 2 \, \mathrm{arctan}\bigl(0.015\lambda\zeta\bigr) 
    \approx 8.6 \times 10^{-3}\,\zeta,
    \;,
\end{equation}
which is the effective angular scale of loops with dimensionless radius $\zeta$ at recombination. 
Similarly we identify the angular scale $\theta_t$ corresponding to the `transition' point $z_t(\zeta) = e^{\zeta/2} - 1$ that we introduced in \eref{eq:Q_zeta0_lcdm2} to parametrize $Q(\zeta,z,0)$.  
Applying \eref{eq:angle_scale_map} gives 
\begin{align}\label{eqn:transition_angle}
    \theta_t 
    = \theta_\mathrm{eff}(\zeta,z_t(\zeta)) 
    = 2\,\mathrm{arctan}\left(2\lambda\right) 
    \simeq 1
    \;,
\end{align}
which is independent of $\zeta$.  
For $\zeta < \mathcal{O}(1)$, we have $\theta_t > \theta_\mathrm{min}(\zeta)$.

Having identified the loop's effective angular scale $\theta_\mathrm{eff}(\zeta,z)$ and special redshift $z_\ast(\zeta,\theta_o)$ where $\theta_o = \theta_\mathrm{eff}$, we construct the kernel function at finite opening angle.  
If a given pair of photons has an opening angle of $\theta_o$, the two photons develop correlated phase shifts upon passing through a loop at redshift $z$ with effective angular scale $\theta_\mathrm{eff}(\zeta,z)$ provided that $\theta_o < \theta_\mathrm{eff}(\zeta,z)$.  
Equivalently, these two photons can only become correlated at late times when $z \leq z_\ast(\zeta,\theta_o)$.  
This observation leads to the following relation for the kernel function at finite opening angle:  
\begin{align}\label{eq:Q_from_Q_zero}
    Q(\zeta,z,\theta_o) 
    & \simeq \begin{cases} 
    Q(\zeta,z,0) 
    & , \qquad 
    z \leq z_\ast(\zeta,\theta_o) \\
    0 
    & , \qquad 
    z > z_\ast(\zeta,\theta_o) \\
    \end{cases} 
    \;.
\end{align}
This relation can be understood as follows.  
From the definition of the kernel function in Eqs.~(\ref{eq:DPhi_from_Nboth})~and~(\ref{eq:DPhi_from_Q}), we can interpret $Q \sim dN_\mathrm{both}/dz$ as the rate of change of the effective number of loops crossed by the two photons.  
Then the correspondence between loop sizes and opening angles, captured by \eref{eq:z_ast_def} and illustrated in \fref{fig:opening_angle}, implies that $dN_\mathrm{both}/dz \rightarrow 0$ for $z > z_\ast$ since effectively loops are smaller in size than the opening angle $\theta_o$; while for $z \leq z_\ast$ since all loops are effectively bigger than the opening angle (and this is also where $\lambda$ comes into play), we basically have autocorrelation ($\theta_o = 0$) from $z = z_\ast$ onwards.  
Note that $Q(\zeta,z,\theta_o)$ is discontinuous at $z = z_\ast(\zeta,\theta_o)$, but there is no discontinuity in the correlation function, which involves an integral over $z$.  
We expect that the kernel function will become insensitive to the opening angle for $\theta_o < \theta_\mathrm{min}(\zeta)$, since $\theta_\mathrm{min}(\zeta)$ corresponds to the smallest loops in the network for a given $\zeta$.

Combining Eqs.~(\ref{eq:DPhi_from_Q})~and~(\ref{eq:Q_from_Q_zero}) lets the two-point correlation function of the birefringence signal be written as
\begin{align}\label{eq:DPhi_master_ini}
    \langle \Delta \Phi(\hat{\gamma}_1) \, \Delta \Phi(\hat{\gamma}_2) \rangle 
    & \simeq 
    \bigl(\Acal \, \alpha_\mathrm{em} \bigr)^2 
    \int_0^\infty \! \! \mathrm{d} \zeta 
    \int_0^{\mathrm{Min}[z_\ast(\zeta,\theta_o),z_\cmb]} \! \! \mathrm{d}z \ 
    Q(\zeta,z,0) \ 
    \chi(\zeta,z)
    \;,
\end{align}
where $Q(\zeta,z,0)$ is given by \eref{eq:Q_zeta0_lcdm2}, and $z_\ast(\zeta,\theta_o)$ is the solution of \eref{eq:z_ast_def}. In order to relax the sudden transition in the correlation function near $z_\cmb$ (owing to the un-physical discontinuity in the kernel function $Q(\zeta,z,\theta_o)$), we approximate the minimum function as 
\begin{align}\label{eq:ztilde_star}
    \mathrm{Min}[z_\ast(\zeta,\theta_o),z_\cmb]
    \simeq 
    \frac{z_\cmb\,z_\ast(\zeta,\theta_o)}{z_\cmb + z_\ast(\zeta,\theta_o)} \equiv \tilde{z}_\ast(\zeta,\theta_o).
\end{align}
With this approximation, the two-point angular correlation function becomes finally 
\begin{align}\label{eq:DPhi_master}
    \langle \Delta \Phi(\hat{\gamma}_1) \, \Delta \Phi(\hat{\gamma}_2) \rangle 
    & \simeq 
    \bigl(\Acal \, \alpha_\mathrm{em} \bigr)^2 
    \int_0^\infty \! \! \mathrm{d} \zeta 
    \int_0^{\tilde{z}_\ast(\zeta,\theta_o)} \! \! \mathrm{d}z \ 
    Q(\zeta,z,0) \ 
    \chi(\zeta,z)
    \;.
\end{align}
\eref{eq:DPhi_master} is one of the key results of our work; it allows the two-point angular correlation function of the CMB birefringence signal to be calculated analytically for an arbitrary model of the axion string network, which is parametrized by $\chi(\zeta,z)$.  
Specific choices of $\chi(\zeta,z)$ lead to different correlation functions, as we illustrate in the next section.  

\section{CMB birefringence from string network models}\label{sec:models}

In this section we consider a few models for the axion string network, corresponding to different choices for the dimensionless loop length distribution function $\chi(\zeta,z)$ from \eref{eq:nu_from_chi}.  
For each model we calculate the signature in axion-induced CMB birefringence.  

\subsection{Uniform loop size}\label{sub:model_1}

We begin with a simple string network model in which all the sub-Hubble scale loops have the same size (with respect to the Hubble scale) at any given time.  
This corresponds to choosing the loop length distribution function as 
\begin{align}\label{eq:chi_model_1}
    \chi(\zeta,z) = \xi_0 \, \delta(\zeta - \zeta_0) 
    \;,
\end{align}
which is independent of redshift $z$.  
The two model parameters, $\zeta_0$ and $\xi_0$, control the comoving string loop radius $r = \zeta_0 / a H$ (see \eref{eq:nu_from_chi}) and the effective number of loops per Hubble volume $\xi_\infty = \xi_0$ (see \eref{eq:xi_from_chi}). The two-point correlation function is calculated by performing the $\zeta$ integral in \eref{eq:DPhi_master} to obtain  
\begin{align}
    \langle \Delta \Phi(\hat{\gamma}_1) \, \Delta \Phi(\hat{\gamma}_2) \rangle 
    & \simeq 
    \xi_0 \, 
    \bigl( \Acal \, \alpha_\mathrm{em} \bigr)^2 
    \int_0^{\tilde{z}_\ast(\zeta_0,\theta_o)} \! \! \mathrm{d}z \ 
    Q(\zeta_0,z,0) 
    \;.
\end{align}
The kernel function at zero opening angle, $Q(\zeta,z,0)$, is well approximated by \eref{eq:Q_zeta0_lcdm2} in the $\Lambda$CDM cosmology. 
Performing the $z$ integral gives\footnote{This expression assumes $\theta_o < \theta_t \approx 1$.  For $\theta_o > \theta_t$ we have $\langle \Delta \Phi^2 \rangle \simeq (\xi_0/3\zeta_0) (\Acal \alpha_\mathrm{em})^2 \log^3(1 + \tilde{z}_\ast)$ instead. }
\begin{align}\label{eq:DPhi_model_1}
    \langle \Delta \Phi(\hat{\gamma}_1) \, \Delta \Phi(\hat{\gamma}_2) \rangle
    & \simeq 
    \frac{\xi_0 \zeta_0}{4} \, 
    \bigl( \Acal \, \alpha_\mathrm{em} \bigr)^2 
    \left( \log\bigl(1+\tilde{z}_\ast(\zeta_0,\theta_o)\bigr) - \dfrac{\zeta_0}{3} \right)
    \;,
\end{align}
where $\tilde{z}_\ast(\zeta_0,\theta_o)$ is given by~\eref{eq:ztilde_star}.  

\begin{figure}[t]
\centering
\includegraphics[width=0.75\columnwidth]{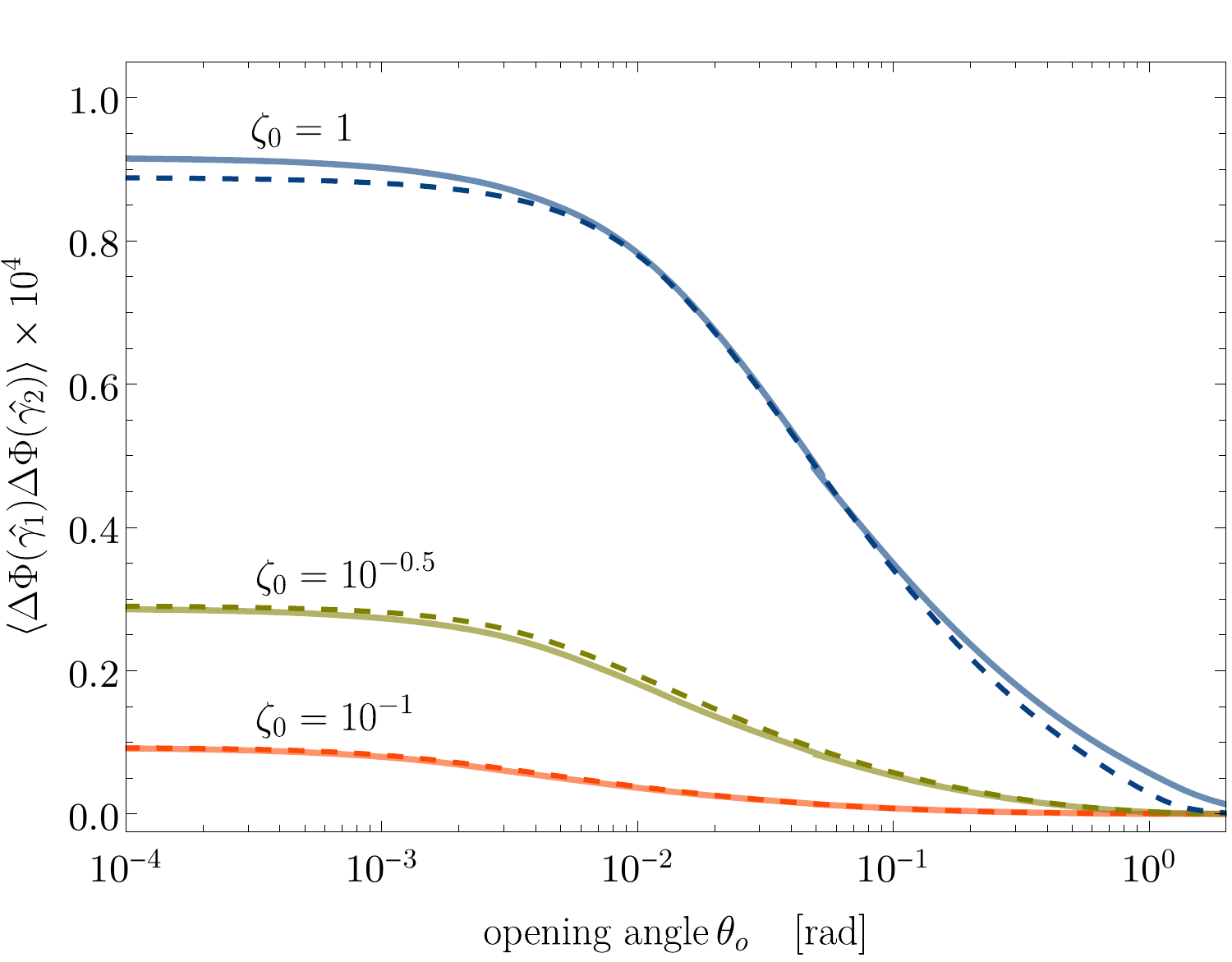}\vspace{0.1cm}\\
\caption{Two-point correlation function for the string network model in \sref{sub:model_1}.  At any time $t$, the network contains only loops of comoving radius $r = \zeta_0/aH$; i.e., the dimensionless loop length distribution is $\chi(\zeta,z) = \xi_0 \, \delta(\zeta-\zeta_0)$, and $\xi_0$ is the effective number of loops per Hubble volume.  The solid curves correspond to a direct numerical integration over the loops' locations and orientations, and the dashed curves show our analytical approximation to these integrals \pref{eq:DPhi_model_1}.  The width of the correlation function is $\sim \theta_\cmb \zeta_0$.  The curves with different $\zeta_0$ are related by an approximate scaling relation, $\langle \Delta \Phi^2 \rangle \approx \zeta_0 \, \mathcal{F}(\theta_o/\zeta_0)$; see \eref{eq:scaling_relations}.  We take $\mathcal{A} = 1$, $\xi_0 = 1$, and the signal scales as $\langle \Delta \Phi(\hat{\gamma}_1) \, \Delta \Phi(\hat{\gamma}_2) \rangle \propto \xi_0 \mathcal{A}^2$. 
}
\label{fig:corr_3zetas}
\end{figure}

The two-point correlation function is shown in \fref{fig:corr_3zetas}.  
The analytic approximation from \eref{eq:DPhi_model_1} is shown by the dashed curves.  
At large angular scales the correlation function decreases, because the correlation function counts the number of loops crossed by both photons \pref{eq:DPhi_from_Nboth} and there are few loops large enough to intersect both photons.  
Decreasing $\theta_o$ increases the correlation function, which now receives contributions from smaller loops.  
On small angular scales, the correlation function approaches a constant, because the smallest loops in the network were already present at recombination.  
Comparing our calculation with the one presented in \rref{Agrawal:2019lkr}, we find an $\mathcal{O}(1)$ difference in the amplitude and shape.  
We also evaluate the correlation function from \eref{eq:DPhi_numerical} where the $6$ integrals are estimated with numerical techniques.  
Our analytic approximation agrees extremely well with the ``direct'' numerical calculation.  
On the one hand, this serves to validate the assumptions that underlie our analytical approximation.  
On the other hand, it implies that the analytic approximations can be used for phenomenological studies, rather than having to perform the time-consuming numerical integration.  

Note that the two-point correlation function \pref{eq:DPhi_model_1} has an overall factor of $\zeta_0$.  
There are two effects that result in this dependence. First, changing $\zeta_0$ changes the area through which photons can pass and accumulate birefringence. Since the correlation depends on this area, this effect tries to make it go like $\propto \zeta_0^2$. Second, changing $\zeta_0$ also changes the number density of strings in the network which go like $\propto 1/\zeta_0$. This is because the total energy density in the network is held fixed. The net result is only a linear dependence on $\zeta_0$. The effect of this linear dependence on the power spectra, will therefore just be an overall re-scaling of the peak amplitude.

\begin{figure}[t]
\centering
\includegraphics[width=0.805\columnwidth]{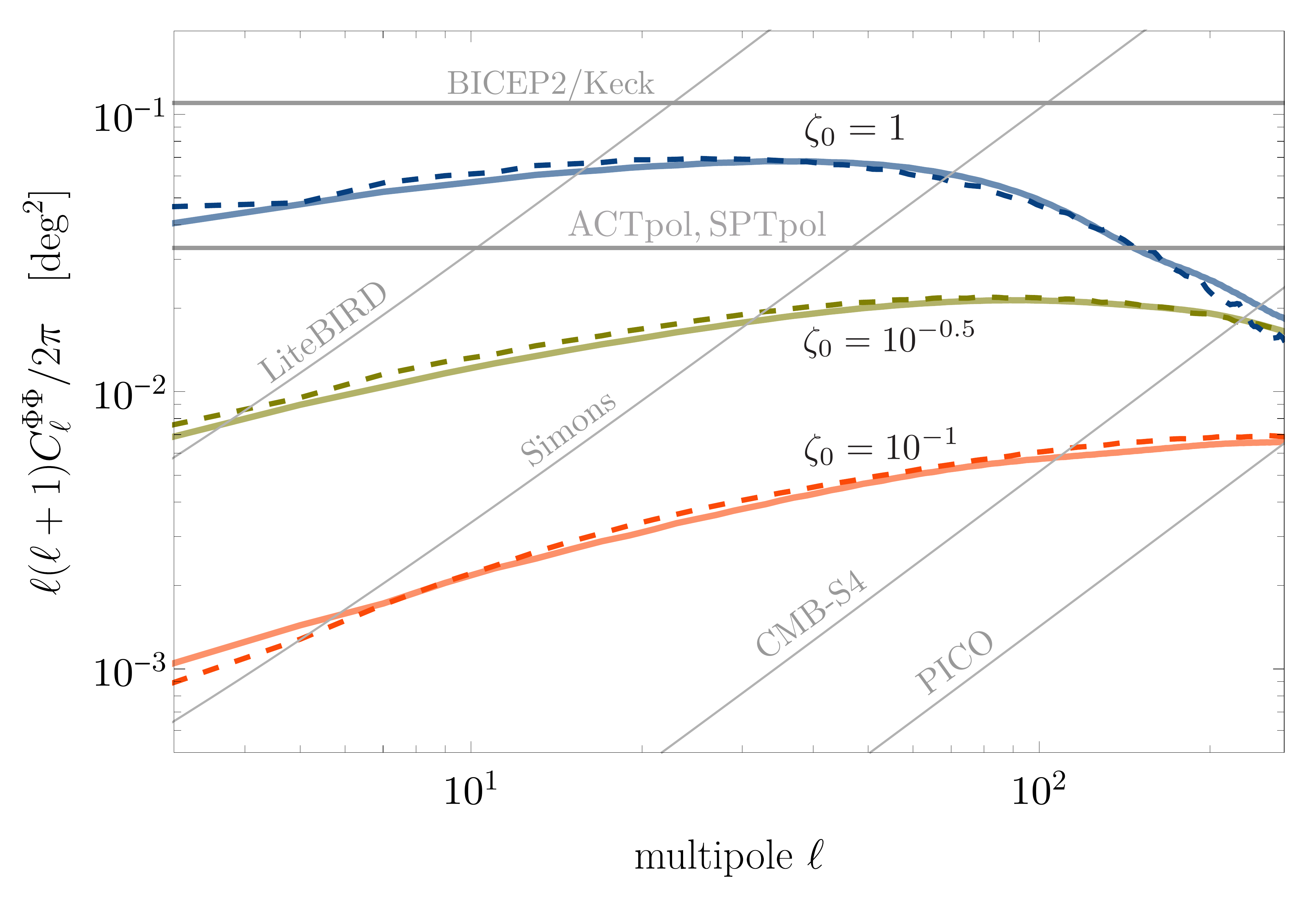}\vspace{0.1cm}\\
\caption{The angular power spectra corresponding to the birefringence correlation functions in \fref{fig:corr_3zetas}.   The thin gray curves show the projected sensitivities of several next-generation CMB surveys (68\% CL), and the current upper limits (assuming a scale-invariant power spectrum) from ACTpol~\cite{Namikawa:2020ffr} (2$\sigma$), SPTpol~\cite{Bianchini:2020osu} (2$\sigma$), and BICEP2/Keck (95\% CL)~\cite{Array:2017rlf}, . The curves peak at a multipole $\ell_p \sim 40/\zeta_0$.  The curves with different $\zeta_0$ are related by an approximate scaling relation, $\ell^2 C_\ell^{\Phi\Phi} \approx \zeta_0 \, \mathcal{G}(\zeta_0 \ell)$; see \eref{eq:scaling_relations}.  We take $\mathcal{A} = 1$, $\xi_0 = 1$, and the signal scales as $C_\ell^{\Phi\Phi} \propto \xi_0 \mathcal{A}^2$.  
}
\label{fig:Cl_model_1}
\end{figure}

We calculate the angular power spectrum using \eref{eq:Cl_def}, and our results appear in \fref{fig:Cl_model_1}.  
We show the predicted power spectra for several values of the dimensionless loop radius $\zeta_0$.
We also overlay the projected sensitivities of next-generation CMB surveys~\cite{Pogosian:2019jbt}.  
These observations will be able to test the presence of an axion string network in our Universe today.  
Testable string network models include those with large loops $\zeta_0 \gtrsim \mathcal{O}(10^{-1})$, abundant strings $\xi_0 \gtrsim \mathcal{O}(1)$, and a sizeable axion-photon coupling $\mathcal{A} \gtrsim \mathcal{O}(10^{-1})$.  
We reserve a detailed analysis of parameter space constraints for future work.  

An interesting feature of axion-string induced birefringence is the presence of a peak in the power spectrum.  
The location of this peak $\ell_p$ is approximately determined by the width of the correlation function when it drops to $1/e$ of its maximum value, $\theta_e$, through the relation $\ell_p \sim \pi / \theta_e$.  
In turn the angular scale $\theta_e$ is determined by the effective angular scale of the loops at the time of recombination \pref{eq:theta_cmb}, which gives $\theta_e \sim 10\theta_\mathrm{min}$ and $\ell_p \sim 40 / \zeta_0$. 
For $\ell < \ell_p$ the spectrum is blue tilted while for $\ell > \ell_p$ it is red tilted.  
This is apparent in \fref{fig:Cl_model_1}.
The larger the loop size $\zeta_0$, the smaller the multipole $\ell_p$ at which we see a peak.  
Therefore, if the network is dominated by Hubble-scale string loops ($\zeta_0 = 1$), $\theta_\mathrm{min} \simeq \theta_\cmb \approx 0.015$, then the peak is around $\ell_p \sim 40$.

From Figs.~\ref{fig:corr_3zetas} and \ref{fig:Cl_model_1} we observe that the two-point correlation function and the angular power spectrum exhibit the approximate scaling relations 
\begin{align}\label{eq:scaling_relations}
    \langle \Delta \Phi(\hat{\gamma}_1) \Delta \Phi(\hat{\gamma}_2) \rangle \approx \zeta_0 \, \mathcal{F}(\theta_o / \zeta_0) 
    \qquad \text{and} \qquad 
    \ell (\ell+1) C_\ell^{\Phi\Phi} / 2\pi \approx \zeta_0 \, \mathcal{G}(\zeta_0 \ell) 
    \;.
\end{align}
The effect of changing the dimensionless loop length $\zeta_0$ is to multiply the function by an overall factor of $\zeta_0$ and scale the coordinate $\theta_o \to \theta_o / \zeta_0$ or $\ell \to \zeta_0 \ell$.  
This behavior can be understood from our analytic approximation.  
Recall from the definition of $z_\ast(\zeta,\theta_o)$ in \eref{eq:z_ast_def} that in the small angle approximation\footnote{The maximum discrepancy is $10\%$ near $\theta_o = 1$, and the approximation improves for smaller $\theta_o$.} $\tan \theta_o/2 \approx \theta_0/2$, and this special redshift $z_\ast$ is only a function of the ratio $\theta_o/\zeta_0$, and we can write $z_\ast(\zeta_0,\theta_o) \approx z_\ast(\theta_o/\zeta_0)$.  
The same relation also applies for $\tilde{z}_\ast(\theta_o/\zeta_0)$.  
This approximation allows the two-point correlation function from \eref{eq:DPhi_model_1} to be written as in \eref{eq:scaling_relations} with $\mathcal{F}(\theta_o/\zeta_0) = (\xi_0/4) \, (\mathcal{A} \, \alpha_\mathrm{em})^2 \, \log\bigl(1 + \tilde{z}_\ast(\theta_o/\zeta_0)\bigr)$ where the $\zeta_0/3$ term is negligible as compared with the logarithm.   

Note that the scaling relations can also be expressed in terms of the characteristic angular scale $\theta_\mathrm{min}$ and the peak multipole $\ell_p$.  
Their relations to $\zeta_0$ are $\theta_\mathrm{min} \approx 8.5 \times 10^{-3} \zeta_0$ from \eref{eq:theta_cmb} and $\ell_p \sim 40 / \zeta_0$.  

\subsection{Mixed loop size}\label{sub:model_2}

The one-scale string network model from the previous subsection provides a good guide for how the axion string-induced birefringence signal depends on the size and abundance of string loops.  
However, a more realistic model would also account for the distribution over possible loop sizes.  
Here we are motivated by the results of a recent numerical simulation~\cite{Gorghetto:2018myk} that studied axion string network evolution.  
The simulation revealed that approximately 20\% of the total string length is present in sub-Hubble scale loops with a logarithmically-spaced distribution of lengths, and the remaining 80\% is present in `infinite' strings.  
In the simulation, an `infinite' string corresponds to a large loops that wraps the periodic boundary conditions of the simulation volume many times.  
For this general class of models, we will parametrize the dimensionless loop length distribution function as 
\begin{align}\label{eq:chi_model_2}
    \chi(\zeta,z) = (1-f_\mathrm{sub}) \, \xi_0 \, \delta(\zeta - \zeta_\mathrm{max}) + f_\mathrm{sub} \, \xi_0 \, \frac{\Theta(\zeta_\mathrm{max} - \zeta) \, \Theta(\zeta - \zeta_\mathrm{min})}{\zeta_\mathrm{max} - \zeta_\mathrm{min}}  
    \;.  
\end{align}
The first term corresponds to the `infinite' strings, which we model as Hubble-scale loops with a comoving radius $r = \zeta_{\mathrm{max}}/aH$ with $\zeta_{\mathrm{max}} = \mathcal{O}(1)$.
The second term corresponds to the logarithimcally-spaced distribution of sub-Hubble scale loops.  
Note that $\chi(\zeta,z)$ is independent of redshift $z$. 
There are four model parameters corresponding to the effective number of strings per Hubble volume $\xi_0$, the minimum dimensionless loop radius $\zeta_\mathrm{min}$, the maximum radius $\zeta_\mathrm{max}$, and the fraction of total string length in sub-hubble loops $f_\mathrm{sub}$. 
The coefficients are chosen such that $\xi_\infty(z) = \xi_0$ using \eref{eq:xi_from_chi}.  

To assess the birefringence signal in this model, the two-point correlation function is calculated from \eref{eq:DPhi_master}: 
\begin{equation}\label{eq:DPhi_model_2}
\begin{split}
    \langle \Delta \Phi(\hat{\gamma}_1) \, \Delta \Phi(\hat{\gamma}_2) \rangle 
    & \simeq 
    (1-f_\mathrm{sub}) \, \xi_0 \, 
    \bigl( \Acal \, \alpha_\mathrm{em} \bigr)^2 
    \int_0^{\tilde{z}_\ast(\zeta_\mathrm{max},\theta_o)} \! \! \mathrm{d}z \ 
    Q(\zeta_\mathrm{max},z,0) \ 
    \\ & \qquad 
    + 
    f_\mathrm{sub} \, \xi_0 \, 
    \bigl( \Acal \, \alpha_\mathrm{em} \bigr)^2 
    \int_{\zeta_\mathrm{min}}^{\zeta_\mathrm{max}} \! \! \mathrm{d}\zeta 
    \int_0^{\tilde{z}_\ast(\zeta,\theta_o)} \! \! \mathrm{d}z \ 
    \frac{Q(\zeta,z,0)}{\zeta_\mathrm{max} - \zeta_\mathrm{min}}
    \;.
\end{split}
\end{equation}
The first term, corresponding to the population of Hubble-scale loops, is equivalent to the integral we encountered in \sref{sub:model_1} and the result is simply \eref{eq:DPhi_model_1} with the replacements $\xi_0 \to (1-f_\mathrm{sub}) \, \xi_0$ and $\zeta_0 \to \zeta_\mathrm{max}$. 
In the second term, evaluating the integrals analytically is more complicated because of the $\zeta$-dependent upper limit of $z$ integration. However, for $\theta_o = 0$ we have $\tilde{z}_\ast(\zeta,0) = z_\cmb$ and the integrals simplify to 
\begin{equation}
\begin{split}
    \langle \Delta \Phi(\hat{\gamma}_1) \, \Delta \Phi(\hat{\gamma}_2) \rangle \Bigr|_{\theta_o = 0}
    & \simeq 
    (1-f_\mathrm{sub}) \, \xi_0 \, 
    \bigl( \Acal \, \alpha_\mathrm{em} \bigr)^2 \ 
    \dfrac{\zeta_\mathrm{max}}{4}\log\left(1+z_\cmb\right) 
    \\ & \qquad 
    + 
    f_\mathrm{sub} \, \xi_0 \, 
    \bigl( \Acal \, \alpha_\mathrm{em} \bigr)^2 \ 
    \dfrac{\zeta_\mathrm{max} + \zeta_\mathrm{min}}{8}\log\left(1+z_\cmb\right) 
    \;,
\end{split}
\end{equation}
where we have neglected additional terms that are $\mathcal{O}(\zeta_\mathrm{max}^2, \zeta_\mathrm{min}^2)$.  
It is interesting to note that the sub-Hubble scale loops contribute parametrically the same as the Hubble-scale loops modulo the different $f_\mathrm{sub}$ dependence.  

\begin{figure}[t]
\centering
\includegraphics[width=1\columnwidth]{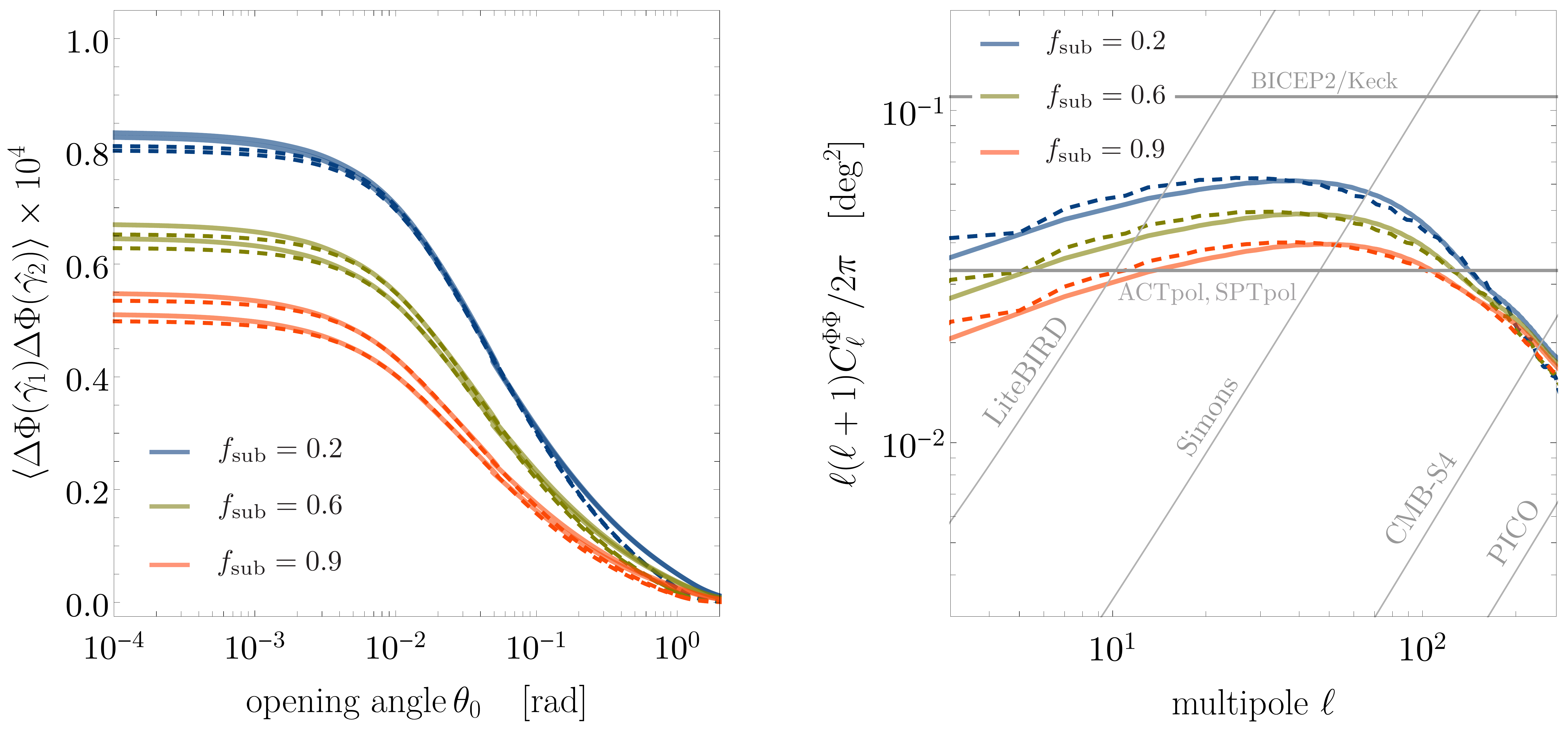}\vspace{0.1cm}\\
\caption{
The two-point correlation functions (left) and angular power spectra (right) for the string network model in \sref{sub:model_2}.  A fraction $1-f_\mathrm{sub}$ of the loops have size $\zeta_\mathrm{max} = 1$, and the remaining fraction $f_\mathrm{sub}$ have logarithmically-distributed sizes between $\zeta_\mathrm{max}$ and $\zeta_\mathrm{min} = 10^{-1}$ (upper set of curves) or $10^{-2}$ (lower set).  We only show $\zeta_\mathrm{min} = 10^{-1}$ on the right panel.  We take $\mathcal{A} = 1$, $\xi_0 = 1$, and the signal scales as $\langle \Delta \Phi(\hat{\gamma}_1) \, \Delta \Phi(\hat{\gamma}_2) \rangle, C_\ell^{\Phi\Phi} \propto \xi_0 \mathcal{A}^2$. }  
\label{fig:Cl_model_2}
\end{figure}

\fref{fig:Cl_model_2} shows the two-point correlation function and the angular power spectrum for mixed-length axion string networks.  
We fix the maximum dimensionless loop radius such that $\zeta_\mathrm{max} = 1$, and we show the results for $\zeta_\mathrm{min} = 10^{-1}$ and $10^{-2}$ as well as $f_\mathrm{sub} = 0.2$, $0.6$, and $0.9$.  
Raising $f_\mathrm{sub}$ reduces the number of Hubble-scale loops with $\zeta = \zeta_\mathrm{max}$ and increases the number of sub-Hubble loops with $\zeta_\mathrm{min} < \zeta < \zeta_\mathrm{max}$.  
Similarly, lowering $\zeta_\mathrm{min}$ spreads the distribution of sub-Hubble loops to smaller sizes while conserving the total amount of strings in this population.  
Both raising $f_\mathrm{sub}$ and lowering $\zeta_\mathrm{min}$ lead to a reduction in the birefringence signal.  
This is because larger loops `capture' more photons and have a stronger impact on the birefringence signal, but raising $f_\mathrm{sub}$ or lowering $\zeta_\mathrm{min}$ reduces the number of large loops.  

In this model as well, we clearly see a broken power law power spectrum (blue tilted at low $\ell$s and red-tilted at high $\ell$s), and correspondingly a peak at $\ell_p \sim 40 / \zeta_\mathrm{max}$ which is not that sensitive to different fractions $f_\mathrm{sub}$. This is because the signal is mostly dependent on the largest loops in the network, which we take to be Hubble-sized, $\zeta_\mathrm{max} = 1$. This also leads roughly to the same scaling behavior of the two point correlation function and the power spectrum as in the previous model from Sec.~\ref{sub:model_1}. Also note that the shape of the power spectrum for the model here is approximately degenerate with that earlier model.

In drawing \fref{fig:Cl_model_2} we have taken $\xi_0 = 1$ since the signal simply scales as $\xi_0$.  
However, recent string network simulations indicate that the average number of strings per Hubble volume could be much larger as a consequence of the scaling violation discussed in \sref{sub:strings}: a logarithimc growth in the total number of strings can accumulate over the course of the cosmic history, reaching $\xi_0 \approx 0.15 \log(f_a / H_0) \approx 10-20$~\cite{Gorghetto:2018myk} today.  
There is a corresponding enhancement to the signal $C_\ell^{\Phi\Phi} \propto \xi_0$, which the reader should bear in mind when studying \fref{fig:Cl_model_2} and evaluating the detection prospects for string-induced birefringence.  

Since this mixed loop length model is motivated in part by recent developments in axion string network simulations, it is interesting to ask how future refinements in our picture of axion string networks will impact their expected signatures in CMB birefringence. 
Future simulations will provide a better picture of the small-scale network properties, such as the population of sub-Hubble scale loops:  What is the length distribution of these loops?  How is it cutoff at small scales?  How does it evolve with time?  
The answers to these questions, however, are not expected to significantly impact the birefringence signal, and our study here demonstrates that it is dominated instead by the largest loops in the network.  
In this sense, the CMB signatures of an axion string network in the universe today are robust against hard-to-calculate and small-scale properties of axion string networks.  

On the other hand, the effective number of strings per Hubble volume, parametrized by $\xi_0$, directly affects the signal which grows proportionally to $\xi_0$.  
For string networks in scaling it is expected that $\xi_0 = \mathcal{O}(1)$, whereas a logarithmic deviation from scaling can lead to $\xi_0 = \mathcal{O}(10)$.  
This would lead to a significant enhancement of the birefringence signal, making detection prospects more favorable.  
At the same time, there is a degeneracy with the axion-photon coupling $g_{a\gamma\gamma} = - \Acal \alpha_\mathrm{em} / \pi f_a$. since the signal also goes as $\mathcal{A}^2$.  
Therefore it is imperative to better understand whether axion string networks (and global strings more generally) maintain scaling or whether they exhibit a possibly-logarithmic deviation from scaling.  
It is necessary to reduce this uncertainty in string network modeling in order to reliably extract information about ultralight axions and their interactions.  

\subsection{String network collapse}\label{sub:model_3}

As a final example, we consider a string network model that collapses, as we described in \sref{sub:strings}.  
If the axion's mass falls into the window $3H_0 \lesssim m_a \lesssim 3H_\cmb$ then the strings will develop domain walls connecting them at some time between recombination and today.  
Moreover if the discrete symmetry breaking implies $N_\mathrm{dw} = 1$ then this string-wall network will collapse into axion particles.
Whereas the birefringence signal ``accumulates'' as photons propagate through the string network, this accumulation is terminated when the string network decays.\footnote{When the string network collapses, the energy that it carries is transferred to the axion field.  
This inhomogeneous and evolving axion field will also contribute to CMB birefringence, and its effect is more closely connected to CMB birefringence from axion dark matter, which has been studied in great detail~ \cite{Fedderke:2019ajk}, rather than birefringence from axion strings.  In the analysis presented here, we neglect the effect of this axion field energy on CMB birefringence, and we provide an extended discussion of this point in \aref{app:biref_washout}.  }
We model the decaying string network with the following dimensionless loop length distribution function: 
\begin{align}\label{eq:chi_model_3}
    \chi(\zeta,z) = \xi_0 \ \delta(\zeta - \zeta_0) \ \Theta(z - z_c) 
    \;,
\end{align}
which generalizes the uniform size loop network model from \sref{sub:model_1}.  
There are three model parameters, where the new parameter $z_c$ is the redshift at which the network collapses.  
We define $z_c$ by the condition $m_a = 3 H(z_c)$, and if $m_a < 3 H_0$ then we take $z_c = 0$ and the network does not collapse.  
This condition resolves to 
\begin{align}\label{eq:zc_from_ma}
    z_c = \left[\left(\frac{(m_a/3H_0)^2 - \Omega_{\Lambda}}{\Omega_{m}}\right)^{1/3}-1 \right]\,\Theta(m_a - 3H_0)\;,
\end{align}
which gives the collapse redshift as a function of the axion's mass. 
We also define the effective collapse angle as $\theta_c = \theta_\mathrm{eff}(\zeta_0,z_c)$ where $\theta_\mathrm{eff}(\zeta_0,z) = 2\,\mathrm{arctan}\left(\lambda\,\zeta_0/(a(z) H(z) s(z))\right)$ is defined by  \eref{eq:angle_scale_map}, and the expression for a matter-dominated universe also works well giving $\theta_c = 2 \mathrm{arctan}[\lambda\,\zeta_0 / 2 (\sqrt{1+z_c}-1)]$.  

The two-point correlation function is calculated from \eref{eq:DPhi_master} which gives 
\begin{align}
    \langle \Delta \Phi(\hat{\gamma}_1) \, \Delta \Phi(\hat{\gamma}_2) \rangle 
    & \simeq 
    \xi_0 \, 
    \bigl( \Acal \, \alpha_\mathrm{em} \bigr)^2 
    \int_{z_c}^{\tilde{z}_\ast(\zeta_0,\theta_o)} \! \! \mathrm{d}z \ 
    Q(\zeta,z_0,0) \, 
    \Theta(\tilde{z}_\ast - z_c) 
    \;,
\end{align}
where the lower limit of integration is now $z_c$.  
Since this approximation to the correlation function vanishes identically for large opening angles (small $\tilde{z}_\ast$), there is a kink at the value of $\theta_o$ where $\tilde{z}_\ast(\zeta_0,\theta_o) = z_c$.  
To avoid unphysical features in the corresponding angular power spectrum, we smooth the kink by replacing $\tilde{z}_\ast \rightarrow (z_c^{3/2} + \tilde{z}^{3/2}_\ast)^{2/3}$.  
Moreover, by writing \eref{eq:chi_model_3} we have assumed that the network disappears as soon as $m_a \approx 3H$, whereas causality requires the collapse to take at least $\mathcal{O}(1)$ Hubble times.  
We expect that the finite duration of collapse will `smooth' the correlation function at large angular scales, and this observation further motivates our heuristic approach above. 
The evaluation of the $z$ integral is straightforward, and we find\footnote{This expression assumes $\theta_o$ and $\theta_c < \theta_t \approx 1$.  For larger values of $\theta_o$ and $\theta_c$ we have instead, $\langle \Delta \Phi^2 \rangle \simeq \xi_0 (\mathcal{A} \alpha_\mathrm{em})^2 [(\zeta_0/4)\log(1+(z_c^{3/2} + \tilde{z}^{3/2}_\ast)^{2/3}) - (1/3\zeta_0) \, \log^3(1+z_c) - \zeta_0^2/12]$ for $\theta_o < \theta_t < \theta_c$, and we have $\langle \Delta \Phi^2 \rangle \simeq (\xi_0/3\zeta_0) (\mathcal{A} \alpha_\mathrm{em})^2 [\log^3(1+(z_c^{3/2} + \tilde{z}^{3/2}_\ast)^{2/3}) - \log^3(1+z_c)]$ for $\theta_t < \theta_o, \theta_c$.}
\begin{align}\label{eq:DPhi_model_3}
    \langle \Delta \Phi(\hat{\gamma}_1) \, \Delta \Phi(\hat{\gamma}_2) \rangle
    & \simeq 
    \frac{\xi_0 \zeta_0}{4}
    \bigl( \Acal \, \alpha_\mathrm{em} \bigr)^2 
    \log\left(\frac{1+(z_c^{3/2} + \tilde{z}^{3/2}_\ast)^{2/3}}{1+z_c}\right) 
    \;.
\end{align}
Note that \eref{eq:DPhi_model_1} for the model in \sref{sub:model_1} is just a special case of this formula with $z_c = 0$.

\begin{figure}[t]
\centering
\includegraphics[width=1.\columnwidth]{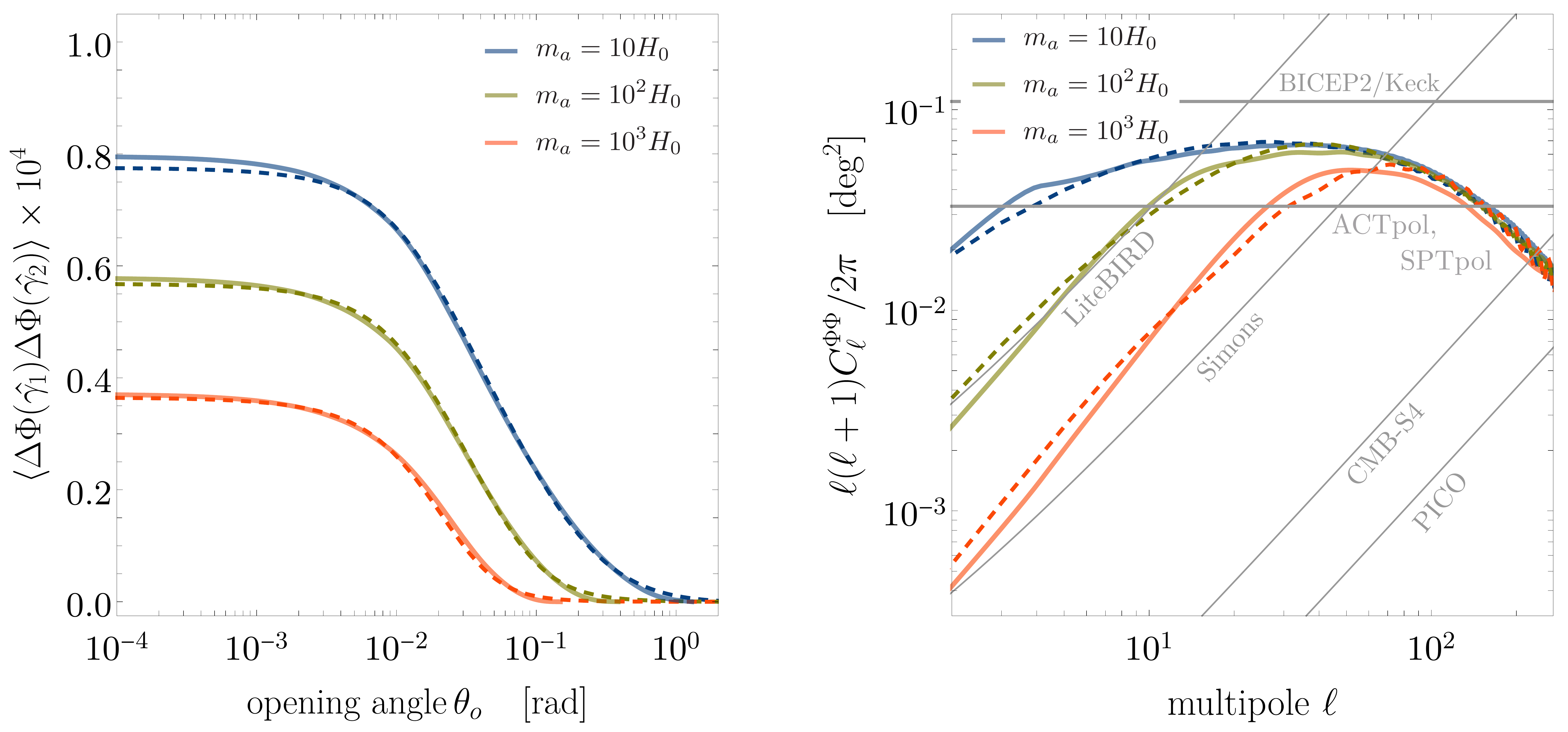}\vspace{0.1cm}\\
\caption{The two-point correlation functions (left) and angular power spectra (right) for the string network model in \sref{sub:model_3}.  At any time, the network only contains loops of comoving radius $r = \zeta_0/aH$, and we take $\zeta_0 = 1$.  The network collapses (assumed instantaneous) at redshift $z_c$, which is controlled by the axion mass through \eref{eq:zc_from_ma}, and we show $m_a / H_0 = 10, 10^2, 10^3$.  We take $\mathcal{A} = 1$, $\xi_0 = 1$, and the signal scales as $\langle \Delta \Phi(\hat{\gamma}_1) \, \Delta \Phi(\hat{\gamma}_2) \rangle, C_\ell^{\Phi\Phi} \propto \xi_0 \mathcal{A}^2$.
}
\label{fig:Cl_model_3}
\end{figure}

We show the correlation function and angular power spectrum in \fref{fig:Cl_model_3}.  
Allowing the string network to collapse before today leads to a suppression of the correlation function at large angular scales.
This can be understood as follows.  
Correlation results from photons passing through a common set of string loops.  
If the opening angle between the photons is large, the loops must also be large to intersect them both.  
In this model the comoving loop radius grows with time as $r = \zeta_0/aH$, so the largest loops aren't present until late times.  
If the network collapses before these large loops have formed, the correlation function is suppressed on large angular scales.  

In terms of the angular power spectrum, this suppression corresponds to $\ell (\ell+1) C_\ell^{\Phi\Phi} \sim \ell^2$ at small $\ell$.  
This introduces a new scale $\ell_{m_a}$ in the power spectrum, controlled by the mass of the axion $\ell_{m_a} \sim \pi/\theta_c(m_a)$. 
For masses not too close to $H_0$, we have $\ell_{m_a} \sim 3 \zeta_0^{-1} (m_a/H_0)^{1/3}$, giving $\ell_{m_a} \sim \zeta_0^{-1} \{7,10,25\}$ respectively for the three masses considered here. Notice that this transition scale falls below the peak at $\ell \approx \ell_p \sim 40 \zeta_0^{-1}$.
This is apparent in the right panel of~\fref{fig:Cl_model_3}.
If axion string-induced birefringence were observed, the presence of this additional scale (other than $\ell_p$) in the power spectrum could be interpreted as evidence that the string network has already decayed.
Additionally, it would be possible to infer the axion's mass based on the angular scale $\ell_{m_a}$.  

\section{Summary and discussion}\label{sec:conclusion}

In the work presented here we have studied CMB birefringence as a test of ultralight axions.  
If the Peccei-Quinn symmetry is broken after inflation, then a network of axion strings is expected to be formed at the corresponding cosmological phase transition.  
Axion strings surviving after recombination leave a distinctive imprint on the polarization pattern of the CMB.  
The evolution of axion string networks is still poorly understood, and the distribution of loop sizes in the universe today varies across different string network models.  
This observation motivated us to study the signatures of different axion string network models on CMB birefringence.  

The key results of the work presented here are summarized as follows.  
\begin{itemize}
    \item  Building upon the loop-crossing model of \rref{Agrawal:2019lkr}, we have developed an analytic formalism for calculating CMB birefringence for different axion string network models.  In particular, the two-point correlation function~\pref{eq:DPhi_from_Q} is expressed as a double-integral over a kernel function $Q(\zeta,z,\theta_o)$~\pref{eq:Q_def} and a variable $\chi(\zeta,z)$~\pref{eq:nu_from_chi} that parametrizes the network model.  We have derived an analytic approximation to the kernel function~\pref{eq:Q_from_Q_zero}, and validated these approximations against ``direct'' numerical integration.  The bottom line is \eref{eq:DPhi_master} from which the two-point correlation function can be calculated for any axion string network by evaluating two integrals on $\zeta$ and $z$.    
    \item  We have applied our analytic formalism to several string network models:  a network in which all the loops have the same size at any time, a network in which there is a log-distributed mix of loop sizes, and a network that collapses between recombination and today.  The choice of models is intended to demonstrate the versatility of our formalism while also being motivated by state-of-the-art string network simulations.  For each model we calculate the two-point angular correlation function of the CMB birefringence signal and the corresponding angular power spectrum.  Comparing with the projected sensitivity of planned telescopes, we conclude that axion-string-induced birefringence will be tested with these next-generation CMB surveys.  We reserve a detailed analysis of parameter space constraints for future work.  
    \item  We have emphasized that a better understanding of axion string networks, and global string networks more generally, is essential if we seek to use CMB birefringence to extract information about the axion's properties and its couplings to matter.  For example, the birefringence signal is generally proportional to $\xi_0 \mathcal{A}^2$ where $\xi_0$ is the effective number of strings per Hubble volume and $\mathcal{A} \propto g_{a\gamma\gamma}$ controls the axion-photon coupling strength.  We would like to interpret a detection of CMB birefringence as a measurement of $\mathcal{A}$, but we could be prevented from doing so, because of this degeneracy and our currently-poor understanding of axion string network evolution, which controls $\xi_0$.  
    \item  For the string network models in Secs.~\ref{sub:model_1}~and~\ref{sub:model_2}, the loop length distribution has a characteristic length scale (dimensionless loop radius, $\zeta_0$ or $\zeta_\mathrm{max}$).  For these models, we find that the correlation function and power spectrum exhibit the approximate scaling relations $\langle \Delta \Phi^2 \rangle \approx \zeta \, \mathcal{F}(\theta_o/\zeta)$ and $\ell^2 C_\ell^{\Phi\Phi} \approx \zeta \, \mathcal{G}(\zeta \ell)$.
    \item  We have pointed out a generic signature of axion-string-induced birefringence that could be used to distinguish this scenario from other sources of CMB birefringence.  From the correspondence between opening angle $\theta_o$ and redshift $z$, which we established in \sref{sub:finite_opening}, it follows that the two-point correlation function approaches a constant for $\theta_o$ below a model-dependent angular scale $\theta_e$. This leads to a peak in the angular power spectrum $\ell(\ell+1)C^{\Phi \Phi}_\ell$ at around $\ell_p \sim \pi/\theta_e$ with the signal being blue tilted and red tilted for $\ell < \ell_p$ and $\ell > \ell_p$ respectively. In the case where the mass of the ultralight axion is within the window $H_0 \lesssim m_a \lesssim H_\cmb$ and the domain wall number $N_\mathrm{dw} = 1$, there is another scale induced in the power spectrum, $\ell_{m_a} \sim \pi/\theta_c$, where $\theta_c$ is the angular extent of the largest loops present in the network at the time of collapse. Below this scale the power spectrum rises like $\ell(\ell+1)C_\ell^{\Phi\Phi} \sim \ell^2$. This provides a unique handle to probe such ultralight axion masses.
\end{itemize}

The work presented here represents a step toward testing an ultralight axion and its associated string network by using CMB birefringence.  
As such, there are several directions for future work.  
Our analytic formalism for describing string network models captures only sub-Hubble scale string loops, and cannot accommodate (formally) ``infinite'' strings whose curvature radius is much larger than the Hubble scale.  
It is important to assess the impact of these infinite strings, since they appear in string network simulations~\cite{Gorghetto:2018myk}, although we expect that their effect on CMB birefringence can be largely captured by Hubble-scale loops.  

Going beyond our study of axion string networks, it would be interesting to explore string-wall networks with stable domain walls ($N_\mathrm{dw} > 1$), and evaluate their distinctive signatures in CMB birefringence.  
We have provided a brief discussion of the extension of the loop crossing formalism for domain walls in \aref{app:domain_walls}.

Whereas we have motivated our study of axion string networks -- and the choice of $\chi(\zeta,z)$ -- based on the qualitative features observed by string network simulations, it would be interesting to calculate the birefringence signal directly from the simulation output.  However, it would be challenging to obtain a realistic estimate of the CMB signal given the limited dynamic range of these simulations.  

Finally, it is well-known that cosmic strings deform spacetime due to their gravitational influence, leading to a conical singularity and the associated phenomenon of gravitational lensing.  
Since axion strings also induce a birefringence of CMB photons through the axion-photon coupling, a cross-correlation of the lensing and birefringence measurements could serve as the ``smoking gun'' of axion string networks in the universe today. 

\subsubsection*{Acknowledgements}
We are grateful to Prateek Agrawal, Anson Hook, and Junwu Huang for illuminating discussions of axion string-induced birefringence; Toshiya Namikawa and Levon Pogosian for clarification of CMB birefringence limits; Ken Olum for valuable discussion of string network dynamics; and we also thank Jose J. Blanco-Pillado, Asier Lopez-Eiguren, Tanmay Vachaspati, and Alex Vilenkin for discussions and comments on the draft.  M.A. is supported by a NASA ATP theory grant NASA-ATP Grant No. 80NSSC20K0518.  

\appendix
\begin{appendix}

\section{Derivation of axion-induced birefringence}\label{app:derive_Delta_Phi}

Here we derive the induced birefringence effect on photons, due to a background of classical axion field. A detailed derivation of this effect appears in~\cite{Fedderke:2019ajk}, however with the simplifying assumption that the axion field does not change in the plane perpendicular to the direction of photon ray. Here we present a derivation that relaxes this assumption and obtain the same result.

From the axion modified Maxwell's equations, we have the following equation of motion in Coulomb gauge ($\nabla\cdot{\bf A} = 0$) for the {\it physical} degree of freedom ${\bf A}$ (i.e. after solving for the non-dynamical component $A^{0}$ of the vector field):
\begin{equation}
\left[\delta^{ij}\Box + \dfrac{\Acal\,\alpha_{em}}{\pi}\hat{\mathcal{L}}^{ij}\right]{A}^{j} = 0
\end{equation}
where $\hat{\mathcal{L}}^{ij}$ is a non-local operator equal to
\begin{eqnarray}
\hat{\mathcal{L}}^{ij} = \dfrac{\dot{a}}{f_a}\epsilon^{ikj}\partial_{k} - \epsilon^{ilj}\dfrac{\partial_{l}a}{f_a}\partial_t - \epsilon^{lkj}\partial_t\partial^{i}\left\{\dfrac{1}{\nabla^2}\dfrac{\partial_{l}a}{f_a}\partial_k\right\} + \left(\dfrac{\Acal\,\alpha_{em}}{\pi}\right)\epsilon^{ilk}\epsilon^{nmj}\dfrac{\partial_{l}a}{f_a}\partial_k\left\{\dfrac{1}{\nabla^2}\dfrac{\partial_{n}a}{f_a}\partial_{m}\right\}\;.
\end{eqnarray}
Let's assume the following ansatz for the vector field
\begin{eqnarray}
{\bf A} = {\bf p}({\bf r},t)e^{iS({\bf r},t)}
\end{eqnarray}
where we shall define the wave-vector and frequency as $\dot{S} \equiv \omega$ and $\nabla S \equiv - {\bf k}$. This gives
\begin{eqnarray}
&&\Box {\bf p} - k_{\mu}k^{\mu}\,{\bf p} + \dfrac{\Acal\,\alpha_{em}}{\pi f_a}\left(\dot{a}\,\nabla\times{\bf p} - \nabla a \times \dot{\bf p} - \text{Re}\left\{e^{-iS}\nabla\,\partial_t\,\dfrac{1}{\nabla^2}\left(\nabla a\cdot\,\nabla\times{\bf p}e^{iS}\right)\right\}\right)\nonumber\\
&&\;\;\;\;\;\;\;\;\;\;\;\;\;
+\left(\dfrac{\Acal\,\alpha_{em}}{\pi f_a}\right)^2\left(\nabla a\times\text{Re}\left\{e^{-iS}\,\nabla\,\dfrac{1}{\nabla^2}\,\left(\nabla a\cdot\,\nabla\times{\bf p}e^{iS}\right)\right\}\right)\nonumber\\
&&-i\Biggl[\left(\partial_{\mu}k^{\mu}\right){\bf p} + 2\,k^{\mu}\partial_{\mu}{\bf p} \nonumber\\
&&\;\;\;\;\;\;\;\;\;\;\;\;\;
- \dfrac{\Acal\,\alpha_{em}}{\pi f_a}\left(\dot{a}\,{\bf k}\times{\bf p} + \omega\,\nabla a \times {\bf p} + \text{Im}\left\{e^{-iS}\nabla\,\partial_t\,\dfrac{1}{\nabla^2}\left(\nabla a\cdot\,\nabla\times{\bf p}e^{iS}\right)\right\}\right)\nonumber\\
&&\;\;\;\;\;\;\;\;\;\;\;\;\;
-\left(\dfrac{\Acal\,\alpha_{em}}{\pi f_a}\right)^2\left(\nabla a\times\text{Im}\left\{e^{-iS}\,\nabla\,\dfrac{1}{\nabla^2}\,\left(\nabla a\cdot\,\nabla\times{\bf p}e^{iS}\right)\right\}\right)\Biggr] = 0\;,
\end{eqnarray}
along with the Coulomb gauge condition $\nabla\cdot{\bf p} = 0 = {\bf k}\cdot{\bf p}$. Now let's consider the eikonal limit in which both ${\bf p}$ and the background axion field $a$ vary slowly in space and time, while $S$ rapidly. Then, for the non-local term we have
\begin{equation}
\dfrac{1}{\nabla^2}\left(\nabla a\cdot\,\nabla\times{\bf p}e^{iS}\right) = \dfrac{-e^{iS}}{|{\bf k}|^2}\left(\nabla a\cdot\,\nabla\times{\bf p} - i\nabla a\cdot\,{\bf k}\times{\bf p}\right) + \mathcal{O}\left(|{\bf k}|^{-3}\right)\;.
\end{equation}
Collecting terms of different orders in powers of ${\bf k}$ and $\omega$, we can solve the system of equations order by order. To leading order, we get $k_{\mu}k^{\mu} \approx 0$ from the real part of the above equation; while from the imaginary part (upon discarding $(\partial_{\mu}k^{\mu}){\bf p}$ under eikonal approximation) we get
\begin{eqnarray}
k^{\mu}\partial_{\mu}{\bf p} &\approx& \dfrac{\Acal\,\alpha_{em}}{2\pi f_a}\left[\left(\dot{a}\,{\bf k} + \omega\nabla a\right)\times{\bf p} + \dfrac{\omega {\bf k}}{|{\bf k}|^2}\,\nabla a\cdot({\bf k}\times{\bf p})\right]\;.
\end{eqnarray}
Making use of vector identities along with the Coulomb gauge condition ${\bf k}\cdot{\bf p} = 0$, this can be further simplified to the following equation dictating the desired birefringence effect:
\begin{eqnarray}
n^{\mu}\partial_{\mu}{\bf p} &\approx& \dfrac{\Acal\,\alpha_{em}}{2\pi f_a}\,\left(n^{\mu}\partial_{\mu}a\right)\hat{\bf k}\times{\bf p}\;.
\end{eqnarray}
Here we have used $\omega = |{\bf k}|$ and $n^{\mu} = k^{\mu}/\omega$ is the null vector. Let's call ${\bf e}_1$ and ${\bf e}_2$ as the two orthonormal unit vectors (${\bf e}_1\cdot{\bf e}_2 = 0$ with $|{\bf e}_1| = |{\bf e}_2| = 1$) in the plane perpendicular to ${\bf k}$ such that ${\bf e}_1\times{\bf e}_2 = {\bf k}$. Then ${\bf p}$ admits the following general solution
\begin{eqnarray}
{\bf p}(\gamma) = p\left[\sin\left(\dfrac{\Acal\,\alpha_{em}}{2\pi f_a}\int^{\gamma}_{\gamma_i}\mathrm{d}\gamma\cdot\partial_{\gamma}a + \varphi\right){\bf e}_1 - \cos\left(\dfrac{\Acal\,\alpha_{em}}{2\pi f_a}\int^{\gamma}_{\gamma_i}\mathrm{d}\gamma\cdot\partial_{\gamma}a + \varphi\right){\bf e}_2\right]
\end{eqnarray}
where $p = |{\bf p}|$ and $\varphi = \tan^{-1}({\bf p}\cdot{{\bf e}_1}/{\bf p}\cdot{{\bf e}_2})|_{\gamma_i}$ are constants set by initial conditions and $\gamma$ is an affine parameter describing the null trajectory. This gives the stated polarization rotation along the photon trajectory $\gamma$:
\begin{equation}
\Delta \Phi = \tan^{-1}({\bf p}\cdot{{\bf e}_1}/{\bf p}\cdot{{\bf e}_2})|_{\gamma_f} - \tan^{-1}({\bf p}\cdot{{\bf e}_1}/{\bf p}\cdot{{\bf e}_2})|_{\gamma_i} = \dfrac{\Acal\,\alpha_{em}}{2\pi f_a}\int^{\gamma_f}_{\gamma_i} \mathrm{d}\gamma\cdot\partial_{\gamma}a\;.
\end{equation}

\subsubsection*{Stokes Parameters}

To begin with, the electric (and magnetic) field from the vector field ${\bf A}$ across the whole sky, can be obtained as usual:
\begin{eqnarray}
{\bf E} &=& - \dot{\bf A} -\nabla A^{0} = -i\omega{\bf p}\,e^{iS} + \mathcal{O}\left(\partial {\bf p},\dfrac{\Acal\,\alpha_{em}}{\pi f_a}\partial a\right)\nonumber\\
{\bf B} &=& \nabla\times{\bf A} = - i{\bf k}\times{\bf p}\,e^{iS} + \mathcal{O}\left(\partial {\bf p}\right)\;.
\end{eqnarray}
The relevant Stokes parameters~\cite{Jackson:1999} are
\begin{eqnarray}
Q &=& \left|{\bf e}_2\cdot{\bf E}\right|^2 - \left|{\bf e}_1\cdot{\bf E}\right|^2 \approx \omega^2 p^2\cos\left(2\Delta\Phi + 2\varphi\right)\nonumber\\
U &=& 2\text{Re}\left\{\left({\bf e}_2\cdot{\bf E}\right)^*\left({\bf e}_1\cdot{\bf E}\right)\right\}\approx \omega^2 p^2\sin\left(2\Delta\Phi + 2\varphi\right)\;,
\end{eqnarray}
using which we can define
\begin{eqnarray}
\mathcal{P} \equiv Q \pm iU = \omega^2 p^2\,e^{\pm i2\varphi}\,e^{\pm i2\Delta\Phi} = \mathcal{P}_0\,e^{\pm i2\Delta\Phi}\;.
\end{eqnarray}
Here $p$, $\varphi$, and $\Delta\Phi$ are fields defined on the 2-sphere, and $\mathcal{P}_0$ is the original (without birefringence) combination of Stokes parameters. This is the same as Eq.~(5.1) of Ref.~\cite{Agrawal:2019lkr}. See their discussion about the relevant effects on CMB polarization.

\section{Birefringence from axion radiation after network collapse}\label{app:biref_washout}

In this appendix, we derive the birefringence due to a classical bath of axion excitations. 
It will be useful to decompose the axion field (here denoted by $\varphi$) into Fourier mode functions $\psi_{\bf k}$ in the following way:
\begin{align}\label{eq:Fourier_decomp}
    \varphi({\bf x},t) = \left(\frac{a(t_c)}{a(t)}\right)^{3/2}\int\frac{\mathrm{d}^3{\bf k}}{(2\pi)^3} \psi_{\bf k}(t)\,e^{-i{\bf k}\cdot{\bf x}}\;.
\end{align}
where $t_c$ is the time at which the network collapses ($3H \simeq m_a$) and $t > t_c$. Here we have extracted the Hubble dilution factor $\sim a^{-3/2}$ out front, leaving the rest of the mode evolution within $\psi_{\bf k}(t)$. The birefringence observed today ($t = t_0$) at Earth (${\bf x} = 0$), in a particular direction $\hat{\bf p}$, due to the axion field is
\begin{align}\label{eq:birefringence_axionbathini}
    \Delta\Phi(\hat{\bf p}) = \frac{g_{\varphi\gamma\gamma}}{2}\,\Delta \varphi = \frac{g_{\varphi\gamma\gamma}}{2}\left[\varphi({\bf x}_c,t_c) - \varphi(0,t_0)\right]
\end{align}
where ${\bf x}_c = (r_c,\hat{\bf p}) = (t_0-t_c,\hat{\bf p})$ is the point on the CMB past light cone (at the `collapse 2-sphere') in the direction $\hat{\bf p} = (\sin\theta\cos\phi, \sin\theta\sin\phi, \cos\theta)$. Note that in writing the above formula we have assumed that there are no non-trivial topological configurations, and we can replace $\int_C \mathrm{d}X^{\mu}\partial_\mu\varphi$ with $\Delta\varphi$. Decomposing the axion field on this collapse 2-sphere on the CMB past light cone
\begin{align}
    \varphi({\bf x}_c,t_c) = \sum^{\infty}_{\ell=0}\sum^{\ell}_{m = -\ell}\bar{a}_{\ell m}(r_c,t_c)\,Y_{\ell m}(\hat{\bf p}),
\end{align}
the axion field today
\begin{align}
    \varphi(0,t_0) = \sum^{\infty}_{\ell=0}\sum^{\ell}_{m = -\ell}\bar{a}_{\ell m}(0,t_0)\,Y_{\ell m}(\hat{\bf p})\;,
\end{align}
and the birefringence field
\begin{align}
    \Delta\Phi(\hat{\bf p}) = \sum^{\infty}_{\ell=0}\sum^{\ell}_{m = -\ell}\bar{a}^{\Phi}_{\ell m}\,Y_{\ell m}(\hat{\bf p}) =  \sum^{\infty}_{\ell=0}\sum^{\ell}_{m = -\ell}\frac{g_{a\gamma\gamma}}{2\pi}\left[\bar{a}_{\ell m}(r_c,t_c) - \bar{a}_{\ell m}(0,t_0)\right]\,Y_{\ell m}(\hat{\bf p})\;,
\end{align}
and equating it to the previous Fourier decomposition gives\footnote{The normalisation is $\int\mathrm{d}^2\hat{\bf p} = 4\pi$, with the orthonormality $\int\mathrm{d}^2\hat{\bf p}\,Y^*_{lm}(\hat{\bf p})\,Y_{l'm'}(\hat{\bf p}) = \delta_{ll'}\,\delta_{mm'}$.}
\begin{align}
    \bar{a}^{\Phi}_{\ell m} = \frac{g_{a\gamma\gamma}}{2}\int\frac{\mathrm{d}^3{\bf k}}{(2\pi)^3} \int \mathrm{d}^2\hat{\bf p}\,Y^*_{\ell m}(\hat{\bf p})\left[\psi_{\bf k}(t_c)\,e^{-i\,r_c\,{\bf k}\cdot\hat{\bf p}} - \left(\frac{a(t_c)}{a(t_0)}\right)^3\psi_{\bf k}(t_0)\right]\;.
\end{align}
Therefore the birefringence power spectrum is
\begin{equation}\label{eq:birefringence_axionbath2}
\begin{split}
    \langle \bar{a}^{\Phi\,*}_{\ell' m'}\,\bar{a}^\Phi_{\ell m}\rangle &= \left(\frac{g_{a\gamma\gamma}}{2}\right)^2\int\frac{\mathrm{d}^3{\bf k}}{(2\pi)^3}\int\frac{\mathrm{d}^3{\bf k'}}{(2\pi)^3} \int \mathrm{d}^2\hat{\bf p}\,\int \mathrm{d}^2\hat{\bf p}'\,Y_{\ell' m'}(\hat{\bf p}')\,Y^*_{\ell m}(\hat{\bf p})
    \\ 
    & \quad \times 
    \Biggl\{\langle \psi^*_{{\bf k}'}(t_c)\,\psi_{\bf k}(t_c)\rangle\,e^{-i\,r_c\,({\bf k}\cdot\hat{\bf p} - {\bf k}'\cdot\hat{\bf p}')} 
    \\ 
    & \qquad \quad 
    - \left(\frac{a(t_c)}{a(t_0)}\right)^{3}\left(\langle \psi^*_{{\bf k}'}(t_c)\,\psi_{{\bf k}}(t_0)\rangle\,e^{i\,r_c\,{\bf k}'\cdot\hat{\bf p}'} + \langle \psi^*_{\bf k'}(t_0)\,\psi_{\bf k}(t_c)\rangle\,e^{-i\,r_c\,{\bf k}\cdot\hat{\bf p}}\right)
    \\ 
    & \qquad \quad 
    + \left(\frac{a(t_c)}{a(t_0)}\right)^{6}\langle \psi^*_{{\bf k}'}(t_0)\,\psi_{\bf k}(t_0)\rangle\Biggr\}
\end{split}
\end{equation}
where the average is taken with respect to the distribution from which the mode functions are picked, and we need to know $\langle \psi^*_{\bf k}(t_1)\psi_{\bf k'}(t_2)\rangle$ in order to proceed with the calculation. Under the assumption 
of statistical homogeneity and isotropy, we can write
\begin{align}\label{eq:def_P}
    \langle \psi^*_{\bf k}(t_1)\psi_{\bf k'}(t_2)\rangle = (2\pi)^3 \, \delta^3({\bf k} - {\bf k'})\,\mathcal{P}_k(t_1,t_2)\;,
\end{align}
where $\mathcal{P}_k(t_1,t_2)$ is defined through this equation and $\mathcal{P}_k(t,t)$ is a real quantity by definition. The birefringence power spectrum can now be evaluated further. From~\eref{eq:birefringence_axionbath2} we have
\begin{equation}\label{eq:birefringence_axionbath_2}
\begin{split}
    \langle \bar{a}^{\Phi\,*}_{\ell'm'}\,\bar{a}^\Phi_{\ell m}\rangle 
    &= \left(\frac{g_{a\gamma\gamma}}{2}\right)^2\int\frac{\mathrm{d}^3{\bf k}}{(2\pi)^3}\int \mathrm{d}^2\hat{\bf p}\,\int \mathrm{d}^2\hat{\bf p}'\,Y_{\ell'm'}(\hat{\bf p}')\,Y^*_{\ell m}(\hat{\bf p})
    \\ & \quad 
    \times \Biggl\{\mathcal{P}_k(t_c,t_c)\,e^{-i\,r_c\,{\bf k}\cdot(\hat{\bf p} - \hat{\bf p}')} 
    \\ & \qquad \quad 
    - \left(\frac{a(t_c)}{a(t_0)}\right)^{3}\left(\mathcal{P}_k(t_c,t_0)\,e^{i\,r_c\,{\bf k}\cdot\hat{\bf p}'} + \mathcal{P}_k(t_0,t_c)\,e^{-i\,r_c\,{\bf k}\cdot\hat{\bf p}}\right)
    \\ & \qquad \quad 
    + \left(\frac{a(t_c)}{a(t_0)}\right)^{6} \mathcal{P}_k(t_0,t_0)\Biggr\}\,
\end{split}
\end{equation}
which, upon using the identities
\begin{align}
    & e^{i{\bf k}\cdot{\bf x}} = \sum^{\infty}_{\ell=0}\,i^\ell\,(2\ell + 1)\,j_\ell(kx)\,P_\ell(\hat{\bf k}\cdot\hat{\bf x})\nonumber\\
    & \int\mathrm{d}^2\hat{\bf p}\,Y_{\ell m}(\hat{\bf p})\,P_{\ell'}(\hat{\bf k}\cdot\hat{\bf p}) = \frac{4\pi}{(2\ell+1)}\,\delta_{\ell,\ell'}\,Y_{\ell m}(\hat{\bf k})\;,
\end{align}
gives the following
\begin{equation}\label{eq:Cl_washout}
\begin{split}
    \langle \bar{a}^{\Phi\,*}_{\ell'm'}\,\bar{a}^\Phi_{\ell m}\rangle &= 4\pi\,\left(\frac{g_{a\gamma\gamma}}{2}\right)^2\Biggl\{\delta_{\ell,\ell'}\,\delta_{m,m'}\int \mathrm{d}k\left[\frac{k^2}{2\pi^2}\mathcal{P}_k(t_c,t_c)\right]j^2_{\ell}\left(k\,r_c\right) \\ 
    &-\left.\delta_{m,0}\,\delta_{m',0}\,\delta_{\ell,0}\,\delta_{\ell',0}\left(\frac{a(t_c)}{a(t_0)}\right)^{3}\int \mathrm{d}k\left[\frac{k^2}{2\pi^2}\left\{\mathcal{P}_k(t_c,t_0) + \mathcal{P}_k(t_0,t_c)\right\}\right]j_{0}\left(k\,r_c\right)\right. \\ 
    &+ \delta_{m,0}\,\delta_{m',0}\,\delta_{\ell,0}\,\delta_{\ell',0}\left(\frac{a(t_c)}{a(t_0)}\right)^{6}\int \mathrm{d}k\left[\frac{k^2}{2\pi^2}\mathcal{P}_k(t_0,t_0)\right]\Biggl\}\;.
\end{split}
\end{equation}
We will explicitly calculate power for $\ell>0$ multipoles here. Since for this we only need the first term, for our purposes we don't even need to study the evolution of the axion field for times after the collapse. 

We consider the situation where the occupation number function of the axion excitations, $n(k)$, is known after collapse. We assume that the time scale of collapse is smaller than the expansion rate of the Universe and so we neglect effects due to Hubble dilution during collapse. We also assume the bath of axions to be a collection of incoherent excitations that satisfy the equipartition condition $\langle |\dot{\psi}_{\bf k}(t_c)|^2 \rangle = \omega_k^2 \langle|\psi_{\bf k}(t_c)|^2\rangle = \omega_k^2\mathcal{P}_{k}(t_c,t_c)$ on account of negligible self interactions. To find the distribution for mode functions at $t_c$, consider the energy density in the axion field at $t_c$
\begin{align}\label{eq:Etot_1}
 \langle \rho_c \rangle
 = \int\frac{\mathrm{d}^3{\bf k}}{(2\pi)^3}\,\omega^2_k\,\mathcal{P}_k(t_c,t_c)
\end{align}
where $\omega_k = \sqrt{{\bf k}^2 + m_a^2}$. On the other hand using the occupation number function $n(k)$, we have
\begin{align}\label{eq:Etot_2}
    \langle \rho_c \rangle = \int\frac{\mathrm{d}^3{\bf k}}{(2\pi)^3}\,n(k)\,\omega_k\;.
\end{align}
Equating~\eref{eq:Etot_1} and~\eref{eq:Etot_2} 
gives
\begin{align}\label{eq:mode_amplitude}
    \mathcal{P}_k(t_c,t_c) = \frac{n(k)}{\omega_k}\;.
\end{align}
Since we are only interested with $\ell>0$ multipoles, we don't pursue the exercise of constructing an initial distribution for $\{\psi_{\bf k},\dot{\psi}_{\bf k}\}$ and then evolving different mode functions. So~\eref{eq:mode_amplitude} would suffice.

\begin{itemize}
    \item \uline{Birefringence power spectrum for $d\mathcal{E}_k/d(\log k) = const.$}
\end{itemize}

\begin{figure}[t]
\centering
\includegraphics[width=0.55\columnwidth]{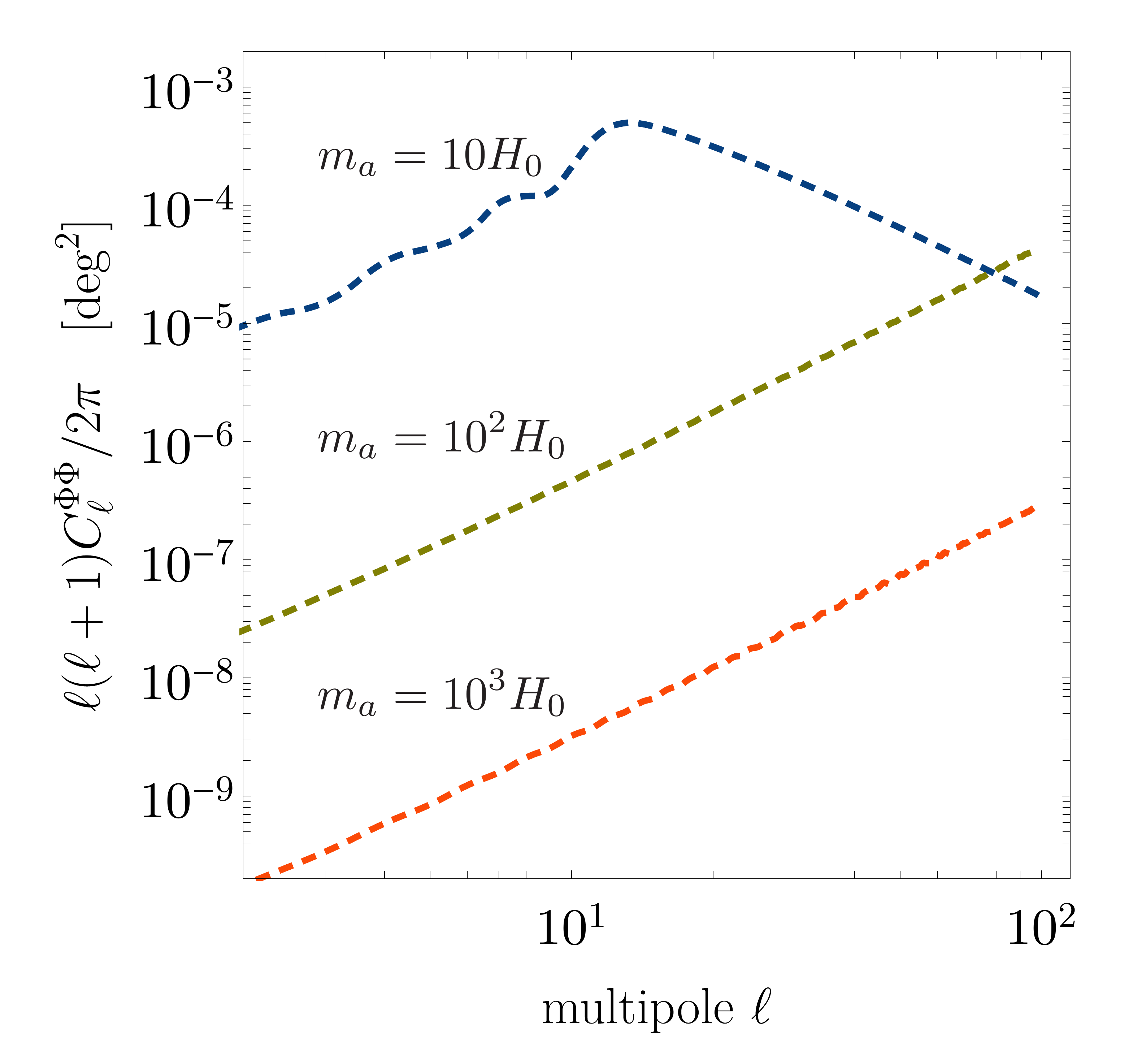}\vspace{0.1cm}\\
\caption{The birefringence power spectrum due to the barely-relativistic bath of axions produced after string network collapse at $3H = m_a$. The curves correspond to the $3$ different masses considered in the main text. We take $\mathcal{A} = 1$, $\xi_0 = 1$, and the signal scales as $C_\ell^{\Phi\Phi} \propto \xi_0 \mathcal{A}^2$.}
\label{fig:Cl_axionbath}
\end{figure}

~\eref{eq:Etot_2} can be re-written as
\begin{align}
    \langle \rho_c \rangle = \int^{\log f_a }_{\log m_a}\mathrm{d}(\log k)\left[\frac{k^3}{2\pi^2}n(k)\right]\omega_k\;,
\end{align}
while the energy density of the string network just before collapse is
\begin{align}
    \langle \rho_c \rangle \approx \pi\xi f_a^2\log\left(\frac{f_a}{H_c}\right)H_c^2
\end{align}
with $H_c \simeq m_a/3$. Here we have plugged in the tension $\mu$ of the string $\pi f_a^2\log\left(f_a/H\right)$. Therefore, conservation of energy itself suggests
\begin{align}\label{eq:occ_numb_func}
   \left[\frac{k^3}{2\pi^2}\,n(k)\right]\omega_k \approx \frac{\pi\xi}{9}m_a^2f_a^2\;.
\end{align}
On the other hand, considering a causally connected volume of size $L^3 \sim H_c^{-3}$ at the time of collapse, the power spectrum $d\mathcal{E}_k/dk$ defined through
\begin{align}
    \langle \rho_c \rangle \simeq H^3_c\int^{\log f_a }_{\log m_a}\mathrm{d}(\log k)\,k\,\frac{d\mathcal{E}_k}{dk},
\end{align}
must scale like $k^{-1}$
\begin{align}
     \frac{d\mathcal{E}_k}{dk} \propto \frac{Lf_a^2}{k} \qquad\qquad \frac{1}{L} \lesssim k \lesssim f_a
\end{align}
where $L = 2\pi/H_c = 6\pi/m_a$. This behaviour is supported by some simulations~\cite{Chang:1998tb,Saurabh:2020pqe}, however with a notable exception~\cite{Hiramatsu:2012gg} which suggests a much faster $\sim k^{-2}$ fall off. Here we shall assume the $k^{-1}$ behavior of the power spectrum $d\mathcal{E}_{k}/dk$. Then, using \eref{eq:occ_numb_func} and \eref{eq:mode_amplitude}, (and after some trivial re-scaling of quantities) we have the following birefringence power spectrum (c.f. \eqref{eq:Cl_washout})
\begin{equation}
\begin{split}
    \langle \bar{a}^{\Phi\,*}_{\ell'm'}\,\bar{a}^\Phi_{\ell m}\rangle 
    & \approx 4\pi\,\left(\frac{\mathcal{A}\alpha_\mathrm{em}}{2\pi}\right)^2\left\{\delta_{\ell,\ell'}\,\delta_{m,m'}\frac{\pi\xi}{9}\int^{\log\left(\frac{f_a}{m_a}\right)}_{0} \mathrm{d}x\,\,\frac{j^2_{\ell}\left(e^{x}\,m_a\,r_c\right)}{(1+e^{2x})}\right.\\
    &-\left.\delta_{m,0}\,\delta_{m',0}\,\delta_{\ell,0}\,\delta_{\ell',0}\left(\frac{a(t_c)}{a(t_0)}\right)^{3}\int \mathrm{d}k\left[\frac{k^2}{2\pi^2}\left\{\mathcal{P}_k(t_c,t_0) + \mathcal{P}_k(t_0,t_c)\right\}\right]j_{0}\left(k\,r_c\right)\right.\\
    &+ \left.\delta_{m,0}\,\delta_{m',0}\,\delta_{\ell,0}\,\delta_{\ell',0}\left(\frac{a(t_c)}{a(t_0)}\right)^{6}\int \mathrm{d}k\left[\frac{k^2}{2\pi^2}\mathcal{P}_k(t_0,t_0)\right]\right\}
\end{split}
\end{equation}
giving
\begin{align}
    C^{\Phi\,\Phi}_{\ell}\Big|_{\ell\ge 1} \approx \xi\frac{(\mathcal{A}\alpha_\mathrm{em})^2}{9}\int^{\log\left(\frac{f_a}{m_a}\right)}_{0} \mathrm{d}x\,\,\frac{j^2_{\ell}\left(e^{x}\,m_a\,r_c\right)}{(1+e^{2x})}\;.
\end{align}
For $\ell \geq 1$ the second and third terms in $\langle \bar{a}^{\Phi\,*}_{\ell'm'}\,\bar{a}^\Phi_{\ell m}\rangle$ do not contribute to $C_\ell^{\Phi\Phi}$.  
\fref{fig:Cl_axionbath} shows this for the three different masses considered in the main text. It is evident that this `washout birefringence' is much smaller as compared to that induced from the string network.

\section{Stable domain wall networks}\label{app:domain_walls}

In this appendix, we consider an axion string-wall network with stable domain walls ($N_\mathrm{dw} > 1$).  
We first derive a constraint on the model parameters by requiring that the defect network's energy density never comes to dominate the Universe.  
Second, we briefly discuss how our analytical formalism (for calculating CMB birefringence) can be extended to accommodate a stable string-wall network.

\textbf{Energy budget considerations.}  
Assuming that the string-wall network is in scaling \cite{Hiramatsu:2012sc,Hiramatsu:2010yn}, the energy density (per physical volume) of the domain wall network goes like
\begin{align}
    \rho_\mathrm{dw} \simeq \xi_\mathrm{dw}\,m_a\,f_a^2\,H
    \;,
\end{align}
where $\sigma \approx m_a f_a^2$ is a wall's surface tension.  
Here $\xi_\mathrm{dw}$ counts the effective number of domain walls in a Hubble volume, which is the ratio of the total domain wall area divided by a Hubble-scale area.
To avoid cosmological constraints on the presence of a string network in our universe today, we impose $\rho_{\mathrm{dw},0} \ll \rho_\mathrm{crit} = 3 M_\mathrm{pl}^2 H_0^2$, which resolves to 
\begin{align}
    \frac{\xi_\mathrm{dw} m_a f_a^2}{3 M_\mathrm{pl}^2 H_0} \simeq 0.3 \left( \frac{\xi_\mathrm{dw}}{1} \right) \left( \frac{m_a}{3 H_\mathrm{cmb}} \right) \left( \frac{f_a}{10^{16} \ \mathrm{GeV}} \right)^2 \ll 1 
    \;.
\end{align}
Note that $m_a < 3 H_\mathrm{cmb} \simeq 9 \times 10^{-29} \ \mathrm{eV}$ for walls that form after recombination.  
This can be easily satisfied; in order to have $\xi_\mathrm{dw} = \mathcal{O}(1)$, we can take $m_a \approx 3 H_\mathrm{cmb}$ and $f_a \ll 10^{16} \ \mathrm{GeV}$.

\textbf{Extending the formalism for domain walls.}  
The formalism is parallel in logic to that of the string network developed in this paper: With $n$ and $A$ being the wall distribution function and area of a wall respectively, we can build the relative wall area distribution function  $\nu_\mathrm{dw} = dn/dA$ for domain wall networks
\begin{align}\label{eq:nu_wall}
    \nu_\mathrm{dw}(A,z) & = \int_0^\infty \! \! \mathrm{d} \zeta \ \chi_\mathrm{dw}(\zeta,z) \, \frac{a(z) H(z)}{A} \ \delta\Bigl(A - \frac{\zeta}{a^2(z) H^2(z)} \Bigr) 
    \;.
\end{align}
where here $\zeta$ is to parameterize the area of walls in units of $H^{-2}$. Here $\chi_\mathrm{dw}$ is the model function that specifies a particular model for the string-wall network, defined analogously to the case of string network $\chi_\mathrm{dw} = d\xi_\mathrm{dw}/d\zeta$. Here
\begin{align}
    \xi_\mathrm{dw}(A,z) =  a^{-1}(z)H^{-1}(z)\int^{A}_{0} \mathrm{d}A'\,A'\,\nu_\mathrm{dw}(A',z) =  \int^{A\,a^2(z)H^2(z)}\mathrm{d}\zeta\; \chi_\mathrm{dw}(\zeta,z)
\end{align}
is the dimensionless quantity that counts the total comoving area of walls (in units of $H^{-2}$) in the network within a Hubble volume, that is made up of walls up to size $A$. Using this, the energy density of all walls with comoving area less than $A$ is recovered as $\rho_\mathrm{dw} = \xi_\mathrm{dw}(A,z)\,a(z)^3\,\sigma\,H(z)$ where $\sigma$ is the wall tension.

With this, and the fact that every wall crossing induces $|\Delta\Phi| =  \mathcal{A}\alpha_\mathrm{em}/N_\mathrm{dw}$, we can write the correlation function following a similar line of reasoning as in the case of string loop network:
\begin{align}
    \langle\Delta\Phi(\hat{\gamma}_1)\,\Delta\,\Phi(\hat{\gamma}_2)\rangle &= \frac{(\mathcal{A}\alpha_\mathrm{em})^2}{N^2_\mathrm{dw}}\int^{z_\cmb}_0\mathrm{d}z\int\mathrm{d}^2\hat{s}\int\mathrm{d}\zeta\frac{a^3s^2H^2}{\zeta\,a_0}\chi_\mathrm{dw}(\zeta,z)\nonumber\\
    &\qquad\qquad\Theta\left(\zeta - a^2H^2d_1^2\right)\,\Theta\left(\zeta - a^2H^2d_1^2\right)\,\Theta\left(l_1\right)\,\Theta\left(l_2\right)\;.
\end{align}
Here we have assumed that $d_i$ and $l_i$ are defined as before (c.f.~\eref{eq:d_gamma}; \fref{fig:loop_geometry}), i.e.  $d_1$ and $d_2$ are the distances of photon crossings from the center of a wall. Finally to map the formalism to that of the string network model developed in this paper, we can simply change variables $\zeta \rightarrow \sqrt{\zeta}$ and recover~\eref{eq:DPhi_numerical} (modulo extra factors of $\zeta$ and $N_\mathrm{dw}$).

\end{appendix}

\bibliographystyle{utphys}
\bibliography{axion_string.bib}

\providecommand{\href}[2]{#2}\begingroup\raggedright\begin{thebibliography}{10}

\bibitem{Wilczek:1977pj}
F.~Wilczek, ``{Problem of Strong p and t Invariance in the Presence of
  Instantons},''
\href{http://dx.doi.org/10.1103/PhysRevLett.40.279}{{\em Phys.Rev.Lett.}
  {\bfseries 40} (1978) 279--282}.

\bibitem{Crewther:1979pi}
R.~Crewther, P.~Di~Vecchia, G.~Veneziano, and E.~Witten, ``{Chiral Estimate of
  the Electric Dipole Moment of the Neutron in Quantum Chromodynamics},''
\href{http://dx.doi.org/10.1016/0370-2693(79)90128-X}{{\em Phys.Lett.}
  {\bfseries B88} (1979) 123}.

\bibitem{Afach:2015sja}
J.~M. Pendlebury {\em et~al.}, ``{Revised experimental upper limit on the
  electric dipole moment of the neutron},''
  \href{http://dx.doi.org/10.1103/PhysRevD.92.092003}{{\em Phys. Rev. D}
  {\bfseries 92} no.~9, (2015) 092003},
  \href{http://arxiv.org/abs/1509.04411}{{\ttfamily arXiv:1509.04411
  [hep-ex]}}.

\bibitem{Preskill:1982cy}
J.~Preskill, M.~B. Wise, and F.~Wilczek, ``{Cosmology of the Invisible
  Axion},''
\href{http://dx.doi.org/10.1016/0370-2693(83)90637-8}{{\em Phys.Lett.}
  {\bfseries B120} (1983) 127--132}.

\bibitem{Abbott:1982af}
L.~Abbott and P.~Sikivie, ``{A Cosmological Bound on the Invisible Axion},''
\href{http://dx.doi.org/10.1016/0370-2693(83)90638-X}{{\em Phys.Lett.}
  {\bfseries B120} (1983) 133--136}.

\bibitem{Dine:1982ah}
M.~Dine and W.~Fischler, ``{The Not So Harmless Axion},''
  \href{http://dx.doi.org/10.1016/0370-2693(83)90639-1}{{\em Phys. Lett. B}
  {\bfseries 120} (1983) 137--141}.

\bibitem{Witten:1984dg}
E.~Witten, ``{Some Properties of O(32) Superstrings},''
  \href{http://dx.doi.org/10.1016/0370-2693(84)90422-2}{{\em Phys. Lett. B}
  {\bfseries 149} (1984) 351--356}.

\bibitem{Svrcek:2006yi}
P.~Svrcek and E.~Witten, ``{Axions In String Theory},''
  \href{http://dx.doi.org/10.1088/1126-6708/2006/06/051}{{\em JHEP} {\bfseries
  06} (2006) 051}, \href{http://arxiv.org/abs/hep-th/0605206}{{\ttfamily
  arXiv:hep-th/0605206}}.

\bibitem{Arvanitaki:2009fg}
A.~Arvanitaki, S.~Dimopoulos, S.~Dubovsky, N.~Kaloper, and J.~March-Russell,
  ``{String Axiverse},''
  \href{http://dx.doi.org/10.1103/PhysRevD.81.123530}{{\em Phys. Rev. D}
  {\bfseries 81} (2010) 123530},
  \href{http://arxiv.org/abs/0905.4720}{{\ttfamily arXiv:0905.4720 [hep-th]}}.

\bibitem{Acharya:2010zx}
B.~S. Acharya, K.~Bobkov, and P.~Kumar, ``{An M Theory Solution to the Strong
  CP Problem and Constraints on the Axiverse},''
  \href{http://dx.doi.org/10.1007/JHEP11(2010)105}{{\em JHEP} {\bfseries 11}
  (2010) 105}, \href{http://arxiv.org/abs/1004.5138}{{\ttfamily arXiv:1004.5138
  [hep-th]}}.

\bibitem{Cicoli:2012sz}
M.~Cicoli, M.~Goodsell, and A.~Ringwald, ``{The type IIB string axiverse and
  its low-energy phenomenology},''
  \href{http://dx.doi.org/10.1007/JHEP10(2012)146}{{\em JHEP} {\bfseries 10}
  (2012) 146}, \href{http://arxiv.org/abs/1206.0819}{{\ttfamily arXiv:1206.0819
  [hep-th]}}.

\bibitem{Hui:2016ltb}
L.~Hui, J.~P. Ostriker, S.~Tremaine, and E.~Witten, ``{Ultralight scalars as
  cosmological dark matter},''
  \href{http://dx.doi.org/10.1103/PhysRevD.95.043541}{{\em Phys. Rev. D}
  {\bfseries 95} no.~4, (2017) 043541},
  \href{http://arxiv.org/abs/1610.08297}{{\ttfamily arXiv:1610.08297
  [astro-ph.CO]}}.

\bibitem{Spector:2016vwo}
{\bfseries ALPS} Collaboration, A.~Spector,
  \href{http://dx.doi.org/10.3204/DESY-PROC-2009-03/Spector_Aaron}{``{ALPS II
  technical overview and status report},''} in {\em {12th Patras Workshop on
  Axions, WIMPs and WISPs}}, pp.~133--136.
\newblock 2017.
\newblock \href{http://arxiv.org/abs/1611.05863}{{\ttfamily arXiv:1611.05863
  [physics.ins-det]}}.

\bibitem{Beyer:2020dag}
K.~A. Beyer, G.~Marocco, R.~Bingham, and G.~Gregori, ``{Axion detection through
  resonant photon-photon collisions},''
  \href{http://dx.doi.org/10.1103/PhysRevD.101.095018}{{\em Phys. Rev. D}
  {\bfseries 101} no.~9, (2020) 095018},
  \href{http://arxiv.org/abs/2001.03392}{{\ttfamily arXiv:2001.03392
  [hep-ph]}}.

\bibitem{Inoue:2008zp}
Y.~Inoue, Y.~Akimoto, R.~Ohta, T.~Mizumoto, A.~Yamamoto, and M.~Minowa,
  ``{Search for solar axions with mass around 1 eV using coherent conversion of
  axions into photons},''
  \href{http://dx.doi.org/10.1016/j.physletb.2008.08.020}{{\em Phys. Lett. B}
  {\bfseries 668} (2008) 93--97},
  \href{http://arxiv.org/abs/0806.2230}{{\ttfamily arXiv:0806.2230
  [astro-ph]}}.

\bibitem{Arik:2008mq}
{\bfseries CAST} Collaboration, E.~Arik {\em et~al.}, ``{Probing eV-scale
  axions with CAST},''
  \href{http://dx.doi.org/10.1088/1475-7516/2009/02/008}{{\em JCAP} {\bfseries
  02} (2009) 008}, \href{http://arxiv.org/abs/0810.4482}{{\ttfamily
  arXiv:0810.4482 [hep-ex]}}.

\bibitem{Anastassopoulos:2017ftl}
{\bfseries CAST} Collaboration, V.~Anastassopoulos {\em et~al.}, ``{New CAST
  Limit on the Axion-Photon Interaction},''
  \href{http://dx.doi.org/10.1038/nphys4109}{{\em Nature Phys.} {\bfseries 13}
  (2017) 584--590}, \href{http://arxiv.org/abs/1705.02290}{{\ttfamily
  arXiv:1705.02290 [hep-ex]}}.

\bibitem{Armengaud:2019uso}
{\bfseries IAXO} Collaboration, E.~Armengaud {\em et~al.}, ``{Physics potential
  of the International Axion Observatory (IAXO)},''
  \href{http://dx.doi.org/10.1088/1475-7516/2019/06/047}{{\em JCAP} {\bfseries
  06} (2019) 047}, \href{http://arxiv.org/abs/1904.09155}{{\ttfamily
  arXiv:1904.09155 [hep-ph]}}.

\bibitem{Aprile:2020tmw}
{\bfseries XENON} Collaboration, E.~Aprile {\em et~al.}, ``{Excess electronic
  recoil events in XENON1T},''
  \href{http://dx.doi.org/10.1103/PhysRevD.102.072004}{{\em Phys. Rev. D}
  {\bfseries 102} no.~7, (2020) 072004},
  \href{http://arxiv.org/abs/2006.09721}{{\ttfamily arXiv:2006.09721
  [hep-ex]}}.

\bibitem{Bauer:2018uxu}
M.~Bauer, M.~Heiles, M.~Neubert, and A.~Thamm, ``{Axion-Like Particles at
  Future Colliders},''
  \href{http://dx.doi.org/10.1140/epjc/s10052-019-6587-9}{{\em Eur. Phys. J. C}
  {\bfseries 79} no.~1, (2019) 74},
  \href{http://arxiv.org/abs/1808.10323}{{\ttfamily arXiv:1808.10323
  [hep-ph]}}.

\bibitem{Raffelt:2006cw}
G.~G. Raffelt, ``{Astrophysical axion bounds},''
  \href{http://dx.doi.org/10.1007/978-3-540-73518-2_3}{{\em Lect. Notes Phys.}
  {\bfseries 741} (2008) 51--71},
  \href{http://arxiv.org/abs/hep-ph/0611350}{{\ttfamily arXiv:hep-ph/0611350}}.

\bibitem{Fortin:2018aom}
J.-F. Fortin and K.~Sinha, ``{X-Ray Polarization Signals from Magnetars with
  Axion-Like-Particles},''
  \href{http://dx.doi.org/10.1007/JHEP01(2019)163}{{\em JHEP} {\bfseries 01}
  (2019) 163}, \href{http://arxiv.org/abs/1807.10773}{{\ttfamily
  arXiv:1807.10773 [hep-ph]}}.

\bibitem{Fortin:2018ehg}
J.-F. Fortin and K.~Sinha, ``{Constraining Axion-Like-Particles with Hard X-ray
  Emission from Magnetars},''
  \href{http://dx.doi.org/10.1007/JHEP06(2018)048}{{\em JHEP} {\bfseries 06}
  (2018) 048}, \href{http://arxiv.org/abs/1804.01992}{{\ttfamily
  arXiv:1804.01992 [hep-ph]}}.

\bibitem{Dessert:2019sgw}
C.~Dessert, A.~J. Long, and B.~R. Safdi, ``{X-ray signatures of axion
  conversion in magnetic white dwarf stars},''
  \href{http://dx.doi.org/10.1103/PhysRevLett.123.061104}{{\em Phys. Rev.
  Lett.} {\bfseries 123} no.~6, (2019) 061104},
\href{http://arxiv.org/abs/1903.05088}{{\ttfamily arXiv:1903.05088 [hep-ph]}}.

\bibitem{Dessert:2019dos}
C.~Dessert, J.~W. Foster, and B.~R. Safdi, ``{Hard X-ray Excess from the
  Magnificent Seven Neutron Stars},''
  \href{http://dx.doi.org/10.3847/1538-4357/abb4ea}{{\em Astrophys. J.}
  {\bfseries 904} no.~1, (2020) 42},
  \href{http://arxiv.org/abs/1910.02956}{{\ttfamily arXiv:1910.02956
  [astro-ph.HE]}}.

\bibitem{Buschmann:2019pfp}
M.~Buschmann, R.~T. Co, C.~Dessert, and B.~R. Safdi, ``{X-ray Search for Axions
  from Nearby Isolated Neutron Stars},''
  \href{http://arxiv.org/abs/1910.04164}{{\ttfamily arXiv:1910.04164
  [hep-ph]}}.

\bibitem{Dessert:2020lil}
C.~Dessert, J.~W. Foster, and B.~R. Safdi, ``{X-ray Searches for Axions from
  Super Star Clusters},'' \href{http://arxiv.org/abs/2008.03305}{{\ttfamily
  arXiv:2008.03305 [hep-ph]}}.

\bibitem{Arvanitaki:2010sy}
A.~Arvanitaki and S.~Dubovsky, ``{Exploring the String Axiverse with Precision
  Black Hole Physics},''
  \href{http://dx.doi.org/10.1103/PhysRevD.83.044026}{{\em Phys. Rev. D}
  {\bfseries 83} (2011) 044026},
  \href{http://arxiv.org/abs/1004.3558}{{\ttfamily arXiv:1004.3558 [hep-th]}}.

\bibitem{Kitajima:2018zco}
N.~Kitajima, J.~Soda, and Y.~Urakawa, ``{Gravitational wave forest from string
  axiverse},'' \href{http://dx.doi.org/10.1088/1475-7516/2018/10/008}{{\em
  JCAP} {\bfseries 10} (2018) 008},
  \href{http://arxiv.org/abs/1807.07037}{{\ttfamily arXiv:1807.07037
  [astro-ph.CO]}}.

\bibitem{Stott:2018opm}
M.~J. Stott and D.~J.~E. Marsh, ``{Black hole spin constraints on the mass
  spectrum and number of axionlike fields},''
  \href{http://dx.doi.org/10.1103/PhysRevD.98.083006}{{\em Phys. Rev. D}
  {\bfseries 98} no.~8, (2018) 083006},
  \href{http://arxiv.org/abs/1805.02016}{{\ttfamily arXiv:1805.02016
  [hep-ph]}}.

\bibitem{Du:2018uak}
{\bfseries ADMX} Collaboration, N.~Du {\em et~al.}, ``{A Search for Invisible
  Axion Dark Matter with the Axion Dark Matter Experiment},''
  \href{http://dx.doi.org/10.1103/PhysRevLett.120.151301}{{\em Phys. Rev.
  Lett.} {\bfseries 120} no.~15, (2018) 151301},
  \href{http://arxiv.org/abs/1804.05750}{{\ttfamily arXiv:1804.05750
  [hep-ex]}}.

\bibitem{Braine:2019fqb}
{\bfseries ADMX} Collaboration, T.~Braine {\em et~al.}, ``{Extended Search for
  the Invisible Axion with the Axion Dark Matter Experiment},''
  \href{http://dx.doi.org/10.1103/PhysRevLett.124.101303}{{\em Phys. Rev.
  Lett.} {\bfseries 124} no.~10, (2020) 101303},
  \href{http://arxiv.org/abs/1910.08638}{{\ttfamily arXiv:1910.08638
  [hep-ex]}}.

\bibitem{Hu:2000ke}
W.~Hu, R.~Barkana, and A.~Gruzinov, ``{Cold and fuzzy dark matter},''
  \href{http://dx.doi.org/10.1103/PhysRevLett.85.1158}{{\em Phys. Rev. Lett.}
  {\bfseries 85} (2000) 1158--1161},
  \href{http://arxiv.org/abs/astro-ph/0003365}{{\ttfamily
  arXiv:astro-ph/0003365}}.

\bibitem{Hui:1998hq}
L.~Hui, ``{Recovery of the Shape of the Mass Power Spectrum from the Lyman
  Alpha Forest},'' \href{http://dx.doi.org/10.1086/307134}{{\em Astrophys. J.}
  {\bfseries 516} (1999) 519--526},
  \href{http://arxiv.org/abs/astro-ph/9807068}{{\ttfamily
  arXiv:astro-ph/9807068}}.

\bibitem{Irsic:2017yje}
V.~Ir\v{s}i\v{c}, M.~Viel, M.~G. Haehnelt, J.~S. Bolton, and G.~D. Becker,
  ``{First constraints on fuzzy dark matter from Lyman-$\alpha$ forest data and
  hydrodynamical simulations},''
  \href{http://dx.doi.org/10.1103/PhysRevLett.119.031302}{{\em Phys. Rev.
  Lett.} {\bfseries 119} no.~3, (2017) 031302},
  \href{http://arxiv.org/abs/1703.04683}{{\ttfamily arXiv:1703.04683
  [astro-ph.CO]}}.

\bibitem{Marsh:2015xka}
D.~J.~E. Marsh, ``{Axion Cosmology},''
  \href{http://dx.doi.org/10.1016/j.physrep.2016.06.005}{{\em Phys. Rept.}
  {\bfseries 643} (2016) 1--79},
  \href{http://arxiv.org/abs/1510.07633}{{\ttfamily arXiv:1510.07633
  [astro-ph.CO]}}.

\bibitem{Grin:2019mub}
D.~Grin, M.~A. Amin, V.~Gluscevic, R.~Hlǒzek, D.~J.~E. Marsh, V.~Poulin,
  C.~Prescod-Weinstein, and T.~L. Smith, ``{Gravitational probes of ultra-light
  axions},'' \href{http://arxiv.org/abs/1904.09003}{{\ttfamily arXiv:1904.09003
  [astro-ph.CO]}}.

\bibitem{Graham:2015ouw}
P.~W. Graham, I.~G. Irastorza, S.~K. Lamoreaux, A.~Lindner, and K.~A. van
  Bibber, ``{Experimental Searches for the Axion and Axion-Like Particles},''
  \href{http://dx.doi.org/10.1146/annurev-nucl-102014-022120}{{\em Ann. Rev.
  Nucl. Part. Sci.} {\bfseries 65} (2015) 485--514},
  \href{http://arxiv.org/abs/1602.00039}{{\ttfamily arXiv:1602.00039
  [hep-ex]}}.

\bibitem{Vilenkin:1982ks}
A.~Vilenkin and A.~E. Everett, ``{Cosmic Strings and Domain Walls in Models
  with Goldstone and PseudoGoldstone Bosons},''
  \href{http://dx.doi.org/10.1103/PhysRevLett.48.1867}{{\em Phys. Rev. Lett.}
  {\bfseries 48} (1982) 1867--1870}.

\bibitem{Peccei:1977hh}
R.~D. Peccei and H.~R. Quinn, ``{CP Conservation in the Presence of
  Instantons},'' \href{http://dx.doi.org/10.1103/PhysRevLett.38.1440}{{\em
  Phys. Rev. Lett.} {\bfseries 38} (1977) 1440--1443}.

\bibitem{Peccei:1977ur}
R.~D. Peccei and H.~R. Quinn, ``{Constraints Imposed by CP Conservation in the
  Presence of Instantons},''
  \href{http://dx.doi.org/10.1103/PhysRevD.16.1791}{{\em Phys. Rev. D}
  {\bfseries 16} (1977) 1791--1797}.

\bibitem{Kibble:1976sj}
T.~W.~B. Kibble, ``{Topology of Cosmic Domains and Strings},''
  \href{http://dx.doi.org/10.1088/0305-4470/9/8/029}{{\em J. Phys. A}
  {\bfseries 9} (1976) 1387--1398}.

\bibitem{Kibble:1980mv}
T.~W.~B. Kibble, ``{Some Implications of a Cosmological Phase Transition},''
  \href{http://dx.doi.org/10.1016/0370-1573(80)90091-5}{{\em Phys. Rept.}
  {\bfseries 67} (1980) 183}.

\bibitem{VilenkinShellard:1994}
A.~Vilenkin and Shellard, {\em {Cosmic Strings and Other Topological Defects}}.
\newblock {Cambridge University Press}, {Cambridge, UK}, 1994.

\bibitem{Kaiser:1984iv}
N.~Kaiser and A.~Stebbins, ``{Microwave Anisotropy Due to Cosmic Strings},''
\href{http://dx.doi.org/10.1038/310391a0}{{\em Nature} {\bfseries 310} (1984)
  391--393}.

\bibitem{Lopez-Eiguren:2017dmc}
A.~Lopez-Eiguren, J.~Lizarraga, M.~Hindmarsh, and J.~Urrestilla, ``{Cosmic
  Microwave Background constraints for global strings and global monopoles},''
  \href{http://dx.doi.org/10.1088/1475-7516/2017/07/026}{{\em JCAP} {\bfseries
  07} (2017) 026}, \href{http://arxiv.org/abs/1705.04154}{{\ttfamily
  arXiv:1705.04154 [astro-ph.CO]}}.

\bibitem{Chang:2019mza}
C.-F. Chang and Y.~Cui, ``{Stochastic Gravitational Wave Background from Global
  Cosmic Strings},'' \href{http://dx.doi.org/10.1016/j.dark.2020.100604}{{\em
  Phys. Dark Univ.} {\bfseries 29} (2020) 100604},
  \href{http://arxiv.org/abs/1910.04781}{{\ttfamily arXiv:1910.04781
  [hep-ph]}}.

\bibitem{Ramberg:2020oct}
N.~Ramberg and L.~Visinelli, ``{The QCD Axion and Gravitational Waves in light
  of NANOGrav results},''
\newblock 12, 2020.
\newblock \href{http://arxiv.org/abs/2012.06882}{{\ttfamily arXiv:2012.06882
  [astro-ph.CO]}}.

\bibitem{Blanco-Pillado:2013qja}
J.~J. Blanco-Pillado, K.~D. Olum, and B.~Shlaer, ``{The number of cosmic string
  loops},'' \href{http://dx.doi.org/10.1103/PhysRevD.89.023512}{{\em Phys. Rev.
  D} {\bfseries 89} no.~2, (2014) 023512},
  \href{http://arxiv.org/abs/1309.6637}{{\ttfamily arXiv:1309.6637
  [astro-ph.CO]}}.

\bibitem{Blanco-Pillado:2017rnf}
J.~J. Blanco-Pillado, K.~D. Olum, and X.~Siemens, ``{New limits on cosmic
  strings from gravitational wave observation},''
  \href{http://dx.doi.org/10.1016/j.physletb.2018.01.050}{{\em Phys. Lett. B}
  {\bfseries 778} (2018) 392--396},
  \href{http://arxiv.org/abs/1709.02434}{{\ttfamily arXiv:1709.02434
  [astro-ph.CO]}}.

\bibitem{Blanco-Pillado:2017oxo}
J.~J. Blanco-Pillado and K.~D. Olum, ``{Stochastic gravitational wave
  background from smoothed cosmic string loops},''
  \href{http://dx.doi.org/10.1103/PhysRevD.96.104046}{{\em Phys. Rev. D}
  {\bfseries 96} no.~10, (2017) 104046},
  \href{http://arxiv.org/abs/1709.02693}{{\ttfamily arXiv:1709.02693
  [astro-ph.CO]}}.

\bibitem{Auclair:2019wcv}
P.~Auclair {\em et~al.}, ``{Probing the gravitational wave background from
  cosmic strings with LISA},''
  \href{http://dx.doi.org/10.1088/1475-7516/2020/04/034}{{\em JCAP} {\bfseries
  04} (2020) 034}, \href{http://arxiv.org/abs/1909.00819}{{\ttfamily
  arXiv:1909.00819 [astro-ph.CO]}}.

\bibitem{Gorghetto:2021fsn}
M.~Gorghetto, E.~Hardy, and H.~Nicolaescu, ``{Observing Invisible Axions with
  Gravitational Waves},'' \href{http://arxiv.org/abs/2101.11007}{{\ttfamily
  arXiv:2101.11007 [hep-ph]}}.

\bibitem{Figueroa:2020lvo}
D.~G. Figueroa, M.~Hindmarsh, J.~Lizarraga, and J.~Urrestilla, ``{Irreducible
  background of gravitational waves from a cosmic defect network: update and
  comparison of numerical techniques},''
  \href{http://dx.doi.org/10.1103/PhysRevD.102.103516}{{\em Phys. Rev. D}
  {\bfseries 102} no.~10, (2020) 103516},
  \href{http://arxiv.org/abs/2007.03337}{{\ttfamily arXiv:2007.03337
  [astro-ph.CO]}}.

\bibitem{Gelmini:2021yzu}
G.~B. Gelmini, A.~Simpson, and E.~Vitagliano, ``{Gravitational waves from
  axion-like particle cosmic string-wall networks},''
  \href{http://arxiv.org/abs/2103.07625}{{\ttfamily arXiv:2103.07625
  [hep-ph]}}.

\bibitem{Fukuda:2020kym}
H.~Fukuda, A.~V. Manohar, H.~Murayama, and O.~Telem, ``{Axion strings are
  superconducting},'' \href{http://arxiv.org/abs/2010.02763}{{\ttfamily
  arXiv:2010.02763 [hep-ph]}}.

\bibitem{Abe:2020ure}
Y.~Abe, Y.~Hamada, and K.~Yoshioka, ``{Electroweak axion string and
  superconductivity},'' \href{http://arxiv.org/abs/2010.02834}{{\ttfamily
  arXiv:2010.02834 [hep-ph]}}.

\bibitem{Agrawal:2020euj}
P.~Agrawal, A.~Hook, J.~Huang, and G.~Marques-Tavares, ``{Axion string
  signatures II: A cosmological plasma collider},''
  \href{http://arxiv.org/abs/2010.15848}{{\ttfamily arXiv:2010.15848
  [hep-ph]}}.

\bibitem{Carroll:1989vb}
S.~M. Carroll, G.~B. Field, and R.~Jackiw, ``{Limits on a Lorentz and Parity
  Violating Modification of Electrodynamics},''
  \href{http://dx.doi.org/10.1103/PhysRevD.41.1231}{{\em Phys. Rev. D}
  {\bfseries 41} (1990) 1231}.

\bibitem{Carroll:1991zs}
S.~M. Carroll and G.~B. Field, ``{The Einstein equivalence principle and the
  polarization of radio galaxies},''
  \href{http://dx.doi.org/10.1103/PhysRevD.43.3789}{{\em Phys. Rev. D}
  {\bfseries 43} (1991) 3789}.

\bibitem{Harari:1992ea}
D.~Harari and P.~Sikivie, ``{Effects of a Nambu-Goldstone boson on the
  polarization of radio galaxies and the cosmic microwave background},''
  \href{http://dx.doi.org/10.1016/0370-2693(92)91363-E}{{\em Phys. Lett. B}
  {\bfseries 289} (1992) 67--72}.

\bibitem{Carroll:1998bd}
S.~M. Carroll, ``{Quintessence and the rest of the world},''
  \href{http://dx.doi.org/10.1063/1.59405}{{\em AIP Conf. Proc.} {\bfseries
  478} no.~1, (1999) 291--294}.

\bibitem{Lue:1998mq}
A.~Lue, L.-M. Wang, and M.~Kamionkowski, ``{Cosmological signature of new
  parity violating interactions},''
  \href{http://dx.doi.org/10.1103/PhysRevLett.83.1506}{{\em Phys. Rev. Lett.}
  {\bfseries 83} (1999) 1506--1509},
  \href{http://arxiv.org/abs/astro-ph/9812088}{{\ttfamily
  arXiv:astro-ph/9812088}}.

\bibitem{Liu:2006uh}
G.-C. Liu, S.~Lee, and K.-W. Ng, ``{Effect on cosmic microwave background
  polarization of coupling of quintessence to pseudoscalar formed from the
  electromagnetic field and its dual},''
  \href{http://dx.doi.org/10.1103/PhysRevLett.97.161303}{{\em Phys. Rev. Lett.}
  {\bfseries 97} (2006) 161303},
  \href{http://arxiv.org/abs/astro-ph/0606248}{{\ttfamily
  arXiv:astro-ph/0606248}}.

\bibitem{Gluscevic:2012me}
V.~Gluscevic, D.~Hanson, M.~Kamionkowski, and C.~M. Hirata, ``{First CMB
  Constraints on Direction-Dependent Cosmological Birefringence from WMAP-7},''
  \href{http://dx.doi.org/10.1103/PhysRevD.86.103529}{{\em Phys. Rev. D}
  {\bfseries 86} (2012) 103529},
  \href{http://arxiv.org/abs/1206.5546}{{\ttfamily arXiv:1206.5546
  [astro-ph.CO]}}.

\bibitem{Contreras:2017sgi}
D.~Contreras, P.~Boubel, and D.~Scott, ``{Constraints on direction-dependent
  cosmic birefringence from Planck polarization data},''
  \href{http://dx.doi.org/10.1088/1475-7516/2017/12/046}{{\em JCAP} {\bfseries
  12} (2017) 046}, \href{http://arxiv.org/abs/1705.06387}{{\ttfamily
  arXiv:1705.06387 [astro-ph.CO]}}.

\bibitem{Abazajian:2016yjj}
{\bfseries CMB-S4} Collaboration, K.~N. Abazajian {\em et~al.}, ``{CMB-S4
  Science Book, First Edition},''
  \href{http://arxiv.org/abs/1610.02743}{{\ttfamily arXiv:1610.02743
  [astro-ph.CO]}}.

\bibitem{Minami:2020odp}
Y.~Minami and E.~Komatsu, ``{New Extraction of the Cosmic Birefringence from
  the Planck 2018 Polarization Data},''
  \href{http://dx.doi.org/10.1103/PhysRevLett.125.221301}{{\em Phys. Rev.
  Lett.} {\bfseries 125} no.~22, (2020) 221301},
  \href{http://arxiv.org/abs/2011.11254}{{\ttfamily arXiv:2011.11254
  [astro-ph.CO]}}.

\bibitem{Aghanim:2018eyx}
{\bfseries Planck} Collaboration, N.~Aghanim {\em et~al.}, ``{Planck 2018
  results. VI. Cosmological parameters},''
\href{http://arxiv.org/abs/1807.06209}{{\ttfamily arXiv:1807.06209
  [astro-ph.CO]}}.

\bibitem{Takahashi:2020tqv}
F.~Takahashi and W.~Yin, ``{Kilobyte Cosmic Birefringence from ALP Domain
  Walls},'' \href{http://arxiv.org/abs/2012.11576}{{\ttfamily arXiv:2012.11576
  [hep-ph]}}.

\bibitem{Fujita:2020ecn}
T.~Fujita, K.~Murai, H.~Nakatsuka, and S.~Tsujikawa, ``{Detection of isotropic
  cosmic birefringence and its implications for axion-like particles including
  dark energy},'' \href{http://dx.doi.org/10.1103/PhysRevD.103.043509}{{\em
  Phys. Rev. D} {\bfseries 103} no.~4, (2021) 043509},
  \href{http://arxiv.org/abs/2011.11894}{{\ttfamily arXiv:2011.11894
  [astro-ph.CO]}}.

\bibitem{Nakagawa:2021nme}
S.~Nakagawa, F.~Takahashi, and M.~Yamada, ``{Cosmic Birefringence Triggered by
  Dark Matter Domination},'' \href{http://arxiv.org/abs/2103.08153}{{\ttfamily
  arXiv:2103.08153 [hep-ph]}}.

\bibitem{Agrawal:2019lkr}
P.~Agrawal, A.~Hook, and J.~Huang, ``{A CMB Millikan experiment with cosmic
  axiverse strings},'' \href{http://dx.doi.org/10.1007/JHEP07(2020)138}{{\em
  JHEP} {\bfseries 07} (2020) 138},
  \href{http://arxiv.org/abs/1912.02823}{{\ttfamily arXiv:1912.02823
  [astro-ph.CO]}}.

\bibitem{Yamaguchi:1998gx}
M.~Yamaguchi, M.~Kawasaki, and J.~Yokoyama, ``{Evolution of axionic strings and
  spectrum of axions radiated from them},''
  \href{http://dx.doi.org/10.1103/PhysRevLett.82.4578}{{\em Phys. Rev. Lett.}
  {\bfseries 82} (1999) 4578--4581},
  \href{http://arxiv.org/abs/hep-ph/9811311}{{\ttfamily arXiv:hep-ph/9811311}}.

\bibitem{Yamaguchi:2002sh}
M.~Yamaguchi and J.~Yokoyama, ``{Quantitative evolution of global strings from
  the Lagrangian view point},''
  \href{http://dx.doi.org/10.1103/PhysRevD.67.103514}{{\em Phys. Rev. D}
  {\bfseries 67} (2003) 103514},
  \href{http://arxiv.org/abs/hep-ph/0210343}{{\ttfamily arXiv:hep-ph/0210343}}.

\bibitem{Hiramatsu:2010yu}
T.~Hiramatsu, M.~Kawasaki, T.~Sekiguchi, M.~Yamaguchi, and J.~Yokoyama,
  ``{Improved estimation of radiated axions from cosmological axionic
  strings},'' \href{http://dx.doi.org/10.1103/PhysRevD.83.123531}{{\em Phys.
  Rev. D} {\bfseries 83} (2011) 123531},
  \href{http://arxiv.org/abs/1012.5502}{{\ttfamily arXiv:1012.5502 [hep-ph]}}.

\bibitem{Hiramatsu:2012gg}
T.~Hiramatsu, M.~Kawasaki, K.~Saikawa, and T.~Sekiguchi, ``{Production of dark
  matter axions from collapse of string-wall systems},''
  \href{http://dx.doi.org/10.1103/PhysRevD.85.105020}{{\em Phys. Rev. D}
  {\bfseries 85} (2012) 105020},
  \href{http://arxiv.org/abs/1202.5851}{{\ttfamily arXiv:1202.5851 [hep-ph]}}.
  [Erratum: Phys.Rev.D 86, 089902 (2012)].

\bibitem{Kawasaki:2014sqa}
M.~Kawasaki, K.~Saikawa, and T.~Sekiguchi, ``{Axion dark matter from
  topological defects},''
  \href{http://dx.doi.org/10.1103/PhysRevD.91.065014}{{\em Phys. Rev. D}
  {\bfseries 91} no.~6, (2015) 065014},
  \href{http://arxiv.org/abs/1412.0789}{{\ttfamily arXiv:1412.0789 [hep-ph]}}.

\bibitem{Hindmarsh:2019csc}
M.~Hindmarsh, J.~Lizarraga, A.~Lopez-Eiguren, and J.~Urrestilla, ``{Scaling
  Density of Axion Strings},''
  \href{http://dx.doi.org/10.1103/PhysRevLett.124.021301}{{\em Phys. Rev.
  Lett.} {\bfseries 124} no.~2, (2020) 021301},
  \href{http://arxiv.org/abs/1908.03522}{{\ttfamily arXiv:1908.03522
  [astro-ph.CO]}}.

\bibitem{Hindmarsh:2021vih}
M.~Hindmarsh, J.~Lizarraga, A.~Lopez-Eiguren, and J.~Urrestilla, ``{Approach to
  scaling in axion string networks},''
  \href{http://arxiv.org/abs/2102.07723}{{\ttfamily arXiv:2102.07723
  [astro-ph.CO]}}.

\bibitem{Gorghetto:2018myk}
M.~Gorghetto, E.~Hardy, and G.~Villadoro, ``{Axions from Strings: the
  Attractive Solution},'' \href{http://dx.doi.org/10.1007/JHEP07(2018)151}{{\em
  JHEP} {\bfseries 07} (2018) 151},
  \href{http://arxiv.org/abs/1806.04677}{{\ttfamily arXiv:1806.04677
  [hep-ph]}}.

\bibitem{Vaquero:2018tib}
A.~Vaquero, J.~Redondo, and J.~Stadler, ``{Early seeds of axion
  miniclusters},'' \href{http://dx.doi.org/10.1088/1475-7516/2019/04/012}{{\em
  JCAP} {\bfseries 04} (2019) 012},
  \href{http://arxiv.org/abs/1809.09241}{{\ttfamily arXiv:1809.09241
  [astro-ph.CO]}}.

\bibitem{Kawasaki:2018bzv}
M.~Kawasaki, T.~Sekiguchi, M.~Yamaguchi, and J.~Yokoyama, ``{Long-term dynamics
  of cosmological axion strings},''
  \href{http://dx.doi.org/10.1093/ptep/pty098}{{\em PTEP} {\bfseries 2018}
  no.~9, (2018) 091E01}, \href{http://arxiv.org/abs/1806.05566}{{\ttfamily
  arXiv:1806.05566 [hep-ph]}}.

\bibitem{Martins:2018dqg}
C.~J. A.~P. Martins, ``{Scaling properties of cosmological axion strings},''
  \href{http://dx.doi.org/10.1016/j.physletb.2018.11.031}{{\em Phys. Lett. B}
  {\bfseries 788} (2019) 147--151},
  \href{http://arxiv.org/abs/1811.12678}{{\ttfamily arXiv:1811.12678
  [astro-ph.CO]}}.

\bibitem{Buschmann:2019icd}
M.~Buschmann, J.~W. Foster, and B.~R. Safdi, ``{Early-Universe Simulations of
  the Cosmological Axion},''
  \href{http://dx.doi.org/10.1103/PhysRevLett.124.161103}{{\em Phys. Rev.
  Lett.} {\bfseries 124} no.~16, (2020) 161103},
  \href{http://arxiv.org/abs/1906.00967}{{\ttfamily arXiv:1906.00967
  [astro-ph.CO]}}.

\bibitem{Klaer:2019fxc}
V.~B. Klaer and G.~D. Moore, ``{Global cosmic string networks as a function of
  tension},'' \href{http://dx.doi.org/10.1088/1475-7516/2020/06/021}{{\em JCAP}
  {\bfseries 06} (2020) 021}, \href{http://arxiv.org/abs/1912.08058}{{\ttfamily
  arXiv:1912.08058 [hep-ph]}}.

\bibitem{Gorghetto:2020qws}
M.~Gorghetto, E.~Hardy, and G.~Villadoro, ``{More Axions from Strings},''
  \href{http://arxiv.org/abs/2007.04990}{{\ttfamily arXiv:2007.04990
  [hep-ph]}}.

\bibitem{Fedderke:2019ajk}
M.~A. Fedderke, P.~W. Graham, and S.~Rajendran, ``{Axion Dark Matter Detection
  with CMB Polarization},''
  \href{http://dx.doi.org/10.1103/PhysRevD.100.015040}{{\em Phys. Rev. D}
  {\bfseries 100} no.~1, (2019) 015040},
  \href{http://arxiv.org/abs/1903.02666}{{\ttfamily arXiv:1903.02666
  [astro-ph.CO]}}.

\bibitem{Farina:2016tgd}
M.~Farina, D.~Pappadopulo, F.~Rompineve, and A.~Tesi, ``{The photo-philic QCD
  axion},'' \href{http://dx.doi.org/10.1007/JHEP01(2017)095}{{\em JHEP}
  {\bfseries 01} (2017) 095}, \href{http://arxiv.org/abs/1611.09855}{{\ttfamily
  arXiv:1611.09855 [hep-ph]}}.

\bibitem{Chang:1998tb}
S.~Chang, C.~Hagmann, and P.~Sikivie, ``{Studies of the motion and decay of
  axion walls bounded by strings},''
  \href{http://dx.doi.org/10.1103/PhysRevD.59.023505}{{\em Phys. Rev. D}
  {\bfseries 59} (1999) 023505},
  \href{http://arxiv.org/abs/hep-ph/9807374}{{\ttfamily arXiv:hep-ph/9807374}}.

\bibitem{Hiramatsu:2012sc}
T.~Hiramatsu, M.~Kawasaki, K.~Saikawa, and T.~Sekiguchi, ``{Axion cosmology
  with long-lived domain walls},''
  \href{http://dx.doi.org/10.1088/1475-7516/2013/01/001}{{\em JCAP} {\bfseries
  01} (2013) 001}, \href{http://arxiv.org/abs/1207.3166}{{\ttfamily
  arXiv:1207.3166 [hep-ph]}}.

\bibitem{Hiramatsu:2010yn}
T.~Hiramatsu, M.~Kawasaki, and K.~Saikawa, ``{Evolution of String-Wall Networks
  and Axionic Domain Wall Problem},''
  \href{http://dx.doi.org/10.1088/1475-7516/2011/08/030}{{\em JCAP} {\bfseries
  08} (2011) 030}, \href{http://arxiv.org/abs/1012.4558}{{\ttfamily
  arXiv:1012.4558 [astro-ph.CO]}}.

\bibitem{Sikivie:1982qv}
P.~Sikivie, ``{Of Axions, Domain Walls and the Early Universe},''
  \href{http://dx.doi.org/10.1103/PhysRevLett.48.1156}{{\em Phys. Rev. Lett.}
  {\bfseries 48} (1982) 1156--1159}.

\bibitem{Pogosian:2019jbt}
L.~Pogosian, M.~Shimon, M.~Mewes, and B.~Keating, ``{Future CMB constraints on
  cosmic birefringence and implications for fundamental physics},''
  \href{http://dx.doi.org/10.1103/PhysRevD.100.023507}{{\em Phys. Rev. D}
  {\bfseries 100} no.~2, (2019) 023507},
  \href{http://arxiv.org/abs/1904.07855}{{\ttfamily arXiv:1904.07855
  [astro-ph.CO]}}.

\bibitem{Namikawa:2020ffr}
T.~Namikawa {\em et~al.}, ``{Atacama Cosmology Telescope: Constraints on cosmic
  birefringence},'' \href{http://dx.doi.org/10.1103/PhysRevD.101.083527}{{\em
  Phys. Rev. D} {\bfseries 101} no.~8, (2020) 083527},
  \href{http://arxiv.org/abs/2001.10465}{{\ttfamily arXiv:2001.10465
  [astro-ph.CO]}}.

\bibitem{Bianchini:2020osu}
{\bfseries SPT} Collaboration, F.~Bianchini {\em et~al.}, ``{Searching for
  Anisotropic Cosmic Birefringence with Polarization Data from SPTpol},''
  \href{http://dx.doi.org/10.1103/PhysRevD.102.083504}{{\em Phys. Rev. D}
  {\bfseries 102} no.~8, (2020) 083504},
  \href{http://arxiv.org/abs/2006.08061}{{\ttfamily arXiv:2006.08061
  [astro-ph.CO]}}.

\bibitem{Array:2017rlf}
{\bfseries BICEP2, Keck Arrary} Collaboration, P.~A.~R. Ade {\em et~al.},
  ``{BICEP2 / Keck Array IX: New bounds on anisotropies of CMB polarization
  rotation and implications for axionlike particles and primordial magnetic
  fields},'' \href{http://dx.doi.org/10.1103/PhysRevD.96.102003}{{\em Phys.
  Rev. D} {\bfseries 96} no.~10, (2017) 102003},
  \href{http://arxiv.org/abs/1705.02523}{{\ttfamily arXiv:1705.02523
  [astro-ph.CO]}}.

\bibitem{Jackson:1999}
J.~D. Jackson, {\em {Classical Electrodynamics}}.
\newblock {John Wiley \& Sons, Inc.}, {New York, NY}, 1999.

\bibitem{Saurabh:2020pqe}
A.~Saurabh, T.~Vachaspati, and L.~Pogosian, ``{Decay of Cosmic Global String
  Loops},'' \href{http://dx.doi.org/10.1103/PhysRevD.101.083522}{{\em Phys.
  Rev. D} {\bfseries 101} no.~8, (2020) 083522},
  \href{http://arxiv.org/abs/2001.01030}{{\ttfamily arXiv:2001.01030
  [hep-ph]}}.

\end{thebibliography}\endgroup

\end{document}